%% file: main.tex
\documentclass[twocolumn,prx,superscriptaddress,longbibliography, floatfix, aps]{revtex4-2}

\usepackage{import}

\usepackage{ulem}
\usepackage{array}[=2016-10-06]
\usepackage{graphicx}% Include figure files
\usepackage{physics}% Include physics
\usepackage{dcolumn}% Align table columns on decimal point
\usepackage{bm}
\usepackage{url}

\usepackage{cmap} 
\usepackage[T2A]{fontenc} 
\usepackage[utf8]{inputenc} 

\usepackage{multirow}
\usepackage[warn]{mathtext}
\usepackage{amssymb}

\usepackage{caption}
\usepackage{subcaption}

\usepackage{textcomp} 
\usepackage{indentfirst} 
\usepackage{amsmath} 
\usepackage{graphicx}

\DeclareGraphicsExtensions{.pdf,.png,.jpg}
\usepackage{pgfplots}
\pgfplotsset{compat=newest}

\usepackage{algpseudocode}

\usepackage{adjustbox}
\usepackage{tabularx}
\usepackage{booktabs}
\newlength{\tablewidth}
\renewcommand{\bibname}{references}

\usepackage[nottoc]{tocbibind}
\usepackage{mathtools}
\usepackage{dsfont}
\usepackage{bbm} % nice 1

\usepackage{hyperref}
\hypersetup{
    colorlinks=true,
    linkcolor=blue,
    filecolor=black,      
    urlcolor=blue,
    citecolor= violet
}
\usepackage{xcolor}
\definecolor{C0}{HTML}{4C72B0}
\definecolor{C1}{HTML}{DD8452}
\definecolor{C2}{HTML}{55A868}
\definecolor{C3}{HTML}{C44E52}
\definecolor{darkgreen}{rgb}{0.1, 0.50, 0.32}

\newcommand{\first}[1]{\textcolor{black}{#1}}
\newcommand{\forth}[1]{\textcolor{black}{#1}}
\newcommand{\second}[1]{\textcolor{black}{#1}}
\newcommand{\minor}[1]{\textcolor{black}{#1}}



\newcolumntype{Y}{>{\centering\arraybackslash}X}

\makeatletter
\def\@fnsymbol#1{\ensuremath{\ifcase#1\or *\or \dagger\or \ddagger\or
   \mathsection\or \mathparagraph\or \|\or **\or \dagger\dagger
   \or \ddagger\ddagger \else\@ctrerr\fi}}
\makeatother

\begin{document}

\title{Interaction-Resilient Scalable Fluxonium Architecture with All-Microwave Gates}

\author{Andrei A. Kugut}
\email[Contact author: ]{Andrei.Kugut@ist.ac.at}
\altaffiliation[\\ Present address: ]{Institute of Science and Technology Austria, 3400 Klosterneuburg, Austria}
\affiliation{National University of Science and Technology ``MISIS'', 119049 Moscow, Russia}
\affiliation{Russian Quantum Center, 143025 Skolkovo, Moscow, Russia}

\author{Grigoriy S. Mazhorin}
% \email{mazhorin.gs@phystech.edu}
\affiliation{National University of Science and Technology ``MISIS'', 119049 Moscow, Russia}

\author{Ilya A. Simakov}
% \email{simakov.ia@phystech.edu}
\affiliation{National University of Science and Technology ``MISIS'', 119049 Moscow, Russia}

\date{\today}% It is always \today, today,

\begin{abstract}

Fluxonium qubits demonstrate exceptional potential for quantum processing; yet, realizing scalable architectures using them remains challenging. Here, we propose a scalable fluxonium-based square-grid design with $\sim63$ ns controlled-Z (CZ) gates, achieving coherent errors below $10^{-4}$, activated via microwave-driven transmon couplers. Furthermore, the architecture natively supports  $\sim70$ ns CZZ gates -- three-qubit operations composed of two CZ gates sharing a common qubit -- which directly implement two-qubit parity checks central to quantum error-correction protocols and reduce the incoherent error by $\sim 35\%$ compared to sequential CZs. A central difficulty in building large-scale systems with all-microwave gates and, therefore, static couplings, is suppressing parasitic interactions that extend beyond nearest neighbors to include next-nearest elements. We address this challenge by introducing several design strategies, including frequency allocation for both qubits and couplers, localization of coupler wavefunctions, and a differential oscillator that suppresses residual long-range interactions. Together, these principles establish an interaction-resilient platform for large-scale fluxonium processors and can be adapted to a wide range of fluxonium layouts.

\end{abstract}

\maketitle

% \tableofcontents
\section{Introduction}
\label{section:intro}
\import{sections/}{introduction}

\section{Architecture}
\label{section:architecture}
\import{sections/}{architecture}

\section{CZ gate principles and wavefunction localization}
\label{section:alternating_couplers}
\import{sections/}{alternating_couplers}

\section{Long-range interactions}
\label{section:parasitic_interactions}
\import{sections/}{parasitic_interactions}

\section{CZZ gates}
\label{section:CZZ_gates}
\import{sections/}{CZZ}

\section{Gate performance}
\label{section:performance}
\import{sections/}{performance}

\section{Conclusion and outlook}
\import{sections/}{conclusion}

% \section*{Data availability}

% The data that support the findings of this study are available from the corresponding author upon reasonable request.

\section*{Acknowledgments}
The authors are grateful to Alexey Ustinov, Ilya Moskalenko and Ilya Besedin for valuable comments on the manuscript.
The work was supported by the Ministry of Science and Higher Education of the Russian Federation in the framework of the Program of Strategic Academic Leadership “Priority 2030” (MISIS Strategic Technology Project Quantum Internet).

\appendix

\section{Differential fluxonium}
\label{appendix:differential_fluxonium}
\import{appendixes/}{differential_fluxonium}

\section{Architecture numerical modeling}
\label{appendix:model_static}
\import{appendixes/}{model_static}

\section{Gate simulation}
\label{appendix:model_dynamic}
\import{appendixes/}{model_dynamic}

\section{Coupler decoherence impact}
\label{appendix:coupler_decoherence_impact}
\import{appendixes/}{coupler_decoherence_impact}

\section{Qubit-Qubit interaction}
\label{appendix:qq_analysis}
\import{appendixes/}{qq_analysis}

\section{Coupler-Spectator ZZ interaction}
\label{appendix:ZZ_CS}
\import{appendixes/}{ZZ_CS}

\section{Virtual transitions and time-independent perturbation theory}
\label{appendix:virtual_transitions}
\import{appendixes/}{virtual_transitions}

\section{Leakage research}
\label{appendix:leakage_research}
\import{appendixes/}{leakage}

\section{Parameters of the architecture}
\label{appendix:parameters}
\import{appendixes/}{parameters}

\section{Fabrication robustness}
\label{appendix:fabrication_robustness}
\import{appendixes/}{fabrication_robustness}

\renewcommand{\bibname}{Reference}
\normalem{}
\bibliographystyle{apsrev4-2}
\bibliography{ref}

\end{document}

%% file: sections/introduction.tex
Superconducting quantum computing has seen steady and reliable progress in recent years, marked by the realization of large and increasingly coherent processors. Key advances include high-fidelity multi-qubit control within scalable architectures capable of supporting quantum algorithms \cite{Arute_2019, Strong_Wu_2021, morvan2024phase, gao2025establishing, google2025observation} and error-correction protocols \cite{Krinner_2022, PhysRevLett.129.030501, google2023suppressing, acharya2024quantumerrorcorrectionsurface, erhard2021entangling, marques2022logical, besedin2025realizinglatticesurgerydistancethree}.
Practically all of these milestones have been achieved using transmon-based devices.

Since their introduction \cite{Koch_Charge_2007}, transmon qubits have become the workhorse of large-scale superconducting quantum processors, owing to their intrinsic protection against charge noise, relatively simple fabrication, and mature control techniques.
The transmon circuit implements a weakly anharmonic oscillator in which the two lowest energy levels encode the qubit. 
The interaction with higher-lying states of the transmons is ubiquitously used to execute fast CZ and fSim gates \cite{Rol2019,Foxen2020}, but also introduces phase errors, residual ZZ coupling, and contributes to limited readout fidelity through ionization \cite{Dumas2024}. Moreover, transmons suffer from crosstalk and frequency crowding in multi-qubit systems, placing constraints on achievable gate fidelity and overall processor performance. Consequently, alternative qubit modalities have been sought. Among them fluxonium \cite{Manucharyan_2009} has emerged as a particularly promising candidate, offering enhanced resilience to noise and greater control flexibility \cite{Bao_2021, Ficheux2021, moskalenko.tunable_FF_gate.2021, moskalenko.gate.2022}.

Fluxonium qubits, due to their large anharmonicity and long coherence times, have already demonstrated single-qubit gate fidelities above 0.9999 \cite{Somoroff_Millisecond_2023, rower.CoolFluxoniumSQGates.2024} and two-qubit gate fidelities exceeding 0.999 \cite{MIT.FTF, Lin.cool_CNOT_on_fluxoniums}. These experimental advances, combined with the rich spectral structure of fluxonium, have inspired the development of various quantum processor blueprints \cite{zhao.scalable_FTF_architecture.2025, zhao.scalable_FTF_architecture_2.2025, Nguyen.Scaling, Heunisch.hybrid_FT_scaling_2025, Ciani.Scaling, chan2026system}. A particularly promising approach 
% for such schemes 
combines two key features: coupling computational qubits via transmon couplers, which inherently suppresses parasitic qubit-qubit interactions \cite{MIT.FTF, lange.transmon_coupler_research.2025, mazhorin2026nativecczgatefluxonium, zhao.scalable_FTF_architecture_2.2025, huang2026explorationfluxoniumparameterscapacitive}, and implementing fast entangling gates by near-resonant driving of non-computational states \cite{MIT.FTF, zhao.scalable_FTF_architecture.2025, singh.FTF_CZ_experimental.2025, Rosenfeld.FOF, Heunisch.hybrid_FT_scaling_2025, Simakov.FFF, mazhorin2026nativecczgatefluxonium}.

A known drawback of this approach is the parasitic ZZ interaction between the driven non-computational state and spectator qubit, leading to coherent two-qubit errors \cite{Heunisch.hybrid_FT_scaling_2025, zwanenburg2026crosstalkmultiqubitfluxoniumarchitectures}. Proposed solutions \cite{zhao.scalable_FTF_architecture.2025, Heunisch.hybrid_FT_scaling_2025, zwanenburg2026crosstalkmultiqubitfluxoniumarchitectures} involve flux-tuning the transmon couplers to partially or completely decouple the targeted qubits from their spectators. However, this method risks introducing flux crosstalk that can detune neighboring fluxoniums from their flux sweet spots, worsening dephasing rates and thereby undermining one of the key advantages of fluxonium.

In this work, we present a scalable fluxonium-based architecture with transmon couplers that addresses the challenge of parasitic interactions while maintaining fixed strong couplings and enabling fast, high-fidelity CZ gates. 
Our design employs a systematic arrangement of four distinct coupler types to minimize parasitic interactions involving leveraged non-computational states. To further suppress long-range effects, we use differential fluxoniums that intrinsically reduce direct parasitic couplings and introduce a differential oscillator, an additional circuit element that provides independent control over residual interactions \cite{mundada.ZZ_suppression_via_bus.2023}.

\begin{table*}[!ht]
    \centering
    \footnotesize
    \begin{tabularx}{\linewidth}{l
    >{\centering\arraybackslash}p{2.6cm}
    >{\centering\arraybackslash}p{3.6cm}
    >{\centering\arraybackslash}p{2.7cm}
    >{\centering\arraybackslash}X
    >{\centering\arraybackslash}p{3.2cm}}
        \toprule
        \toprule
        Proposed by & Coupling & Two-qubit gate & Multi-qubit gates & Nearest neighbors $ZZ$ interaction & Scaling approach \\
        \midrule
            Nguyen \textit{et al.}~\cite{Nguyen.Scaling}
            & Fixed \quad\quad\quad\quad\textit{(direct capacitive and inductive)}
            & Cross-resonant CNOT or AC-Stark CZ \quad \quad $\tau\sim\!200\,\text{ns}$, $\varepsilon<10^{-5}$
            & ---
            & complete cancellation
            & Frequency allocation, multipath coupling \textit{(static long-range interactions are \quad\quad\quad not studied)} \\
        \midrule
            Huang, Andersen~\cite{huang2026explorationfluxoniumparameterscapacitive}
            & Fixed \quad\quad\quad\quad\textit{(direct capacitive)}
            & Cross-resonant CNOT $\tau\sim\!163\,\text{ns}$, $\varepsilon<10^{-4}$
            & ---
            & $|\zeta| \sim 50\,\text{kHz}$
            & Frequency allocation \textit{(static long-range interactions are \quad\quad\quad not studied)} \\
        \midrule
        Zhao \textit{et al.}~\cite{zhao.scalable_FTF_architecture.2025}
            & Tunable \quad\quad\textit{(tunable transmon or\quad\quad\quad\quad\quad double transmon)}
            & Microwave-activated CZ $\tau\sim\!50\,\text{ns}$, $\varepsilon<10^{-4}$
            & ---
            & $|\zeta| < 1\,\text{kHz}$
            & DC-flux induced decoupling, \quad \quad frequency allocation \\
        \midrule
        Zhao, Xu, Xue~\cite{zhao.scalable_FTF_architecture_2.2025}
            & Tunable\quad\quad\quad \textit{(tunable transmon)}
            & Microwave-activated CZ $\tau\sim\!50\,\text{ns}$, $\varepsilon<10^{-4}$
            & C$^{\otimes 2}$Z, C$^{\otimes 3}$Z, C$^{\otimes 4}$Z, $\tau\sim\!100/250/300\,\text{ns}$ \quad \quad for $\varepsilon\sim10^{-3}$
            & ---
            & DC-flux induced decoupling \quad\quad\quad\quad \textit{(not analyzed)}, \quad \quad \quad frequency allocation \\
        \midrule
        \textbf{This work}
            & Fixed \quad\quad\quad\textit{(tunable transmon)}
            & Microwave-activated CZ $\tau\sim63\,\text{ns}$, $\varepsilon<10^{-4}$
            & CZZ \quad\quad \quad$\tau\sim70\,\text{ns}$, $\varepsilon\sim10^{-4}$
            & $|\zeta| < 5\,\text{kHz}$
            & Frequency allocation, additional passive elements and connections\\
        \bottomrule
        \bottomrule
    \end{tabularx}
    \caption{Comparison of fluxonium-based architectures. Each gate is characterized by its coherent error $\varepsilon$ and duration $\tau$, excluding flux-induced coupler activation time where applicable.}
    \label{tab:architecture_comparison}
\end{table*}

Together, these design strategies establish an interaction-resilient architecture in which all qubits remain at their flux sweet spots throughout operation. Moreover, this architecture naturally supports three-qubit CZZ operations with durations and coherent error rates comparable to those of conventional single-CZ gates. This capability directly facilitates quantum algorithms and error-correction protocols, such as the surface code \cite{tasler.CZZ_for_surface.2025}, on present-day noisy devices.

\first{In Table~\ref{tab:architecture_comparison}, we compare the main properties of our architecture with alternative blueprints for fluxonium square-grid layouts. The table also lists a \textit{scaling approach} -- a brief description of the features that help suppress long-range parasitic interactions. Such features are crucial for guaranteeing the scalability of a quantum processor, yet they are sometimes overlooked in the literature.}

%% file: sections/architecture.tex
\begin{figure*}[!htb]
    \center{\includegraphics[width=\linewidth]{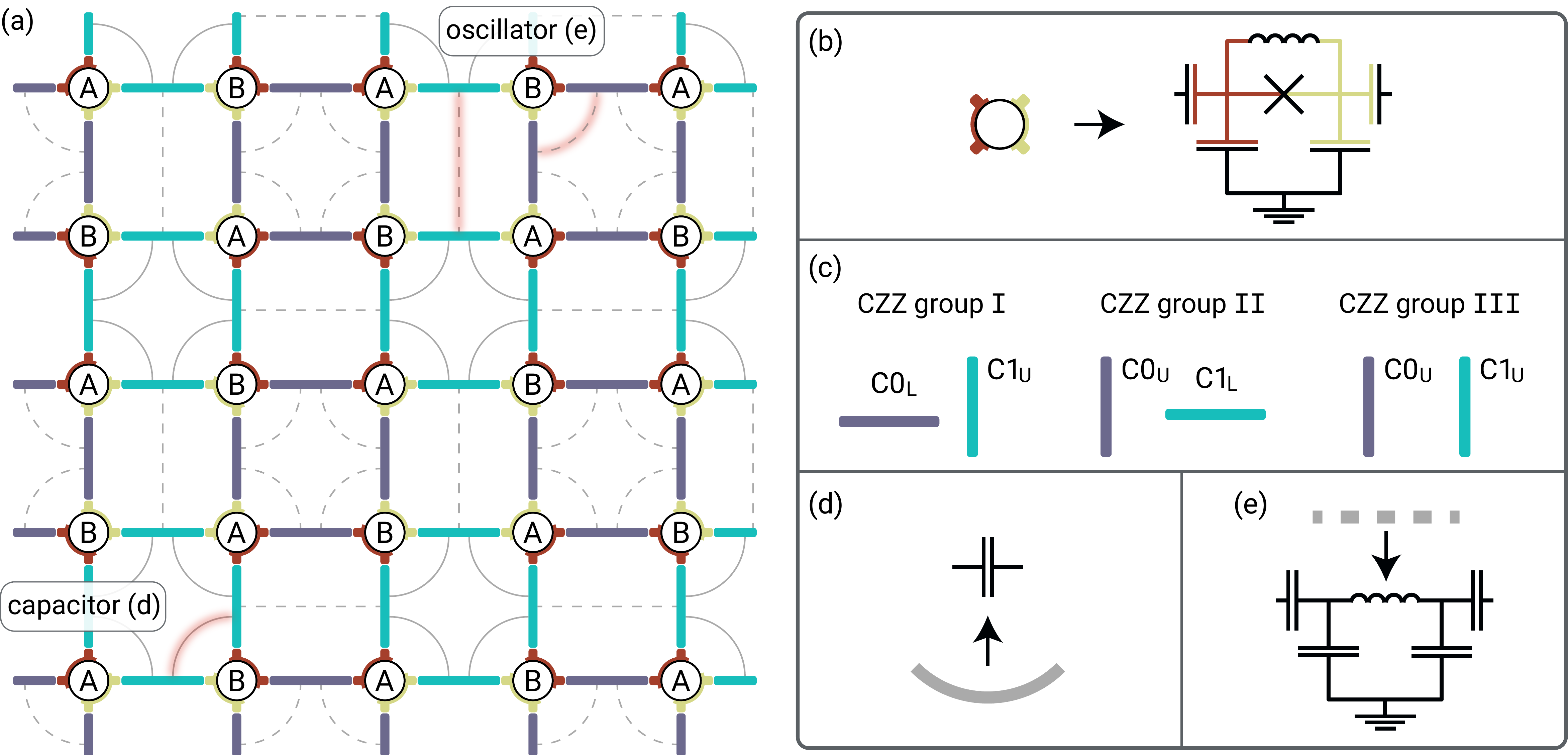}}
    \caption{(a) Schematic of the fluxonium-based architecture. The circles represent fluxonium qubits, and the colored lines indicate transmon couplers. (b) Electrical circuit of the differential fluxonium. The colored islands highlight the connection edges within the scheme. (c) Four types of transmon couplers. Two principal types, (0) and (1), are shown in different colors, while two subtypes, (U) and (L), correspond to vertical and horizontal orientations. Couplers are further divided into three groups that enable CZZ gates. (d) Direct capacitive couplings. (e) Electrical circuit of the differential oscillators, which mediate coupler–coupler interactions and suppresses long-range parasitic coupling.}
    \label{fig:architecture_scheme}
\end{figure*}

The processor schematic is shown in Fig.~\ref{fig:architecture_scheme}. The architecture forms a square grid with fluxonium qubits located at the nodes, and capacitively connected via transmon couplers along the edges. Each element has an individual control line for both dc flux tuning and microwave driving. The fluxonium qubits are biased at the half-flux-quantum sweet spot throughout operation. The transmon couplers can be either tunable or fixed-frequency, with the choice determined primarily by fabrication capabilities and calibration convenience. Without loss of generality, we consider tunable transmons.

The native two-qubit gate in this architecture is the CZ operation. The physical mechanism underlying the gate is simple yet effective: the strong capacitive interaction between the fluxonium qubits and the transmon coupler causes the coupler’s $|0\rangle \leftrightarrow |1\rangle$ transition frequency to split into four distinct values, each corresponding to one of the computational basis states of the adjacent qubits. This effect is analogous to the dispersive frequency shift of a resonator coupled to a qubit. Driving the coupler with a resonant $2\pi$ microwave pulse at a selected transition frequency accumulates a conditional phase of $\pi$ on the corresponding computational state, implementing a CZ gate up to single-qubit $Z$ rotations. In this way, the two-qubit operation effectively reduces to a well-studied single-qubit operation on the transmon. A notable advantage of this approach is that all elements remain at their flux sweet spots during gate execution, thereby ensuring optimal coherence properties.

A drawback of strong static coupling is the emergence of unwanted interactions between non-target elements within the processor. These can be classified into five categories: (i) short-range coupling between nearest-neighbor qubits, (ii) short-range coupling between nearest-neighbor couplers, (iii) long-range coupling between a qubit and its next-nearest qubit, (iv) long-range coupling between a coupler and its next-nearest qubit, and (v) long-range coupling between a coupler and its next-nearest coupler. In this work, we propose design strategies to suppress all these parasitic interactions.

To ensure a high level of qubit-qubit crosstalk suppression, we arrange the fluxoniums into two distinct frequency groups, denoted A and B in Fig.~\ref{fig:architecture_scheme}(a), following a checkerboard pattern. 
The target frequencies for each group are listed in Table~\ref{tab:frequency_table}. This frequency detuning provides robust isolation not only between nearest neighbors but also, remarkably, between next-nearest neighbors, as confirmed by numerical simulations presented in Appendix~\ref{appendix:qq_analysis}.

\begin{table}[!h]
    \centering
    \begin{tabularx}{\linewidth}{l 
    >{\centering\arraybackslash}X 
    >{\centering\arraybackslash}X 
    >{\centering\arraybackslash}X 
    >{\centering\arraybackslash}X}
        \toprule
        \toprule
        \multicolumn{5}{c}{$\mathbf{Fluxoniums}$} \\[0.2cm]
        \quad\quad\quad\quad\quad\quad\quad\quad & \multicolumn{2}{c}{A} & \multicolumn{2}{c}{B} \\
        \midrule     
        $f_{01}$ (MHz)  & \multicolumn{2}{c}{313} & \multicolumn{2}{c}{360} \\
        $f_{12}$ (GHz)  & \multicolumn{2}{c}{5.45} & \multicolumn{2}{c}{5.35} \\
        $f_{03}$ (GHz)  & \multicolumn{2}{c}{8.35} & \multicolumn{2}{c}{8.45} \\
        \midrule
        \multicolumn{5}{c}{$\mathbf{Transmons}$} \\[0.2cm]
        & $\mathrm{C0_L}$ & $\mathrm{C0_U}$ & $\mathrm{C1_L}$ & $\mathrm{C1_U}$ \\
        \midrule
        $f_{01}$ (GHz) & 4.9 & 6.1 & 8 & 8.72 \\
        $\alpha$ (MHz) & -162 & -159 & -157 & -156 \\
        \midrule
        \multicolumn{5}{c}{$\mathbf{Oscillators}$} \\[0.2cm]
        & \multicolumn{2}{c}{$\mathcal{O}_0$} & \multicolumn{2}{c}{$\mathcal{O}_1$} \\
        \midrule
        $f$ (GHz) & \multicolumn{2}{c}{4.2} & \multicolumn{2}{c}{6.5} \\
        \bottomrule
        \bottomrule
        
    \end{tabularx}
    \caption{Key frequency parameters of the elements.}
    \label{tab:frequency_table}
\end{table}

Crosstalk suppression between nearest-neighbor couplers is achieved by employing coupler types with distinct frequencies, determined by the number of connections each qubit has in the lattice, with a minimum separation of 700 MHz. These couplers are classified according to their frequency offset from the fluxonium $|1\rangle \leftrightarrow |2\rangle$ and $|0\rangle \leftrightarrow |3\rangle$ transitions, as summarized in Table~\ref{tab:frequency_table}. Couplers whose frequencies lie closer to the $|1\rangle \leftrightarrow |2\rangle$ ($|0\rangle \leftrightarrow |3\rangle$) transition impart the conditional phase via the $|11\rangle$ ($|00\rangle$) computational state, defining C1-type (C0-type) devices. Each type is further divided into upper (U) and lower (L) subtypes, depending on whether the coupler frequency lies above or below the nearest fluxonium transition. The spatial arrangement of these four coupler types is shown in Fig.~\ref{fig:gate_structure}(a) and discussed in detail in Section~\ref{section:alternating_couplers}.

Additionally, parasitic long-range capacitive couplings are suppressed by employing single-mode differential fluxonium qubits with weak capacitive coupling between islands, as illustrated schematically in Fig.~\ref{fig:architecture_scheme}(b). This design effectively cancels direct parasitic couplings between couplers adjacent to opposite islands, as analyzed in Appendix~\ref{appendix:differential_fluxonium}.

Next, we introduce a differential oscillator, along with an additional direct capacitive coupling, to address long-range interactions between a coupler and a spectator qubit ($\mathcal{O}_0$-type) or another coupler ($\mathcal{O}_1$-type), as depicted in Fig.~\ref{fig:architecture_scheme}(d, e). This differential oscillator can be implemented as a capacitively grounded Josephson junction array with capacitive outer terminations. As an independent component, it offers additional flexibility in circuit parameter optimization, enabling the effective suppression of long-range interactions in the device; the resulting suppression mechanism is discussed in Section~\ref{section:parasitic_interactions}.

As a notable result, the architecture natively supports a two-qubit CZ operation implemented locally on the couplers without requiring system spectrum tuning. This locality permits the simultaneous execution of CZ operations, even when they share a computational qubit. As shown in Fig.~\ref{fig:architecture_scheme}(c), we identify three distinct groups of coupler pairs that enable such operations and thereby introduce CZZ gates. A detailed description and numerical simulations of these gates are presented in Section~\ref{section:CZZ_gates} and Section~\ref{section:performance} correspondingly.

%% file: sections/alternating_couplers.tex
\begin{figure*}[t]
    \center{\includegraphics[width=\linewidth]{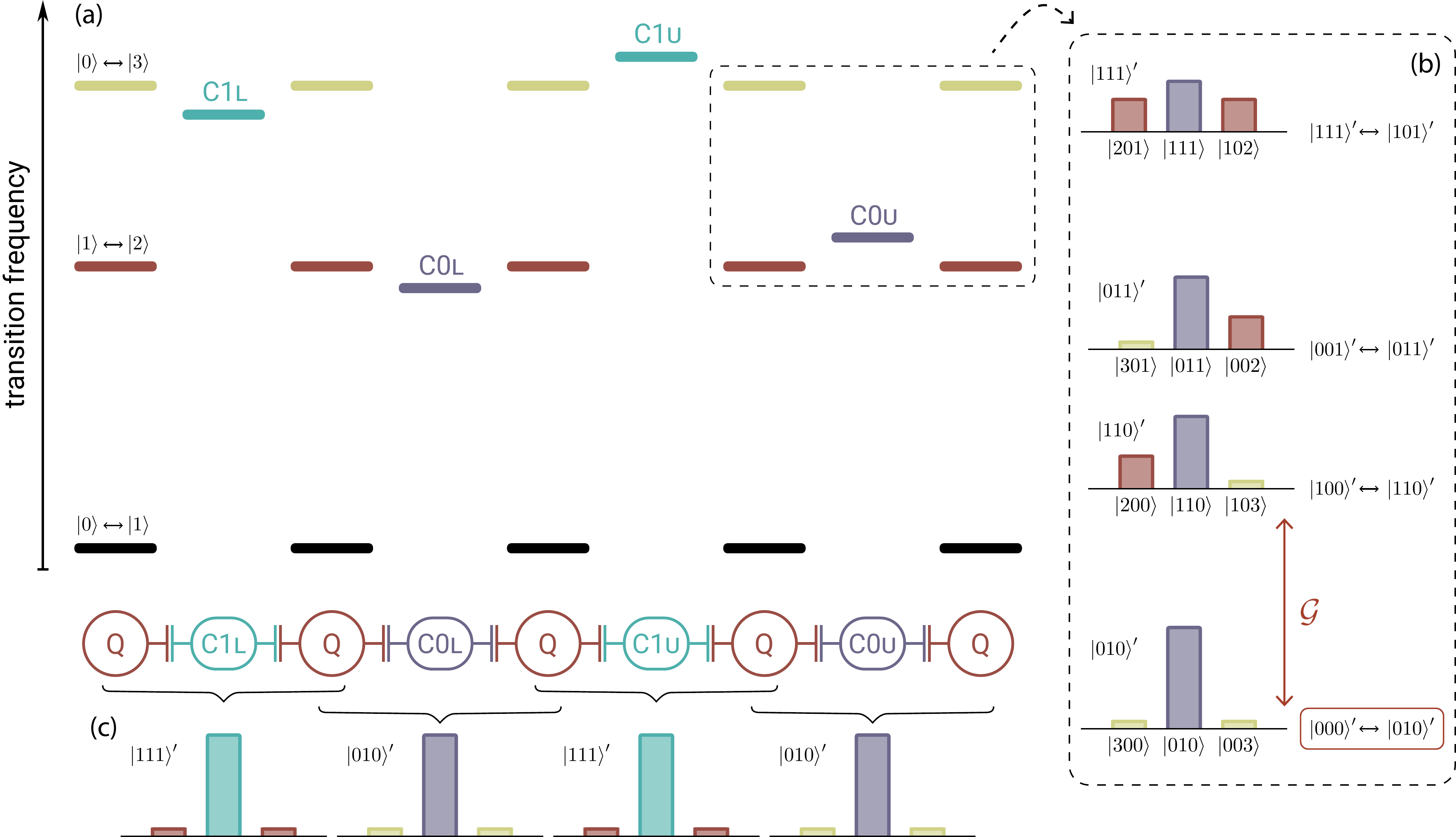}}
    \caption{Coupler frequency allocation and wavefunction localization. (a) Arrangement of the four types of transmon couplers’ $|0\rangle \leftrightarrow |1\rangle$ transitions relative to the fluxonium qubits' $|1\rangle \leftrightarrow |2\rangle$ and $|0\rangle \leftrightarrow |3\rangle$ transitions. (b) Characteristic state-dependent splitting of excited state of the $\mathrm{C0_U}$ coupler type, together with the corresponding wavefunctions illustrating state delocalization. The central bar corresponds to the coupler-localized component, while side bars indicate delocalization onto neighboring fluxonium state. Arrows emphasize the localization of the target state. Additionally we show the gap parameter $\mathcal{G}$. (c) Circuit schematic illustrating the qubit and couplers associated with the transition scheme in (a), including the localized target state for each coupler type.}
    \label{fig:gate_structure}
\end{figure*}

Implementation of a CZ gate via a microwave drive on a transmon coupler mediating two computational qubits requires engineering a state-dependent spectrum, where the coupler's $|0\rangle \leftrightarrow |1\rangle$ transition frequency is dispersively shifted by two adjacent qubits and thereby depends on the four computational states. The corresponding interaction can be described by the Hamiltonian:
\begin{equation}
    \renewcommand{\arraystretch}{2.2}
    \begin{array}{lll}
        \displaystyle \hat{H}_{\mathrm{int}}/h = \frac{1}{4}\left(\chi_\mathrm{A}^{}\, \hat{\sigma}_z^{\mathrm{A}} +
        \chi_\mathrm{B}^{}\,
        \hat{\sigma}_z^{\mathrm{B}}\right)\cdot \hat{\sigma}_z^{\mathrm{C}},
    \end{array}
    \label{eq:basic}
\end{equation}
where A and B denote the qubits, C denotes the coupler, and $\chi_{\mathrm{A, B}}^{}$ determine the dispersive shifts. Among the split coupler transitions, one is selected as the target transition for gate execution, while the others are referred to as side transitions and are kept undisturbed (in the following, corresponding excited coupler states are called target and side as well). The frequency separation between the target transition and its nearest side transition -- hereafter denoted as the gap $\mathcal{G}$ -- determines the achievable gate speed under fixed coherent errors. In the present design, system parameters are tuned to yield $\mathcal{G} \sim 100~\mathrm{MHz}$, enabling fast $(\sim 63~\mathrm{ns})$ two-qubit gates with coherent errors below $10^{-4}$. 

The splitting of the coupler spectrum arises from interactions with the $|1\rangle \leftrightarrow |2\rangle$ and $|0\rangle \leftrightarrow |3\rangle$ transitions of the adjacent qubits. Consequently, the coupler transitions with the largest $\mathcal{G}$ correspond to the computational states $|11\rangle$ and $|00\rangle$. Either of these transitions can be chosen as the target, imparting a conditional $\pi$ phase and thereby realizing the logical operations $U^{11}_{\pi}$ and $U^{00}_{\pi}$, respectively.  These operations are physically distinct yet logically equivalent, related by
\begin{equation}
    U_{\mathrm{CZ}} = U^{11}_{\pi} = -Z_1 Z_2 \cdot U^{00}_{\pi},
    \label{eq:CZ_C0_to_C1_transformation}
\end{equation}
where $Z_i$ are phase rotations, which can be implemented virtually. This mapping enables uniform CZ realization while retaining flexibility in coupler design.

Building a large-scale, fixed-coupling system requires sufficiently strong coupler-qubit interactions to generate a pronounced state-dependent spectrum. However, strong coupling also risks hybridization between the target excited state of a transmon and the high-energy fluxonium states. Such delocalized dressed states can extend across the qubit lattice, leading to long-range parasitic interactions that affect neighboring elements during two-qubit operations.

To address this challenge, we employ fluxoniums whose $|1\rangle \leftrightarrow |2\rangle$ and $|0\rangle \leftrightarrow |3\rangle$ transitions are sufficiently well separated in frequency to place the coupler frequency substantially closer to one of them than to the other. This configuration yields drastically different hybridization degrees for distinct excited coupler states. The guiding design principle is to select the coupler’s target transition such that its excited state remains minimally perturbed by adjacent fluxonium transitions.

Specifically, for a C0-type coupler, second-order perturbation theory applied to a Q-C-Q system indicates that spectral proximity between the coupler's $|0\rangle \leftrightarrow |1\rangle$ transition and the fluxoniums' $|1\rangle \leftrightarrow |2\rangle$ transition induces strong interactions between states: $\ket{111} \sim \ket{201},\,\ket{102}$; $\ket{011}\sim\ket{102}$, and $\ket{110}\sim\ket{201}$, while the state $\ket{010}$ remains unaffected due to a parity selection rule arising from the half-flux-quantum symmetry of the fluxonium. Consequently, the dressed state $\ket{010}'$ remains well localized (with $0.98-0.99$ population at $\ket{010}$) and serves as the target state, whereas the side states $\ket{110}'$, $\ket{011}'$, $\ket{111}'$ become considerably hybridized ($0.78-0.88$ population at original bare states) and acquire sufficient frequency shifts to yield $\mathcal{G} \approx 100\,\mathrm{MHz}$, as illustrated in Fig.~\ref{fig:gate_structure}(b). Nonetheless, the $\ket{010}$ state weakly interacts with $\ket{300}$ and $\ket{003}$ states. Although these interactions are suppressed by large detunings, they can still generate weak long-range parasitic effects, which are discussed in Section \ref{section:parasitic_interactions}.

C1-type couplers operate analogously: the spectral proximity between the coupler's $|0\rangle \leftrightarrow |1\rangle$ transition and the qubits' $|0\rangle \leftrightarrow |3\rangle$ transition induces strong interactions among the states $\ket{101}\sim\ket{301},\,\ket{103}$; $\ket{011}\sim\ket{301}$, and $\ket{110}\sim\ket{103}$, while the dressed state $\ket{111}'$ is chosen as the target.

The resulting localized states for all kinds of couplers are summarized in Fig.~\ref{fig:gate_structure}(c).

%% file: sections/parasitic_interactions.tex
A central challenge in architectures with strong static coupling is the suppression of long-range interactions between non-nearest-neighbor elements. While interactions between next-nearest qubits are largely negligible (see Appendix \ref{appendix:qq_analysis}), long-range interactions affecting the couplers remain more problematic.

Despite the localization of the target state wavefunctions discussed in the previous section, two classes of long-range interactions persist and can significantly degrade CZ gate performance. The first is related to the residual ZZ interaction between a coupler and a spectator qubit, while the second involves crosstalk and ZZ interactions between nearest identical couplers.

To address these effects systematically, we first analyze the C-Q-C-Q configuration, which captures the essential physics of coupler–spectator qubit interaction, and subsequently extend the discussion to more complex configurations involving interactions between nearest identical couplers.

\subsection{Сoupler-spectator qubit ZZ interaction}

\begin{figure}[h!]
    \center{\includegraphics[width=7cm]{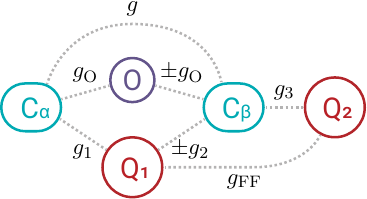}}
    \caption{Graphical representation of the effective Hamiltonian for the universal C-Q-C-Q system, covering all possible coupler combinations, where $\alpha, \beta \in \{0_{\mathrm{U}}, 0_{\mathrm{L}}, 1_{\mathrm{U}}, 1_{\mathrm{L}}\} $ and $\alpha \neq \beta$. Dashed lines indicate capacitive couplings.}
    \label{fig:CQOCQ_model}
\end{figure}

We now consider a model C-Q-C-Q system equipped with an additional element -- the differential oscillator -- designed to provide an extra degree of control over the inter-element interaction. A schematic diagram of the system, including all relevant effective couplings, is shown in Fig.~\ref{fig:CQOCQ_model}.
The sign of the coupling constants $g_\mathrm{O}$ and $g_2$ depends on the type of mutual connection between the two couplers via the differential fluxonium qubit: symmetric ($+$), when they connect to the same island, or antisymmetric ($-$), when they connect to different islands, as indicated by the coloring of qubit islands in Fig.~\ref{fig:architecture_scheme}(a).

The coupler-spectator ZZ interaction corresponds to the shift in the target coupler’s transition frequency that depends on the spectator qubit’s state. Quantifying this interaction requires accounting for two facts: neighboring couplers are never excited simultaneously (see Section \ref{section:CZZ_gates}), and the states of the qubits adjacent to an excited coupler are predetermined by the coupler type. For the C-Q-C-Q chain, this leads to the following expression:
\begin{equation}
\zeta_{\mathrm{CS}}^{(\mathrm{C}x)} = f_{\ket{1x01}'} - f_{\ket{0x01}'} - f_{\ket{1x00}'} + f_{\ket{0x00}'},
\label{eq:ZZ_definition}
\end{equation}
where $x=0$ ($x=1$) corresponds to a $\mathrm{C0}$ ($\mathrm{C1}$) coupler, and $f_{\ket{\cdot}'}$ denotes the frequency of the indicated dressed state.

In Appendix~\ref{subappendix:ZZ_CS_equation} we derive an analytical approximation for the $\zeta_{\mathrm{CS}}$ term between coupler $\mathrm{C_{\alpha}}$ and spectator qubit $\mathrm{Q_2}$ within the subcircuit model shown in Fig.~\ref{fig:CQOCQ_model}:
\begin{equation}
    \begin{split}
         \zeta_{\mathrm{CS}}^{(\mathrm{C}x)} & \approx (-1)^x \cdot \frac{(g_3 \, n^{\mathrm{Q_2}}_{p} n^{\mathrm{C}_\beta}_{01})^2}{f^{\mathrm{C}_\alpha}_{01} - f^{\mathrm{Q_2}}_{p}} \cdot \left( \frac{n^{\mathrm{C}_\alpha}_{01} n^{\mathrm{C}_\beta}_{01}}{f^{\mathrm{C}_\alpha}_{01} - f^{\mathrm{C}_\beta}_{01}} \right)^2 \times \\
        & \times \left(g \,\pm\, g_1 g_2 \frac{(n^{\mathrm{Q_1}}_{t})^2}{f^{\mathrm{C}_\alpha}_{01} - f^{\mathrm{Q_1}}_{t}} \,\pm \,g_{\mathrm{O}}^2 \frac{(n^{\mathrm{O}})^2}{f^{\mathrm{C}_\alpha}_{01} - f^{\mathrm{O}}}\right)^2
    \end{split}
    \label{eq:ZZ_equation}
\end{equation}
Here $n_{ij}$ are matrix elements of the Cooper-pair number operator, and $f_{ij}$ are the corresponding $\ket{i} \leftrightarrow \ket{j}$ transition frequencies. The indices $p$ and $t$ depend on the $\mathrm{C}_{\alpha}$ coupler type: for $\alpha = 0_{\mathrm{U}/\mathrm{L}}$, we set $p=12$ and $t=03$, whereas for $\alpha = 1_{\mathrm{U}/\mathrm{L}}$, there are $p=03$ and $t=12$. This expression applies to all configuration-specific variations of the C-Q-C-Q subcircuit present in the proposed processor architecture, making it a universal description of coupler-spectator ZZ interactions.

The first notable feature of Eq.~(\ref{eq:ZZ_equation}) is the inverse-square dependence of $\zeta_{\mathrm{CS}}$ on the coupler frequency difference. This naturally suppresses coupler-spectator ZZ interactions in mixed C0/C1-type coupler configurations,  where the frequency detuning is large (3.1 GHz and 2.62 GHz). These configurations are implemented along the horizontal and vertical lines of the processor grid. In contrast, same-type coupler pairs, which form same-colored angles, exhibit smaller detunings (1.2 GHz and 0.72 GHz), requiring additional suppression of $\zeta_{\mathrm{CS}}$.

This suppression leverages the second key property of Eq.~(\ref{eq:ZZ_equation}): The additivity of the terms in the final parentheses. Both the direct coupling and the oscillator are adjustable, allowing their terms to be tuned to cancel the fixed qubit term contribution.

For clarity, consider the antisymmetric connection case. In the $\mathrm{C1}/\mathrm{C1}$-type configurations ($\alpha,\beta=1_{\mathrm{U/L}}$), where $f^{\mathrm{Q_1}}_{t} = f^{\mathrm{Q_1}}_{12} < f^{\mathrm{C_{\alpha}}}_{01}$ the qubit term is negative. Hence, cancellation can be achieved by increasing $g$ through enhanced capacitive coupling between the couplers. Conversely, in the $\mathrm{C0}/\mathrm{C0}$-type configurations, the condition $f^{\mathrm{Q_1}}_{t} = f^{\mathrm{Q_1}}_{03} > f^{\mathrm{C_{\alpha}}}_{01}$ yields a positive qubit term, requiring a compensating negative contribution. This can only be achieved by introducing an additional degree of freedom, such as a differential oscillator $\mathcal{O}_0$ with a frequency $f^{\mathrm{O}} < f^{\mathrm{C_{\alpha}}}_{01}$.

To validate the applicability of these findings to our architecture, we analyze a 13-qubit numerical model (Appendix~\ref{appendix:model_static}) that incorporates significant long-range parasitic capacitive couplings and all possible C-Q-C-Q configurations. For antisymmetric coupler configurations, we compute $\zeta_{\mathrm{CS}}$ by diagonalizing the corresponding Q-C-Q-C-Q subcircuits. Including the third qubit is essential for minimizing $\zeta_{\mathrm{CS}}$ because, in the full lattice, a suppression element placed between $\mathrm{C_{\alpha}}$ and $\mathrm{C_{\beta}}$ influences both $\zeta_{\mathrm{C_{\alpha}S}}$ and $\zeta_{\mathrm{C_{\beta}S}}$, each involving distinct spectators (the edge qubits in the Q-C-Q-C-Q chain). The resulting $\zeta_{\mathrm{CS}}$ values for four representative subcircuits are summarized in Table~\ref{tab:zz_antisym_results} (hereafter, we refer to coupler-specific configurations of the form Q-$\mathrm{C_{\alpha}}$-Q-$\mathrm{C_{\beta}}$-Q as $\mathrm{C_{\alpha}}|\mathrm{C_{\beta}}$). Our simulations show that $\zeta_{\mathrm{CS}}$ is suppressed below 40 kHz, which corresponds to negligible gate errors on the order of $10^{-6}$ (see Appendix~\ref{subappendix:ZZ_error}). We further study the fabrication robustness of the described suppression mechanism in Appendix |\ref{appendix:fabrication_robustness}.

\begin{table}[!h]
    \centering
    \begin{tabularx}{\linewidth}{ 
    l 
    >{\centering\arraybackslash}X
    c}
    
        \toprule
        \toprule
        
        & $\zeta_{\mathrm{CS}}$, kHz & Circuit scheme \\
        % & bare & suppressed &  \\

        \midrule
        
        & &  
        \multirow{4}{*}{
            \begin{minipage}{.15\textwidth}
                \includegraphics[width=2 cm]{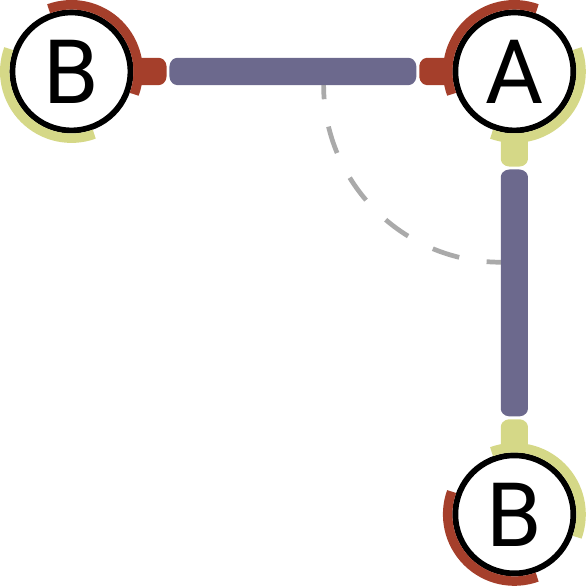}
            \end{minipage}
            }
        \\
        $\mathbf{C0_U}$|$\mathrm{C0_L}$ & $158 \longrightarrow 35$ & \\[0.1cm]
        & & \\[0.1cm]
        $\mathrm{C0_U}$|$\mathbf{C0_L}$ & $-29 \longrightarrow -39$ & \\
        & & \\
        
        \midrule

        & &  
        \multirow{4}{*}{
            \begin{minipage}{.15\textwidth}
                \includegraphics[width=2 cm]{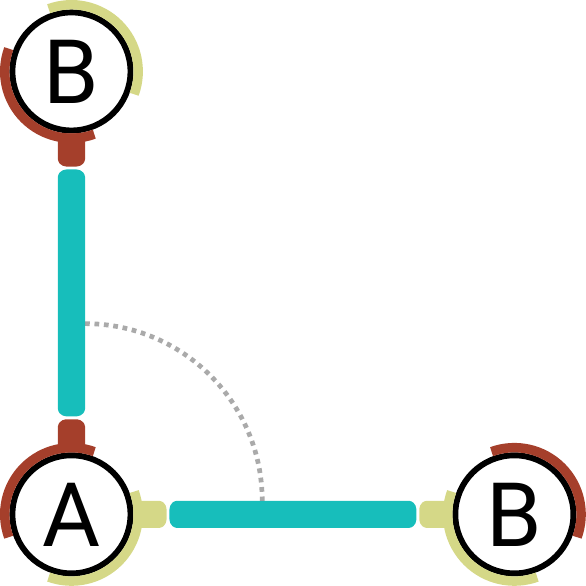}
            \end{minipage}
            }
        \\
        $\mathbf{C1_U}$|$\mathrm{C1_L}$ & $-28 \longrightarrow -2$& \\[0.1cm]
        & & \\[0.1cm]
        $\mathrm{C1_U}$|$\mathbf{C1_L}$ & $104 \longrightarrow 3$& \\
        & & \\
        
        \midrule
        
        & & 
        \multirow{4}{*}{
            \begin{minipage}{.2\textwidth}
                \includegraphics[width=3.5106 cm]{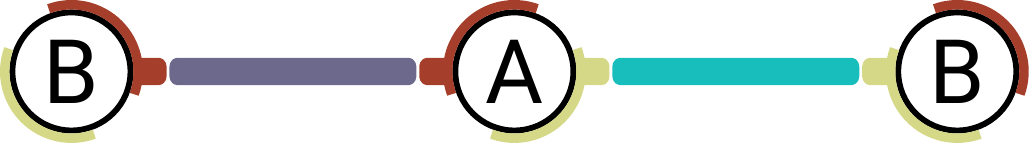}
            \end{minipage}
            }
        \\
        $\mathbf{C0_L}$|$\mathrm{C1_L}$ & $-1$ & \\[0.1cm]
        $\mathrm{C0_L}$|$\mathbf{C1_L}$ & $13$ & \\
        & & \\

        \midrule
        
        & & 
        \multirow{4}{*}{
            \begin{minipage}{.2\textwidth}
                \includegraphics[width=0.49 cm]{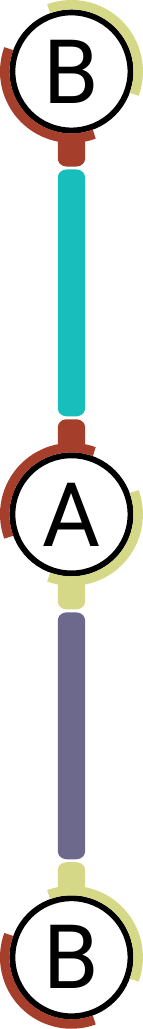}
            \end{minipage}
            }
        \\
        & & \\[0.1cm]
        $\mathbf{C0_U}$|$\mathrm{C1_U}$ & $6$ & \\[1.2cm]
        $\mathrm{C0_U}$|$\mathbf{C1_U}$ & $-6$ & \\
        & & \\[0.1cm]
        & & \\

        \bottomrule
        \bottomrule
        
    \end{tabularx}
    \caption{Coupler-spectator ZZ interaction in four representative three-qubit systems with antisymmetric connection. In each row, the affected coupler is indicated in bold. The arrows illustrate the change in the ZZ coupling strength upon introducing a suppression element—either a differential oscillator or direct capacitive coupling.}
    \label{tab:zz_antisym_results}
\end{table}

The symmetric coupler configuration presents a more complex challenge. Unlike the antisymmetric case, it introduces significant long-range parasitic capacitive couplings. In this regime, $\zeta_{\mathrm{CS}}$ is influenced by strong interactions between nearest identical couplers. To isolate these effects, symmetric configurations are deliberately designed to occur only within square sublattices, forming so-called "strongly interconnected squares" which are analyzed in detail in the following subsection.

\subsection{The problem of strongly interconnected squares}

\begin{figure}[h!]
    \center{\includegraphics[width=7cm]{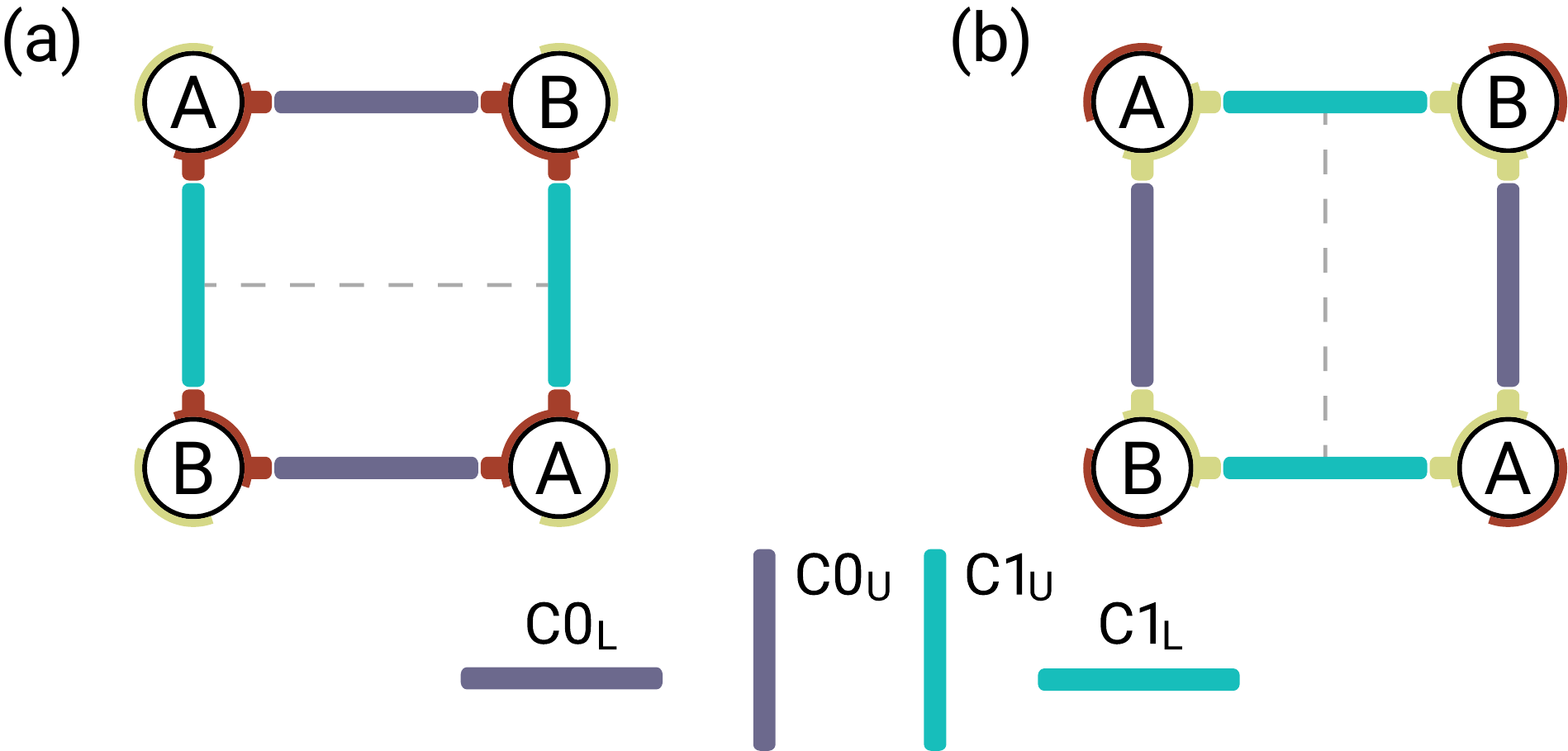}}
    \caption{Schematic of two types of strongly interconnected square sublattices. (a) Squares composed of $\mathrm{C1_U}$ and $\mathrm{C0_L}$ couplers feature red-colored fluxonium islands at their corners. (b) Squares formed by $\mathrm{C1_L}$ and $\mathrm{C0_U}$ couplers instead have yellow-colored islands. Dashed lines indicate differential oscillators suppressing parasitic interactions between the identical couplers.}
    \label{fig:strongly_interconnected_squares_kinds}
\end{figure}

As noted previously, interactions between nearest identical couplers cannot be neglected. The situation becomes particularly challenging when all circuit elements are connected symmetrically. In the proposed blueprint, such symmetric configurations arise exclusively within square sublattices. These squares occur in two different types, illustrated in Fig.~\ref{fig:strongly_interconnected_squares_kinds}.
The first composed of $\mathrm{C1_U}$ and $\mathrm{C0_L}$ couplers, and the second of $\mathrm{C1_L}$ and $\mathrm{C0_U}$ couplers. These structures, which we refer to as strongly interconnected squares, represent the most demanding scenario for suppressing parasitic interactions in the processor. To mitigate these unwanted interactions, we introduce a differential oscillator $\mathcal{O}_1$, connecting opposite C1-type couplers.

Numerical diagonalization of the full four-qubit square Hamiltonian is computationally intractable because of the large Hilbert-space dimension required to capture all relevant interactions. To overcome this challenge, we decompose each strongly interconnected square into simpler C-Q-C-Q-C subcircuits, which we diagonalize individually (see Appendix~\ref{appendix:model_static}). These reduced systems accurately reproduce the dominant coupling mechanisms between identical couplers that give rise to both coupler–spectator and coupler–coupler parasitic interactions. We verify that extending a C-Q-C-Q-C chain with an additional qubit does not qualitatively alter the results, thereby validating the subcircuit simplification.

To examine long-range interactions within these C-Q-C-Q-C subcircuits, we evaluate three major metrics: (1) the coupler-spectator ZZ interaction, which, in the C-Q-C-Q-C basis, takes the form
\begin{equation}
    \begin{split}
        \zeta_{\mathrm{CS}}^{(\mathrm{C}x)} = f_{\ket{1x010}'} - f_{\ket{0x010}'} - f_{\ket{1x000}'} + f_{\ket{0x000}'},
        \\
        \zeta_{\mathrm{CS}}^{(\mathrm{C}x)} = f_{\ket{010x1}'} - f_{\ket{010x0}'} - f_{\ket{000x1}'} + f_{\ket{000x0}'},
    \end{split}
\label{eq:ZZ_CS_definition}
\end{equation}
for the left and right couplers, respectively (analogous to Eq.~(\ref{eq:ZZ_definition})); (2) the ZZ interaction between the target transitions of the identical couplers,
\begin{equation}
\zeta_{\mathrm{CC}}^{(\mathrm{C}x)} = f_{\ket{1x0x1}'} - f_{\ket{0x0x1}'} - f_{\ket{1x0x0}'} + f_{\ket{0x0x0}'};
\label{eq:ZZ_CC_definition}
\end{equation}
and (3) the hybridization parameter
\begin{equation}
D(\psi_1,\psi_2) = \tfrac{1}{2}\abs{\bra{\psi_1}\ket{\psi_2}'}^2 + \tfrac{1}{2}\abs{\bra{\psi_2}\ket{\psi_1}'}^2 ,
\label{eq:D_hybridization_definition}
\end{equation}
which quantifies the degree of state mixing. We compute $D$ for three $(\psi_1, \psi_2)$ pairs of coupler states defined in the C$x$-Q-C-Q-C$x$ basis as: 
\begin{equation}
    \renewcommand{\arraystretch}{1.2}
    \begin{array}{lll}
        \ket{1x0x0} \,\&\, \ket{0x0x1}\quad : \quad \text{target-target}
        \\
        \ket{1x0 \Bar{x} 0} \,\&\, \ket{0x0 \Bar{x} 1}\quad : \quad \text{target-side}
        \\
        \ket{1 \Bar{x} 0x0} \,\&\, \ket{0 \Bar{x} 0x1}\quad : \quad \text{side-target},
    \end{array}
\end{equation}
where $\Bar{x}=(x+1)\,\,\text{mod}\,\,2$.

\begin{figure}[t]
    \center{\includegraphics[width=\linewidth]{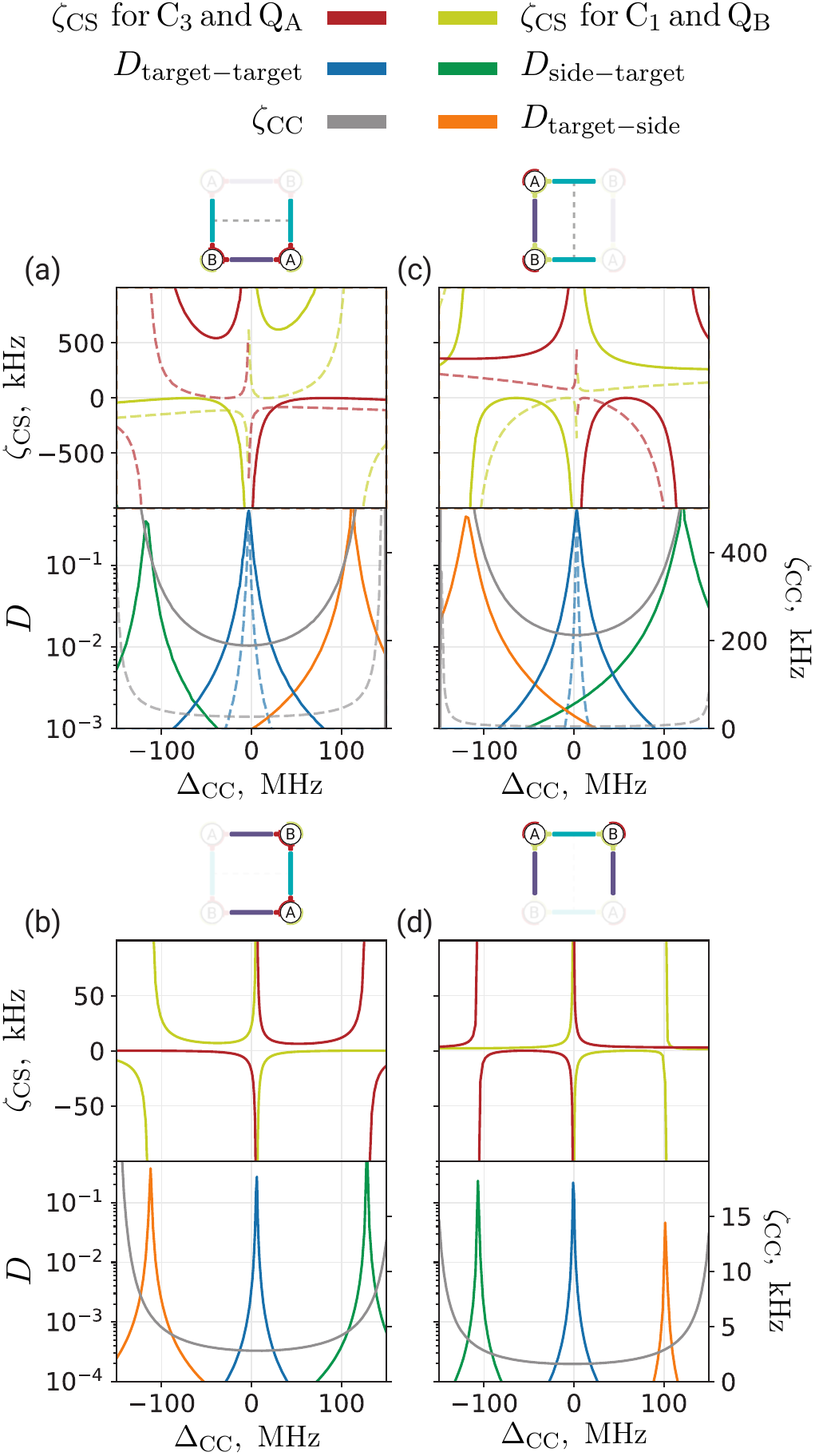}}
    \caption{Parasitic interactions between edge couplers in $\mathrm{C_{1}}$-$\mathrm{Q_A}$-$\mathrm{C_2}$-$\mathrm{Q_B}$-$\mathrm{C_3}$ subcircuits of strongly interconnected squares under variable detuning $\Delta_{\mathrm{CC}}$. Dashed curves show simulations involving the oscillator $\mathcal{O}_1$. Couplers are numbered for reference. (a, b) Full characterization for subcircuits with $\mathrm{C1_U}$ and $\mathrm{C0_L}$ couplers; (c, d) analogous results for those with $\mathrm{C1_L}$ and $\mathrm{C0_U}$.}
    \label{fig:strongly_interconnected_squares_plots}
\end{figure}

Fig.~\ref{fig:strongly_interconnected_squares_plots} presents the behavior of these three metrics for all four possible C-Q-C-Q-C configurations arising from both square types, plotted as a function of the frequency detuning $\Delta_{\mathrm{CC}}$ between opposite couplers. A dramatic increase in the coupler-spectator ZZ interaction occurs whenever the detuning leads to a resonance between target and side states of the opposite couplers ($\Delta_{\mathrm{CC}}\approx 0$ and $\Delta_{\mathrm{CC}}\approx \pm\mathcal{G}$). These resonant regions exhibit significant hybridization of the overlapping states, indicating the presence of strong coupler-coupler interaction and its correlation with the coupler-spectator ZZ.

That suggests a strategy to avoid strong parasitic interactions: get out of the central peak by setting a small detuning of the identical couplers achievable either through circuit design parameters or via tuning the coupler frequency in situ. While for C0-type couplers, this peak is intrinsically narrow enough to realize that, the C1-type coupler case requires an additional interaction suppression. To achieve it, we introduce a differential oscillator, which substantially weakens the central resonance, narrowing the peak and simultaneously suppressing both coupler–spectator and coupler–coupler interactions. This effect is illustrated in Fig.~\ref{fig:strongly_interconnected_squares_plots}(a), where the dashed and the solid curves represent the interaction metrics in absence and in presence of $\mathcal{O}_1$ correspondingly.

The same approach proves effective in suppressing crosstalk between identical couplers. To quantify this parasitic effect, we treat it as leakage from one coupler to another that occurs during CZ gate operation, and estimate its rate (see Appendix~\ref{appendix:leakage_model} for numerical details). The numerical simulation indicates that a detuning of 40 MHz yields a crosstalk error rate below $10^{-5}$. Since the same detuning is sufficient to suppress the ZZ interactions discussed earlier, we adopt 40 MHz as the optimal detuning value for identical couplers in our architecture.
 
The resulting $\zeta_{\mathrm{CS}}$ values and their reduction due to the $\mathcal{O}_1$ oscillator in C1-type coupler cases are summarized in Tab.~\ref{tab:zz_sym_results}. 
\begin{table}[!h]
    \centering
    \begin{tabularx}{\linewidth}{ 
    l 
    >{\centering\arraybackslash}X
    c}
    
        \toprule
        \toprule
        
        & $\zeta_{\mathrm{CS}}$, kHz & Circuit scheme \\
        % & bare & suppressed &  \\

        \midrule

        & &  
        \multirow{4}{*}{
            \begin{minipage}{.15\textwidth}
                \includegraphics[width=2 cm]{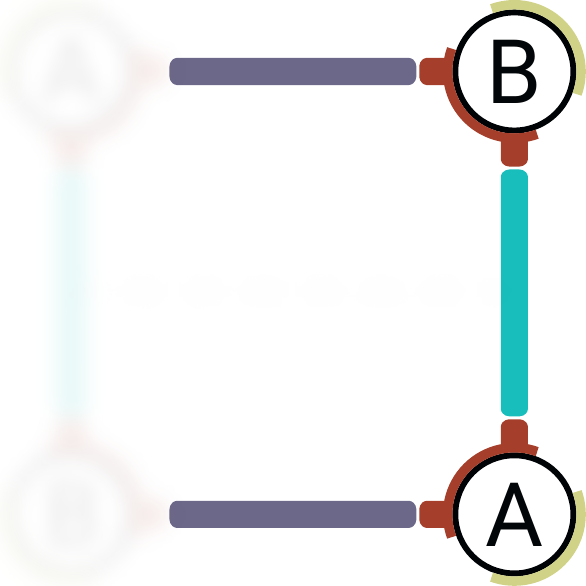}
            \end{minipage}
            }
        \\
        $\mathbf{C0_L}$|$\mathrm{C1_U}$|$\mathrm{C0_L}$ & $-2$& \\[0.1cm]
        & & \\[0.1cm]
        $\mathrm{C0_L}$|$\mathrm{C1_U}$|$\mathbf{C0_L}$ & $7$& \\
        & & \\

        \midrule

        & &  
        \multirow{4}{*}{
            \begin{minipage}{.15\textwidth}
                \includegraphics[width=2 cm]{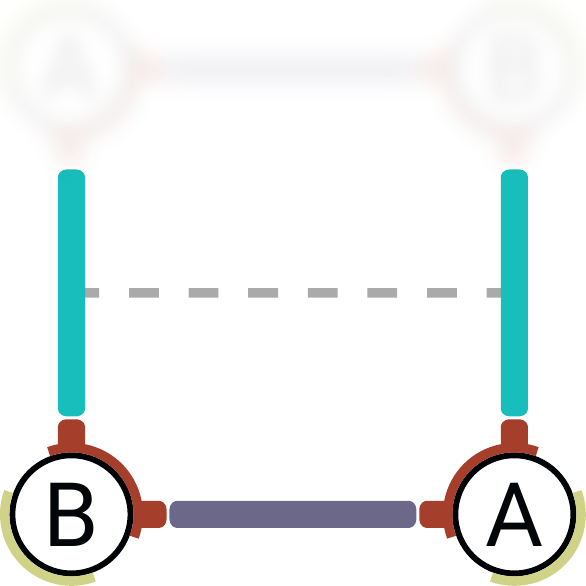}
            \end{minipage}
            }
        \\
        $\mathbf{C1_U}$|$\mathrm{C0_L}$|$\mathrm{C1_U}$ & $578 \longrightarrow 54$& \\[0.1cm]
        & & \\[0.1cm]
        $\mathrm{C1_U}$|$\mathrm{C0_L}$|$\mathbf{C1_U}$ & $-33 \longrightarrow -42$& \\
        & & \\

        \midrule

        & &  
        \multirow{4}{*}{
            \begin{minipage}{.15\textwidth}
                \includegraphics[width=2 cm]{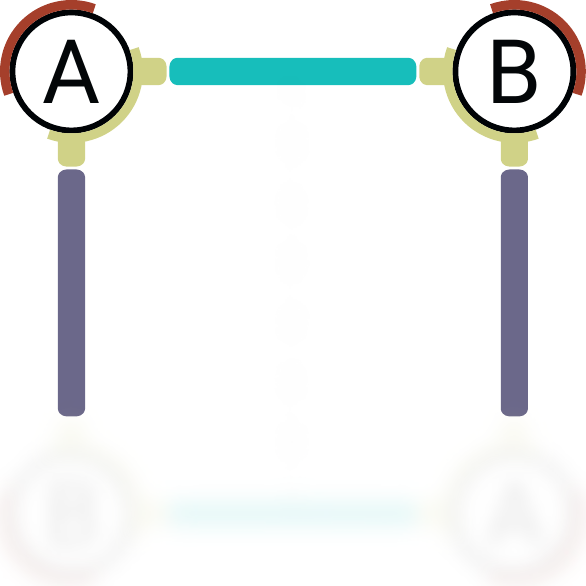}
            \end{minipage}
            }
        \\
        $\mathbf{C0_U}$|$\mathrm{C1_L}$|$\mathrm{C0_U}$ & $<1$& \\[0.1cm]
        & & \\[0.1cm]
        $\mathrm{C0_U}$|$\mathrm{C1_L}$|$\mathbf{C0_U}$ & $4$& \\
        & & \\

        \midrule

        & &
        \multirow{4}{*}{
            \begin{minipage}{.15\textwidth}
                \includegraphics[width=2 cm]{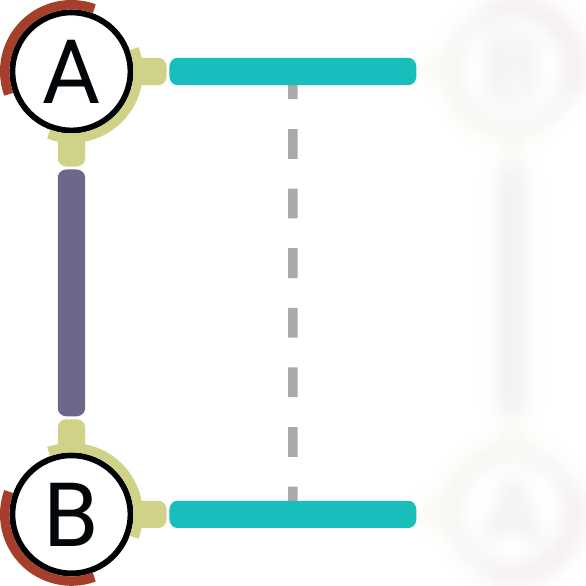}
            \end{minipage}
            }
        \\
        $\mathbf{C1_L}$|$\mathrm{C0_U}$|$\mathrm{C1_L}$ & $428 \longrightarrow 119$& \\[0.1cm]
        & & \\[0.1cm]
        $\mathrm{C1_L}$|$\mathrm{C0_U}$|$\mathbf{C1_L}$ & $-35 \longrightarrow -62$& \\
        & & \\
        
        \bottomrule
        \bottomrule
        
    \end{tabularx}
    \caption{Coupler-spectator ZZ interaction in C-Q-C-Q-C subcircuits that represent strongly interconnected squares. In each row, the affected coupler is indicated in bold. The arrows illustrate the change in $\zeta_{\mathrm{CS}}$ upon adding an $\mathcal{O}_1$ oscillator.}
    \label{tab:zz_sym_results}
\end{table}
The residual ZZ interaction corresponds to gate errors on the order of $10^{-5}$ and lower. Notably, without the additional $\mathcal{O}_1$ oscillator, $\zeta_{\mathrm{CS}}$ for C1-type couplers can exceed 300 kHz, leading to errors above $10^{-4}$, which would dominate the performance of the entire processor. This clearly demonstrates that incorporating differential oscillators within strongly interconnected squares is essential for maintaining low error rates.

In addition, we examine C-Q-C-Q-C subcircuits with identical couplers located outside the strongly interconnected squares. Owing to the geometry of the architecture, both qubits in these subcircuits have an antisymmetric coupler connection configuration, which naturally suppresses ZZ interactions between the identical couplers to below 1 kHz. Nevertheless, robust crosstalk suppression still requires small detuning. Fortunately, since the opposite identical couplers in the strongly interconnected squares are already detuned by 40 MHz, the necessary detuning can be automatically achieved by arranging the modified couplers in alternating sequences, maintaining gate leakage errors below $10^{-5}$.

\forth{To conclude the discussion of long-range interactions, we address the parasitic effects that can appear in the full architecture circuit. While coupler-coupler crosstalks are completely resolved by the additional detuning, ZZ interactions may undergo modifications. The latter are linear in the frequency shifts arising from spectrum dressing, which can be approximately described by perturbation theory, whose orders are additive. Consequently, any ZZ interaction can be split into a sum of terms that are obtainable independently by considering the corresponding paths between the interacting systems. To build intuition about this, one can adopt the virtual-transition framework described in Appendices~\ref{subappendix:ZZ_CS_equation} and~\ref{appendix:virtual_transitions}.}

\forth{Consequently, in the full architecture circuit, each perturbation term that impacts $\zeta_{\mathrm{CS}}$ and $\zeta_{\mathrm{CC}}$ is either accounted for by the aforementioned subcircuits or is at least two orders higher -- its virtual-photon path must cover at least one additional element. This two-order difference should introduce a suppression factor. Since even a factor of 5 pushes the worst unsuppressed $\zeta_{\mathrm{CS}}$ and $\zeta_{\mathrm{CC}}$ values down to the scale of the suppressed ones, we assume these terms to be negligible.}

\forth{A direct consequence of this assumption is that any parasitic interaction of longer range than those covered by the considered subcircuits is negligible as well -- in fact, even more so. For instance, the ZZ interaction between a coupler and the qubit one site beyond a spectator is four perturbation orders higher than $\zeta_{\mathrm{CS}}$, and the same holds for cross-talk between the nearest identical couplers after the described detuning.}

In summary, all parasitic long-range interactions in the proposed architecture -- whether coupler–spectator, coupler-coupler, or crosstalk -- can be either intrinsically minimized or systematically mitigated. This ensures robust two-qubit gate performance within large-scale fluxonium-based processors. \second{Further discussion of the fabrication robustness is presented in Appendix \ref{appendix:fabrication_robustness}.}

%% file: sections/CZZ.tex
A notable feature of the proposed architecture is the ability to execute fast CZZ operations -- two simultaneous CZ gates sharing a common control qubit. The idea of parallel two-qubit gates stems from the microwave-activated CZ, in which the conditional phase is accumulated through a $2\pi$ Rabi oscillation leveraging the excited state of an individual coupler as an ancillary level. Since all elements remain constantly interconnected, multiple couplers can be driven in parallel, allowing distinct target transitions to accumulate phase concurrently.

\begin{figure}[t]
    \center{\includegraphics[width=\linewidth]{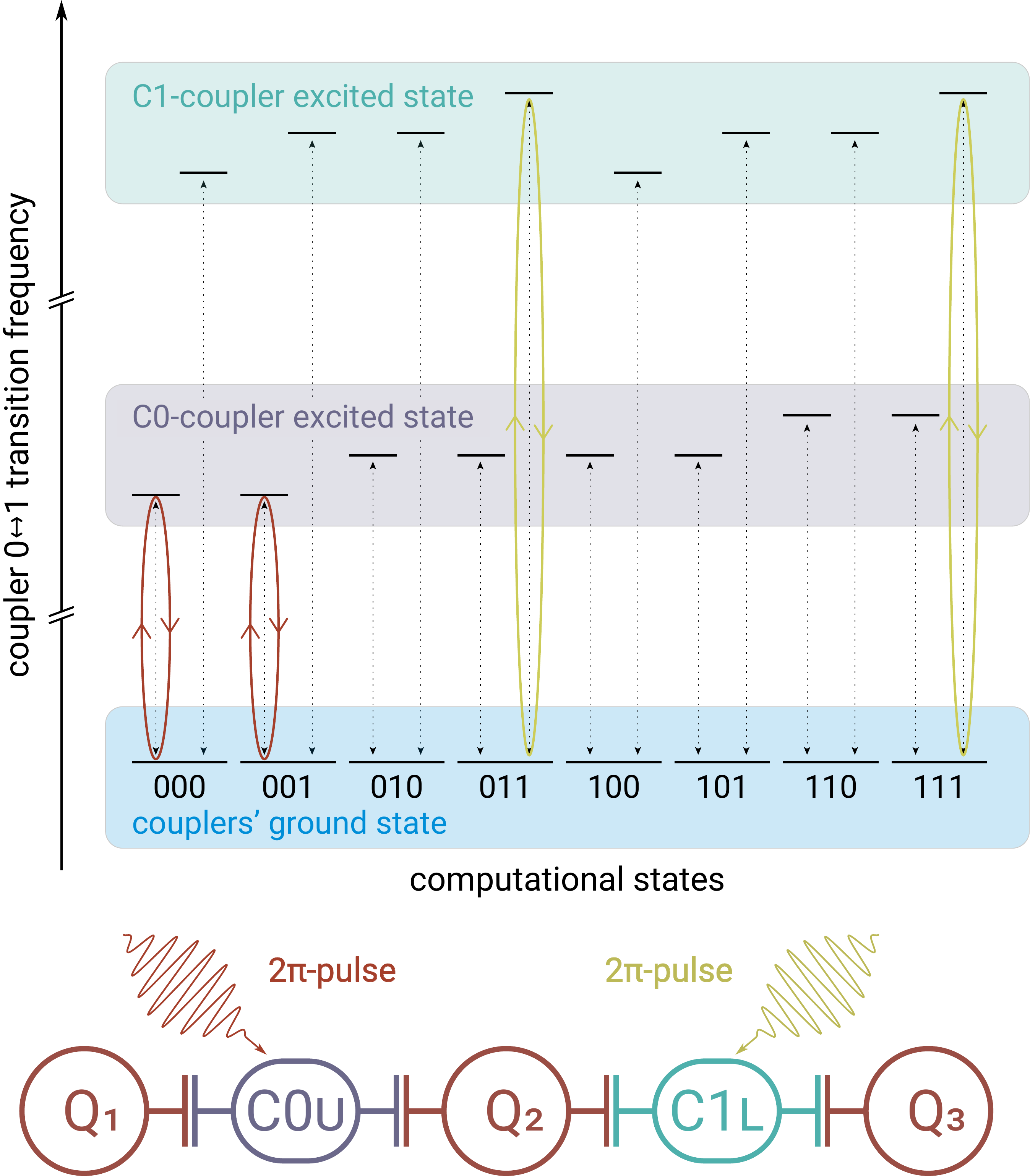}}
    \caption{State-dependent spectrum of the $\mathrm{C0_U}$ and $\mathrm{C1_L}$ couplers and their population oscillations under simultaneous $2\pi$-pulse drives in a three-qubit circuit.}
    \label{fig:simultaneous_CZ_transitions}
\end{figure}

The main obstacle to realizing high-fidelity parallel CZ operations across all neighboring qubit pairs is the strong static ZZ interaction (ranging from 400 kHz to 3 MHz) between the target states of nearest-neighboring couplers, which induces unwanted correlations during concurrent drives. This limitation, however, does not apply to couplers of different types ($\mathrm{C0}$ and $\mathrm{C1}$). Since these two classes acquire their conditional phases on distinct computational states ($|00\rangle$ for C0 and $|11\rangle$ for C1), their target states are never excited simultaneously. This, in principle, eliminates the ZZ interaction, enabling high-fidelity parallel CZ operations using pairs of different type couplers. The mechanism of this CZZ operation is illustrated schematically in Fig.~\ref{fig:simultaneous_CZ_transitions}.

Nonetheless, CZZ gates have one implementation pitfall: the effective spectral gap may shrink, reducing the achievable gate speed. During parallel driving of neighboring couplers, this can occur when the excitation of one coupler reduces the gap-building interaction of another.

\begin{figure}[ht!]
    \center{\includegraphics[width=165 pt]{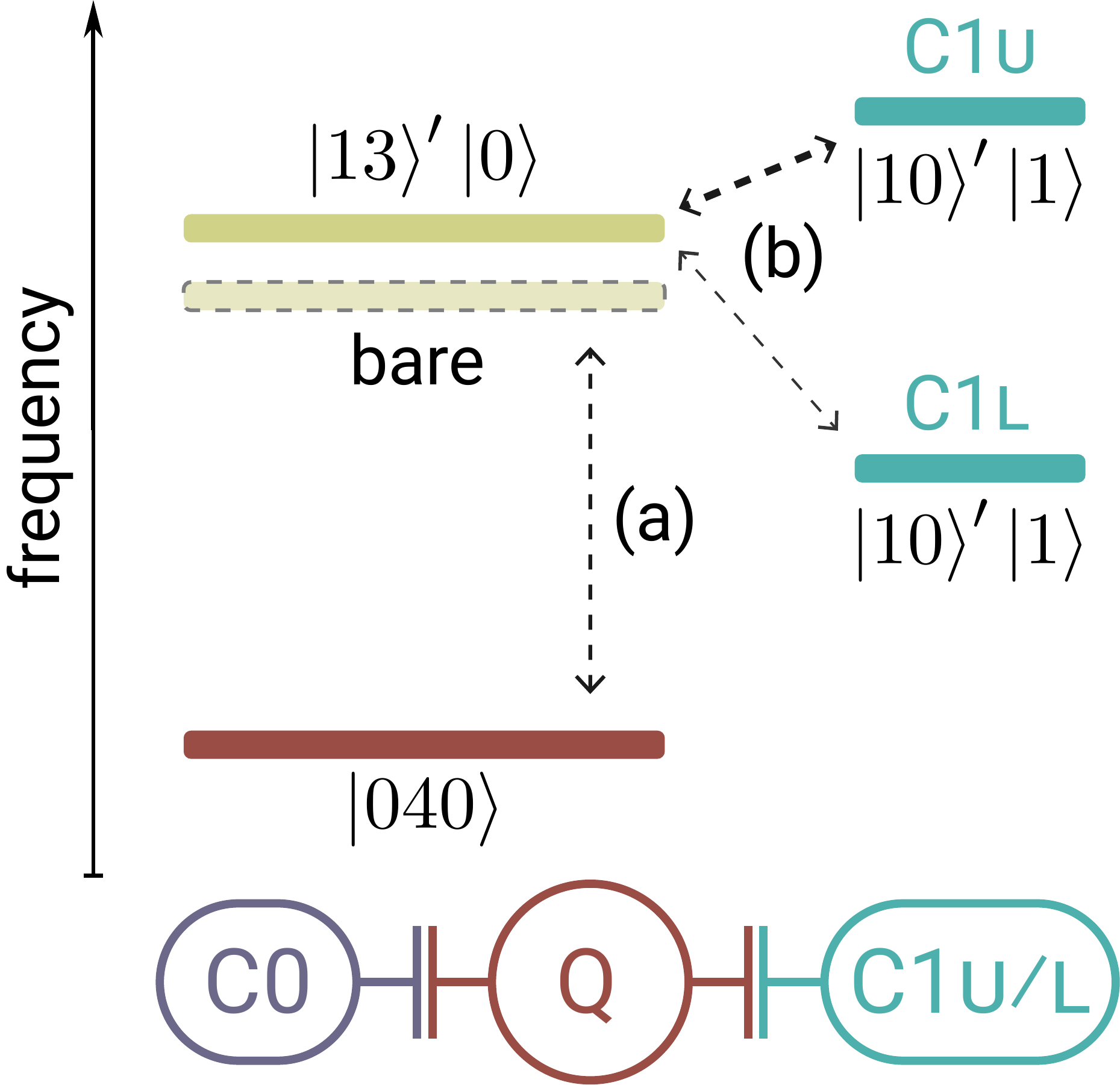}}
    \caption{The energy-level structure of C0-Q-C1 model. The interaction (a) pushes up the bare state $\ket{13}$ in C0-Q subsystem thereby modifying the gap-building interaction (b) with the C1 couplers, which increases for $\mathrm{C1_U}$ and decreases for $\mathrm{C1_L}$.}
    \label{fig:simultaneous_CZ_pitfall}
\end{figure}

To illustrate this effect, we consider the C0-Q-C1 system shown in Fig.~\ref{fig:simultaneous_CZ_pitfall}. In this configuration, when the C0 coupler is excited, the spectral gap of the C1 coupler is determined primarily by its interaction with the
$|10\rangle' \leftrightarrow |13\rangle'$ transition of the diagonalized C0-Q subsystem. The issue arises from the strong interaction between the $|13\rangle$ and $|04\rangle$ states within this subsystem, which shifts the frequency of that transition. Under our circuit parameters (Appendix \ref{appendix:parameters}), the $|04\rangle$ state always lies below $|13\rangle$ in frequency, pushing it up and thereby increasing the $|10\rangle' \leftrightarrow |13\rangle'$ transition frequency, as illustrated in Fig.~\ref{fig:simultaneous_CZ_pitfall}. As a consequence, the gap-building interaction is strengthened for $\mathrm{C1_U}$ and weakened for $\mathrm{C1_L}$, leading to the $\mathrm{C1_L}$ gap narrowing during the CZZ gate. Furthermore, the magnitude of this effect decreases as the C0-coupler frequency increases, enlarging the separation between $|13\rangle$ and $|04\rangle$. Thus, while the gap reduction is modest for the $\mathrm{C0_U}|\mathrm{C1_L}$ configuration (from $\sim105$ MHz to $\sim85$ MHz), it becomes substantial for $\mathrm{C0_L}|\mathrm{C1_L}$, where the gap drops to $\sim32$ MHz, rendering this configuration unsuitable for CZZ operation.

In contrast, the gap of the C0 coupler remains stable because it is set by the interaction with the $|11\rangle' \leftrightarrow |21\rangle'$ transition of the Q-C1 subsystem, which does not undergo any additional shifts. Indeed, the $|21\rangle$ state in this subsystem is strongly detuned from the $|30\rangle$ and $|50\rangle$ states, while its interaction with the $|40\rangle$ state is forbidden by parity.

The primary benefit of the CZZ gate is the two-fold acceleration of a CZ gate pair, which suppresses errors arising from qubit decoherence. Consequently, we anticipate that CZZ gates will serve as a valuable resource for implementing quantum algorithms on current noisy devices. 

A prominent example is the surface code, where stabilizer extraction requires four sequential CZ operations between an ancilla and its neighboring data qubits. A native CZZ gate naturally replaces a pair of consecutive CZ operations sharing the same ancilla, allowing two parity interactions to be executed simultaneously. Consequently, the four-CZ stabilizer measurement cycle is reduced to two CZZ operations, approximately halving the entangling-gate depth. This shorter correction cycle decreases the accumulation of decoherence-induced errors. Recent circuit-level simulations suggest that this reduction in entangling-gate depth can translate into substantially lower logical error rates and higher fault-tolerance thresholds for surface-code implementations employing native CZZ gates~\cite{tasler.CZZ_for_surface.2025}.

% A prominent example is the surface code, whose stabilizer measurement circuit requires four sequential CZ operations sharing a common ancilla qubit. Implementing CZZ gates enables two of these interactions to be performed in parallel, effectively halving the time required for the correction cycle. This reduction in circuit depth mitigates the effects of decoherence and thereby lowers the overall logical error rate.

%% file: sections/performance.tex
To assess the architecture performance, we focus on the native two-qubit operations. Specifically, we numerically simulate CZ and CZZ gates in six realistic three-qubit Q–C–Q–C–Q subcircuits, which encompass all representative coupler configurations (including $\mathcal{O}_0$ oscillator, where required) and capture the essential spectator interactions.

We simulate the time dynamics of each three-qubit subcircuit under truncated Gaussian drive pulses applied to the relevant couplers, numerically integrating the Schrödinger equation (Appendix~\ref{appendix:model_dynamic}). The resulting operation matrix $U$ in the three-qubit computational basis is used to evaluate the coherent CZ gate error with the standard equation \cite{pedersenFidelityQuantumOperations2007}:
\begin{equation}
    \varepsilon = 1 - \frac{\Tr(U^{\dagger}U) + \abs{\Tr(U^{\dagger}_{ideal} U)}^2}{d\,(d+1)},
    \label{eq:error}
\end{equation}
where $d= 2^3$ is the dimension of the computational Hilbert space and $U_\text{ideal}$ denotes the ideal operation matrix for a CZ or CZZ gate. For each configuration, the drive parameters (pulse duration, amplitude, and frequency) are optimized to minimize $\varepsilon$. Then, we decompose the resulting coherent error into leakage $\varepsilon_{\mathrm{leak}}$ and phase $\varepsilon_{\mathrm{\varphi}}$ contributions. For the CZ gate with a diagonal unitary matrix, these components can be defined as:
\begin{equation}
  \varepsilon_{\mathrm{leak}} = 1 - \frac{\Tr(U^{\dagger}U) + (\sum_m \abs{U_{mm}})^2}{d\,(d+1)},
    \label{eq:error_leak}
\end{equation}
mainly accounting for nonunitarity errors (population leakage), and
\begin{equation}
  \varepsilon_{\varphi} = \frac{d^2 - \abs{\sum_m U_{mm}^\text{ideal} (U_{mm}/\abs{U_{mm}})}^2}{d\,(d+1)},
    \label{eq:error_phi}
\end{equation}
characterizing phase errors of the normalized diagonal elements of the resulting gate matrix.

This decomposition reflects the nature of the CZ gate, which ideally accumulates a $\pi$-phase on a selected basis state while conserving its populations. Accordingly, $\varepsilon_{\mathrm{\varphi}}$ captures deviations in the acquired phases, whereas  $\varepsilon_{\mathrm{leak}}$ quantifies population leakage irrespective of phase. The resulting gate durations and error rates are summarized in Table~\ref{tab:cz_errors}.
\begin{table}[!h]
    \centering
    \begin{tabularx}{\linewidth}{ 
    l
    >{\centering\arraybackslash}X
    >{\centering\arraybackslash}X
    >{\centering\arraybackslash}X
    >{\centering\arraybackslash}X
    c}
    
        \toprule
        \toprule
        
        CZ $\&$ CZZ & \multicolumn{4}{c}{Gate errors, $10^{-5}$} & Duration, ns \\
        & $\varepsilon_{\mathrm{\varphi}}$ & $\varepsilon_{\mathrm{leak}}$ & $\varepsilon_{\mathrm{ME}}^{\mathrm{C}}$ & $\varepsilon_{\mathrm{ME}}^{\mathrm{C}\&\mathrm{Q}}$ & \\
        \midrule
        $\mathbf{C0_U}$|$\mathrm{C0_L}$ & 0.3 & 5.8 & 26 & 129 & 63.2 \\[0.1cm]
        $\mathrm{C0_U}$|$\mathbf{C0_L}$ & 0.4 & 2.7 & 22 & 119 & 59.2 \\
        \midrule
        $\mathbf{C1_U}$|$\mathrm{C1_L}$ & 0.3 & 4.2 & 26 & 144 & 66.7 \\[0.1cm]
        $\mathrm{C1_U}$|$\mathbf{C1_L}$ & 1.5 & 1.5 & 23 & 135 & 63.0 \\
           
        \midrule
        $\mathbf{C0_U}$|$\mathrm{C1_U}$ & $<$0.1 & 3.6 & 23 & 125 & 62.2 \\[0.1cm]
        $\mathrm{C0_U}$|$\mathbf{C1_U}$ & $<$0.1 & 2.4 & 23 & 137 & 64.3 \\[0.1cm]
        $\mathbf{C0_U}$|$\mathbf{C1_U}$ & 1.4 & 10.5 & 55 & 160 & 61.8$\&$64.2 \\
        \midrule
        $\mathbf{C0_L}$|$\mathrm{C1_U}$ & $<$0.1 & 1 & 21 & 121 & 61.9 \\[0.1cm]
        $\mathrm{C0_L}$|$\mathbf{C1_U}$ & 5.1 & 2.8 & 28 & 144 & 65 \\[0.1cm]
        $\mathbf{C0_L}$|$\mathbf{C1_U}$ & 4.6 & 8.2 & 65 & 172 & 71.1$\&$75.5 \\
        \midrule
        $\mathbf{C0_U}$|$\mathrm{C1_L}$ & $<$0.1 & 7.1 & 27 & 133 & 65 \\[0.1cm]
        $\mathrm{C0_U}$|$\mathbf{C1_L}$ & 5 & 1.7 & 26 & 135 & 61.9 \\[0.1cm]
        $\mathbf{C0_U}$|$\mathbf{C1_L}$ & 6.9 & 10.1 & 86 & 191 & 70.3$\&$72.2 \\
        \midrule
        $\mathbf{C0_L}$|$\mathrm{C1_L}$ & $<$0.1 & 2.2 & 22 & 122 & 61.5 \\[0.1cm]
        $\mathrm{C0_L}$|$\mathbf{C1_L}$ & 0.1 & 2.5 & 23 & 132 & 61.8 \\[0.1cm]
        $\mathbf{C0_L}$|$\mathbf{C1_L}$ & 69.2 & 20.3 & 763 & 863 & 80$\&$100\\
        
        \bottomrule
        \bottomrule
        
    \end{tabularx}
    \caption{Error rates and pulse durations for CZ and CZZ operations. In each row, the operating couplers are shown in bold — one in the case of a CZ gate and two for a CZZ gate. The coherent error is decomposed into the phase error $\varepsilon_{\mathrm{\varphi}}$ and the leakage error $\varepsilon_{\mathrm{leak}}$. Incoherent errors $\varepsilon_{\mathrm{ME}}^{\mathrm{C}}$ account for coupler decoherence, while $\varepsilon_{\mathrm{ME}}^{\mathrm{C\&Q}}$ include qubit decoherence as well ($T^{\mathrm{C}}_1= T_{\varphi}^{\mathrm{C}}=50 \, \mu s$, $T^{\mathrm{Q}}_1= 300\, \mu s$ and $T_{\varphi}^{\mathrm{Q}}=100 \, \mu s$).}
    \label{tab:cz_errors}
\end{table}
\minor{Gates in $\mathrm{C0_L}$|$\mathrm{C1_U}$, $\mathrm{C0_U}$|$\mathrm{C1_L}$, and $\mathrm{C0_L}$|$\mathrm{C1_L}$ exhibit deviating error rates, which we explain below.}

Appendix~\ref{subappendix:ZZ_error} shows that the phase error rate $\varepsilon_{\varphi}$ is dominated by the parasitic ZZ interaction between the coupler and the spectator, making it a sensitive indicator of ZZ suppression efficiency. \minor{In particular, the long-range parasitic capacitive couplings in the $\mathrm{C0_L}$|$\mathrm{C1_U}$ and $\mathrm{C0_U}$|$\mathrm{C1_L}$ configurations induce large coupler-spectator ZZ rates for C1 couplers ($\zeta_{\mathrm{CS}}\sim210\ \text{kHz}$), which consequently cause exceptionally high phase error rates.} However, in a complete lattice including $\mathcal{O}_1$ oscillators, this issue is resolved as part of the strongly interconnected square problem (see Section~\ref{section:parasitic_interactions} B), resulting in a strong mitigation of these phase errors.

Leakage errors primarily originate from the residual population of the target and the side coupler states. Despite the strong coupling of many degrees of freedom in the three-qubit system, leakage to complex hybridized or oscillator-associated states remains negligible. This is ensured through proper coupler-frequency configuring based on the analysis described in Appendix~\ref{appendix:leakage_model}. Appendix~\ref{appendix:leakage_channels} further identifies characteristic leakage pathways that should be avoided through design or calibration.

Beyond the coherent dynamics, we simulate CZ gates in the presence of decoherence using the same optimized drive pulses. To account for decoherence effects, we solve the Lindblad equation with decay and dephasing jump operators associated with the qubits and the coupler target transitions, choosing representative coherence times $T^{\mathrm{C}}_1 = T_{\mathrm{\varphi}}^{\mathrm{C}} = 50 \, \mu\text{s}$ for the transmon couplers and $T^{\mathrm{Q}}_1 = 300 \, \mu\text{s}$, and $T_{\mathrm{\varphi}}^{\mathrm{Q}} = 100 \, \mu\text{s}$ for fluxonium qubits.
Finally, we compute the Pauli transfer matrix (PTM) $R$ of the process in the three-qubit basis (Appendix~\ref{appendix:model_dynamic}) and evaluate the incoherent CZ gate error using the expression \cite{nielsen2002simple}:
\begin{equation}
    \varepsilon_{\mathrm{ME}} = \frac{d^2-\Tr(R\,R^{\dagger}_\text{ideal})}{d\,(d+1)}
    \label{eq:error_me}
\end{equation}
where $R_\text{ideal}$ is a PTM representation of the target operator. 

We compare two scenarios: decoherence in the couplers only and combined decoherence in both couplers and qubits; the resulting errors are listed in Table~\ref{tab:cz_errors}. Although qubits possess significantly longer coherence times, their decoherence dominates the total error, while coupler decoherence produces a comparatively minor contribution. This effect is studied in detail in Appendix~\ref{appendix:coupler_decoherence_impact}, where we show that in a general Q-C-Q system, the coupler and qubit coherence times corresponding to the same error rate are a factor of $<1/11$ for relaxation and $<1/6$ for dephasing.

Finally, we analyze CZZ operations and observe two main effects. \minor{First, spectral gap reduction limits gate fidelity in the $\mathrm{C0_L}|\mathrm{C1_L}$ configuration making it unsuitable for CZZ operations as discussed in Section ~\ref{section:CZZ_gates}}. Second, the leakage rate of a CZZ gate exceeds the combined leakage of the two corresponding CZ gates. We attribute this additional leakage to side transitions unique to the CZZ and two-photon processes involving side transitions induced by the combined drives.

Despite these disadvantages in the coherent regime, the CZZ gate outperforms two sequential CZs in the presence of decoherence. While coupler decoherence affects both schemes equally due to their identical number of $2\pi$ rotations ($\varepsilon_{\mathrm{ME}}^{\mathrm{C}}$ error), qubit decoherence contributes to the CZZ error at only half the rate of the sequential scheme. Consequently, the total decoherence error $\varepsilon_{\mathrm{ME}}^{\mathrm{C\&Q}}$ for the CZZ gate is substantially reduced relative to two sequential CZ operations: a $39\%$ reduction for $\mathrm{C0_U}|\mathrm{C1_U}$, $35\%$ reduction for $\mathrm{C0_U}|\mathrm{C1_L}$, and $29\%$ reduction for $\mathrm{C0_L}|\mathrm{C1_U}$.

%% file: sections/conclusion.tex
In this work, we introduced a scalable square-grid architecture of fluxonium qubits interconnected by transmon couplers and developed a systematic framework to overcome the parasitic interactions inherent in microwave-activated two-qubit gate designs with fixed couplings. Through a combination of frequency allocation, the introduction of four spectrally distinct coupler types, and strategic layout optimization, we localized the relevant excited states of the coupler and effectively eliminated unwanted interactions among neighboring elements. To suppress residual long-range coupler–spectator and coupler-coupler interactions, we utilize a differential oscillator element that provides additional degrees of freedom for coupling optimization. These approaches \forth{guarantee suppression of crosstalk and the CZ gate phase-error rate down to a negligible level} and establish a versatile hardware toolbox for suppressing both short- and long-range parasitic couplings without sacrificing the benefits of all-microwave control.

Our numerical simulations demonstrate that this architecture supports high-fidelity CZ gates with coherent errors below $10^{-4}$ and enables CZZ gates with comparable durations and error rates around $10^{-4}$. \second{Robustly maintaining these fidelities in the presence of fabrication-induced fluctuations of $E_{\mathrm{J}}$ requires a realistic $\sigma(\delta E_{\mathrm{J}}/E_{\mathrm{J}}) < 0.5\%$.} Importantly, CZZ entangling operations maintain fidelity advantages under realistic decoherence, achieving up to a $39\%$ reduction in total error compared with sequential CZ gates. Taken together, these results outline a clear path toward interaction-resilient fluxonium processors that remain robust to crosstalk and parasitic couplings even in densely connected lattices. The demonstrated principles are broadly applicable to other fixed-coupling superconducting qubit systems and provide a practical foundation for scalable, high-performance quantum computing with microwave-only control.

%% file: appendixes/differential_fluxonium.tex
In this appendix, we examine two essential properties of the differential fluxoniums used in this work. First, although the circuit contains two superconducting islands -- formally providing two degrees of freedom -- the differential fluxonium supports only a single physical mode; no secondary mode appears. Second, the effective mutual capacitance between elements antisymmetrically connected to a differential fluxonium remains low.

\begin{figure}[h!]
    \center{\includegraphics[width=\linewidth]{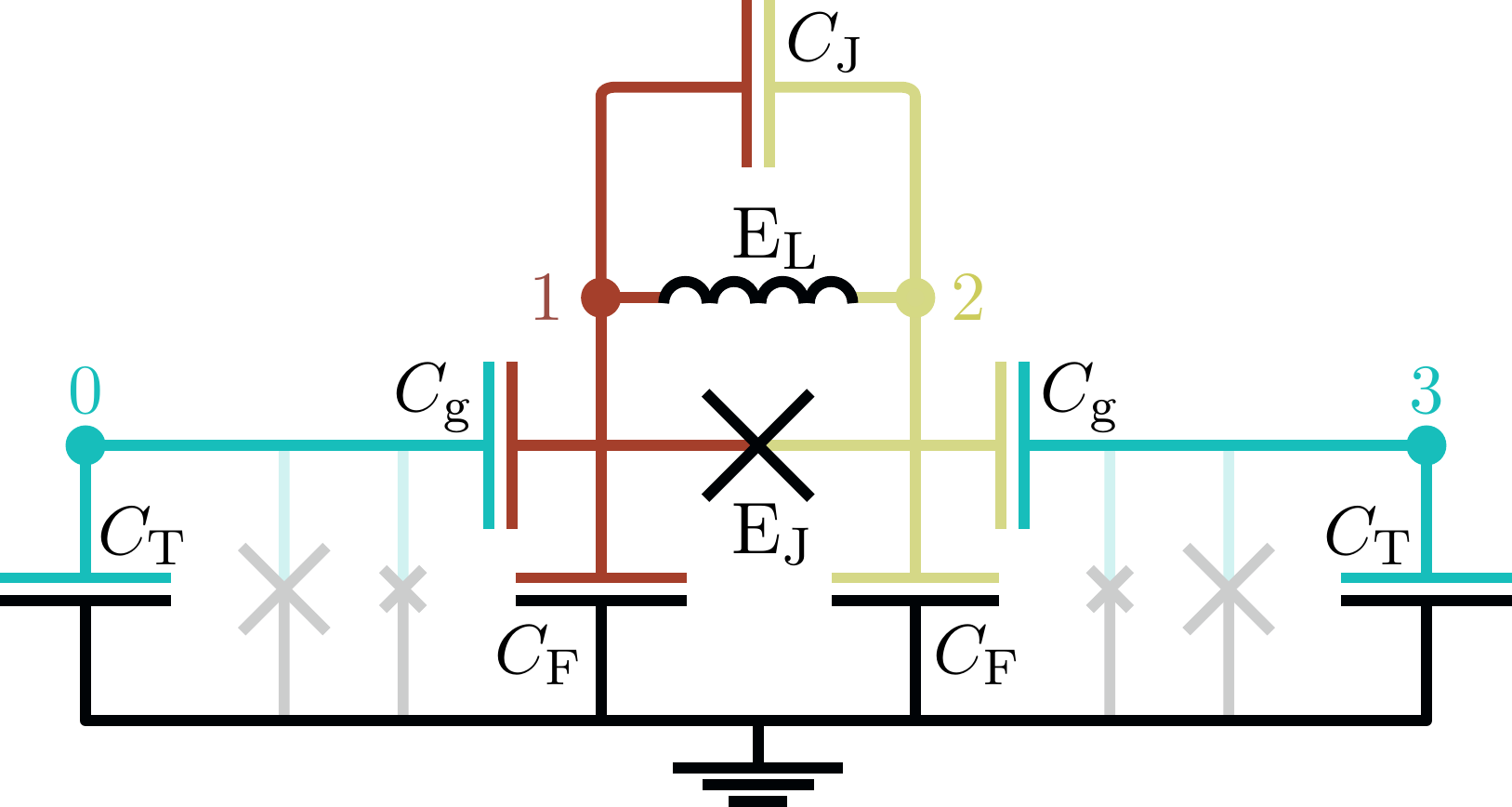}}
    \caption{Circuit consisting of two tunable transmon couplers (islands 0 and 3) coupled antisymmetrically to a differential fluxonium (islands 1 and 2). The capacitance between the transmons is omitted, as it is expected to be negligible.}
    \label{fig:CQC_circuit}
\end{figure}

To analyze these properties, we consider the circuit shown in Fig.~\ref{fig:CQC_circuit}, which consists of two couplers capacitively connected to the distinct islands of a differential fluxonium (antisymmetric connection).

The corresponding capacitance matrix is:
\begin{equation}
\mathbf{C} = 
\begin{pmatrix}
C_{\mathrm{T}}+C_{\mathrm{g}} & -C_{\mathrm{g}} & 0 & 0  \\
-C_{\mathrm{g}} & \Tilde{C}_{\mathrm{F}} + C_{\mathrm{J}} & -C_{\mathrm{J}} & 0  \\
0 & -C_{\mathrm{J}} & \Tilde{C_{\mathrm{F}}} + C_{\mathrm{J}} & -C_{\mathrm{g}} \\
0 & 0 & -C_{\mathrm{g}} & C_{\mathrm{T}}+C_{\mathrm{g}}  \\
\end{pmatrix},
\end{equation}
where $\Tilde{C}_{\mathrm{F}} = C_{\mathrm{F}} + C_{\mathrm{g}}$. Naturally, the Josephson junction's effective capacitance, $C_{\mathrm{J}}$, is much smaller than all other capacitances in the system, while the transmon capacitance $C_{\mathrm{T}}$ is chosen to greatly exceed the self-capacitance of a single fluxonium island (Table~\ref{tab:capacitance_table}). Thus, in the following analysis, we assume $C_{\mathrm{J}} \ll \Tilde{C}_{\mathrm{F}}, C_{\mathrm{g}} \ll C_{\mathrm{T}}$.

The corresponding capacitive energy matrix is:
\begin{equation}
    \mathcal{E}=4 \,e^2\,\mathbf{S}\mathbf{C}^{-1}\mathbf{S}^{\mathrm{T}},
    \label{eq:capacitive_energy}
\end{equation}
where $\mathbf{S}$ is the standard variable transformation for differential elements:
\begin{equation}
    \mathbf{S} = 
    \begin{pmatrix}
    1 & 0 & 0 & 0  \\
    0 & 1 & -1 & 0  \\
    0 & 1 & 1 & 0 \\
    0 & 0 & 0 & 1  \\
    \end{pmatrix}
    \label{eq:S_matrix}
\end{equation}

In the transformed basis, the Hamiltonian in the subspace of indices 1 and 2 takes the form:
\begin{equation}
    \renewcommand{\arraystretch}{2.2}
    \begin{array}{ll}
    
        \displaystyle \hat{H} = \frac{1}{2}\mathcal{E}_{11} \hat{n}_1^2 + \frac{1}{2}\mathrm{E_L} \hat{\varphi}_1^2 - \mathrm{E_J} \cos{\hat{\varphi}_1} \,+\, \frac{1}{2}\mathcal{E}_{22}  \hat{n}_2^2,
        
    \end{array}
\end{equation}
The first degree of freedom corresponds to a fluxonium mode, while the second behaves as a free particle and does not influence the system \cite{chitta2022numerical_methods}. Consequently, despite involving two superconducting islands, the differential fluxonium effectively hosts only a single physical mode.

The capacitive couplings between the transmons and the fluxonium mode are:
\begin{equation}
    \mathcal{E}_{01} = - \mathcal{E}_{13} \approx 4 e^2\, \frac{C_{\mathrm{g}}}{C_{\mathrm{T}}\,\Tilde{C}_{\mathrm{F}}},
\end{equation}
while the effective direct transmon-transmon coupling is:
\begin{equation}
    \mathcal{E}_{03} \approx \frac{C_{\mathrm{g}}\,C_{\mathrm{J}}}{C_{\mathrm{T}}\,\Tilde{C}_{\mathrm{F}}} \cdot \mathcal{E}_{01} \ll \mathcal{E}_{01}.
    \label{eq:antisym_coupling}
\end{equation}
Under the architecture parameters of Table~\ref{tab:capacitance_table}, Eq.\ref{eq:antisym_coupling} shows that $\mathcal{E}_{03}$ is three orders of magnitude smaller than $\mathcal{E}_{01}$, demonstrating that antisymmetric connection strongly suppresses direct capacitive coupling between the couplers. We also note that this antisymmetric connection produces opposite signs for  $\mathcal{E}_{01}$ and $\mathcal{E}_{13}$, a property essential for Eq.~(\ref{eq:ZZ_equation}).

Our architecture also incorporates differential oscillators. We propose implementing them using a capacitively grounded Josephson junction array. The corresponding circuit is identical to Fig.~\ref{fig:CQC_circuit}, except that single Josephson junction is removed ($E_{\mathrm{J}} = 0$). Therefore, all properties discussed in this appendix apply directly to these differential oscillators as well.

%% file: appendixes/model_static.tex
\begin{figure}[ht!]
    \center{\includegraphics[width=\linewidth]{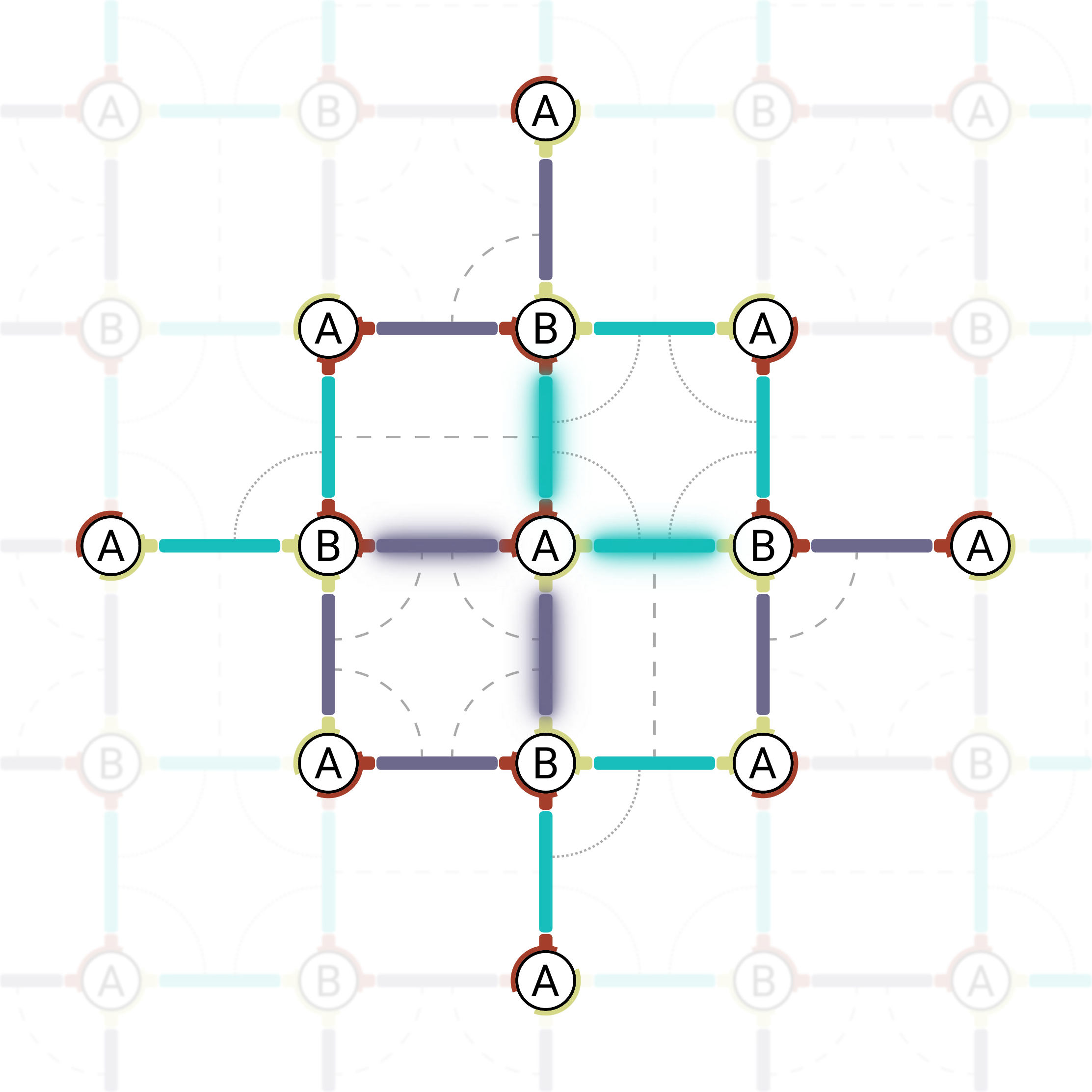}}
    \caption{The 13-qubit circuit used to analyze long-range interactions.It includes all possible spectators for the highlighted couplers, thereby capturing every type of coupler-spectator ZZ interaction in the grid, and contains all representative pairs of nearest identical couplers.}
    \label{fig:architecture_model}
\end{figure}

To thoroughly examine all parasitic interactions within our architecture, we build a numerical model of a 13-qubit system (Fig. \ref{fig:architecture_model}). 
This configuration encompasses all next-nearest neighbors of the central qubit, ensuring that every spectator of the associated couplers is represented. Since these couplers span all possible coupler types, the represented circuit in Fig. \ref{fig:architecture_model} captures every forms of the coupler-spectator ZZ interaction. Moreover, it incorporates all characteristic configurations of nearest identical coupler pairs.

To quantify the described model, we define capacitances for the individual elements (Table~\ref{tab:capacitance_table}) to fit the relevant capacitive energies. Next, we combine these capacitances in the 58-node matrix $\mathcal{C}$ corresponding to the circuit in Fig.~\ref{fig:architecture_model} and compute the capacitive energy matrix $\mathcal{E}$ using the equation (\ref{eq:capacitive_energy}), where the transfer matrix $\mathbf{S}$ implements the standard variable transformation for all differential elements (fluxoniums and oscillators).

The key advantage of this approach is that it explicitly accounts for all direct capacitive couplings, including the long-range ones that are essential for accurately describing long-range parasitic interactions.

However, the exponential growth of computational complexity makes it impossible to diagonalize the full Hamiltonian of the built 13-qubit system. We, therefore, reduce it to subspaces corresponding to the Q-C-Q-C-Q and C-Q-C-Q-C subcircuits of the lattice, within which the Hamiltonian takes the generalized form:
\begin{equation}
    \renewcommand{\arraystretch}{3}
    \begin{array}{ll}
        \displaystyle \hat{H} = \sum_{i \in \mathrm{F}} \left(\frac{1}{2}\,\mathcal{E}_{ii} \, \hat{n}_i^2  + \frac{1}{2}\,\mathrm{E_L}^{(i)} \, \hat{\varphi}_i^2 \,-\, \mathrm{E_J}^{(i)} \, \cos{\hat{\varphi}_i}\right)
        
        \\\displaystyle  \quad\,+\, \sum_{i \in \mathrm{T}} \left(\frac{1}{2}\,\mathcal{E}_{ii} \, \hat{n}_i^2  \,-\, \mathrm{E_J}^{(i)} \, \cos{\hat{\varphi}_i}\right)
        
        \\\displaystyle  \quad\,+\, \sum_{i \in \mathrm{O}} \left(\frac{1}{2}\,\mathcal{E}_{ii} \, \hat{n}_i^2  + \frac{1}{2}\,\mathrm{E_L}^{(i)} \, \hat{\varphi}_i^2 \right)
        
        \\\displaystyle \quad\,+\, \sum_{i < j}{\mathcal{E}_{ij} \, \hat{n}_i \hat{n}_j},
    \end{array}
    \label{eq:universal_circuit_Ham}
\end{equation}
where $\hat{n}$ is a Cooper-pair number operator, $\hat{\varphi}$ is a phase operator, and $\mathrm{E_L}$, $\mathrm{E_J}$ are the individual element energies given in Table~\ref{tab:energy_table}. The indices refer to the involved degrees of freedom: F – fluxoniums, T – transmons, O – oscillators.

The justification for this reduction differs among the cases discussed in Sec.~\ref{section:parasitic_interactions}. In particular, for an antisymmetric connection mediating the elements whose interaction is examined, the strong suppression of the direct long-range couplings (see Appendix~\ref{appendix:differential_fluxonium}) ensures that the contributions of unaccounted elements appear at higher order in perturbation theory and are therefore negligible. This can be understood in terms of virtual photon transition diagrams (see Appendix~\ref{appendix:virtual_transitions}), which must be longer to incorporate additional elements lacking direct couplings.

For the subcircuits of strongly interconnected squares, we explicitly verify that adding an additional qubit to the C–Q–C–Q–C subcircuits does not qualitatively affect the interactions discussed in Section \ref{section:parasitic_interactions}. Meanwhile, neglecting the fourth coupler constitutes an assumption justified by the effective suppression of coupler–coupler interactions.

To diagonalize the reduced Hamiltonians, we apply several stages of further reductions. First, we individually diagonalize distinct elements, truncating the obtained spectra to 6 levels for fluxoniums, 4 levels for transmons, and 3 levels for oscillators. We then combine these components according to the following rules:
\begin{enumerate}
    \item For Q-C-Q-C-Q circuits, we diagonalize C-Q-C and truncate the resulting spectrum to 70 levels, then we diagonalize the three-part Q-CQC-Q system;

    \item For Q-C-(Q,O)-C-Q circuits with an oscillator O between the couplers, we diagonalize C-(Q,O)-C with 209-level truncation, and then Q-CQOC-Q;

    \item For C-Q-C-Q-C circuits, we diagonalize Q-C-Q with 139-level truncation and then C-QCQ-C;

    \item For C-Q-(C,O)-Q-C circuits, we diagonalize Q-(C,O)-Q with 144-level truncation and then diagonalize C-QCOQ-C.
    
\end{enumerate}

In the described stages, we carry $\hat{n}$ operators for each element, thereby including all couplings.

%% file: appendixes/model_dynamic.tex
As discussed in Section \ref{section:performance}, we simulate CZ gate dynamics in Q-C-Q-C-Q subcircuits. We begin by diagonalizing the corresponding static Hamiltonians $H_0$ (see Appendix~\ref{appendix:model_static}), truncating the resulting spectra to 180 levels for $\mathrm{C0_U}|\mathrm{C0_L}$ configuration and 200 for all others (C0-type couplers require fewer levels due to the lower frequencies). These truncations significantly accelerate the simulations while preserving the key physical features, such as the gap shrinking and the considerable Stark shift of a target transition frequency induced by the second excited levels of couplers \cite{mazhorin2025impact}.

We then construct time-dependent Hamiltonians by adding flux drives applied to the couplers. To facilitate further computations, we switch to the interaction picture and apply the rotating wave approximation (RWA), which is justified by the high coupler frequencies. The resulting Hamiltonians take the form:
\begin{equation}
    \renewcommand{\arraystretch}{2}
    \begin{array}{ll}
        \displaystyle \hat{H}^{\mathrm{RWA}}_{\mathrm{int}}(t) = \sum_{n = \mathrm{C_1},\mathrm{C_2}}\frac{1}{2}\,\mathcal{\epsilon}_n(t)\,
        e^{+i\hat{H}_0t} \times
        \\ \quad\quad\quad\quad\quad\quad\,\, \displaystyle \left(\varphi_n^{\mathrm{up}}\,e^{+i f_n t} + \varphi_n^{\mathrm{low}}\,e^{-i f_n t}\right)\, e^{-i\hat{H}_0t},
    \end{array}
    \label{eq:time-dependent_sim_H}
\end{equation}
where $\varphi_n^\text{up}$ and $\varphi_n^\text{low}$ are the upper and lower triangle parts of the corresponding transmon phase operator, $f_n$ is the drive frequency, and $\mathcal{\epsilon}_n$ is the pulse envelope. To suppress leakage into the coupler side transitions, we choose this envelope to be truncated Gaussian:
\begin{equation}
    \displaystyle \mathcal{\epsilon}(t) = \frac{A}{1-e^{-\frac{\tau^2}{8\sigma^2}}} \cdot \left( e^{-\frac{(t - \tau/2)^2}{2\sigma^2}} - e^{-\frac{\tau^2}{8\sigma^2}} \right),
    \label{eq:Gauss_envelope}
\end{equation}
where $A$ is the amplitude, $\tau$ is the pulse duration, and $\sigma=\tau/\sqrt{-8 \ln{0.6}}$ determines the truncation point, chosen such that $60\%$ of the standard Gaussian tails are removed.

To compute an operation matrix $U$ for certain drive pulse parameters in the noiseless case, we numerically solve the Schrödinger equation with the described Hamiltonian (\ref{eq:time-dependent_sim_H}), setting computational basis states $\ket{\psi_i(0)}$ as initial conditions. Then, we project the solutions $\ket{\psi_i(\tau)}$ on the computational basis to get the operation matrix elements as $U_{ij}=\bra{\psi_i(0)}\ket{\psi_j(\tau)}$.

Since our goal is a CZ gate, we further process $U$:  fix the global phase so that $U_{00}$ becomes real and compensate for single-qubit Z-rotations, achieving zero phases on $\ket{100}$, $\ket{010}$, and $\ket{001}$ states (the last step intrinsically involves transforming $U^{00}_{\pi}$ to $U^{11}_{\pi}$).

To incorporate decoherence, we compute the process superoperator $R$ for the same Hamiltonian. Setting matrices of computational superoperator basis as initial conditions, we numerically solve the Lindblad equation:
\begin{equation}
    \dot{\rho} = - \frac{i}{\hbar}\comm{\hat{H}^{\mathrm{RWA}}_{\mathrm{int}}(t)}{\rho} + \sum_i \hat{L}_i \rho \hat{L}_i^{\dagger} - \frac{1}{2}\left\{\hat{L}_i^{\dagger} \hat{L}_i, \rho\right\},
    \label{eq:GKSL}
\end{equation}
where $L_i$ are the jump operators specified in Eq. (\ref{eq:model_jump_operators}). These operators are constructed within a $2^5$-dimensional subspace consisting of the dressed computational states of the three data qubits along with the transitions of both couplers. The remaining dimensions are padded with zeros. We also simplify the operators associated with the couplers by neglecting terms related to dephasing and decay from side coupler transitions.

Finally, we transform the computed superoperator $R$ into the Pauli basis and compensate for all required single-qubit Z-rotations, as in the coherent case.

%% file: appendixes/coupler_decoherence_impact.tex
One of the main advantages of fluxonium qubits is their comparatively long characteristic coherence times. However, CZ operations in our architecture require exciting a high-frequency transmon coupler, whose coherence times are typically modest. In this appendix, we demonstrate that coupler decoherence contributes far less to CZ-gate error than qubit decoherence. Therefore, employing a transmon as a coupler does not compromise the intrinsic coherence advantage of fluxonium qubits.

To examine non-coherent CZ errors induced by different decoherence channels, we analyze a simplified model consisting of two qubits (A and B) and a coupler (C), each element treated as a two-level system. Under the RWA in the target transition frame, the time-dependent Hamiltonian for this model, incorporating the drive applied to the coupler, takes the form:
\begin{equation}
    \renewcommand{\arraystretch}{2.2}
    \begin{array}{lll}
        \displaystyle H(t)/h = \mathcal{G} \,\left( \hat{\sigma}^-_{\mathrm{A}} \hat{\sigma}^+_{\mathrm{A}} + \hat{\sigma}^-_{\mathrm{B}}  \hat{\sigma}^+_{\mathrm{B}}\right)\cdot \hat{\sigma}^+_{\mathrm{C}} \hat{\sigma}^-_{\mathrm{C}}
        
        \\\displaystyle \quad \quad \quad \,\,\,+ \, \frac{1}{2} \mathcal{\epsilon}(t) \cdot \left( \hat{\sigma}^-_{\mathrm{C}} \, e^{-i \Delta t} + \hat{\sigma}^+_{\mathrm{C}} \, e^{i \Delta t} \right),
    \end{array}
    \label{eq:basic_model_ham}
\end{equation}
where $\hat{\sigma}^-=\ket{0}\bra{1}$, $\hat{\sigma}^+=\ket{1}\bra{0}$, $\mathcal{G}$ is a coupler energy gap, $\Delta$ is the drive detuning, and $\mathcal{\epsilon}(t)$ is the Gaussian drive envelope defined in Eq.~(\ref{eq:Gauss_envelope}). 

We set $\mathcal{G}$ to 100 MHz and the pulse duration to 66 ns (similarly to the longest CZ operation in the complete model). Then, we optimize the remaining drive parameters by minimizing the coherent error $\varepsilon_{\mathrm{SE}}$ defined in Eq.~(\ref{eq:error}) following the procedure described earlier.

Once the optimal parameters are fixed, we solve the Lindblad equation using the Hamiltonian (\ref{eq:basic_model_ham}), varying the coherence times of both qubits (Q) and the coupler (C). The corresponding Lindblad jump operators representing decay and pure dephasing are defined as:
\begin{equation}
    \renewcommand{\arraystretch}{2.2}
    \begin{array}{lll}
        \displaystyle \hat{L}_{1}^{\mathrm{A,\,B}} = \frac{1}{2\sqrt{T_1^{\mathrm{Q}}}} \, \left(\hat{\sigma}_x^{\mathrm{A,\,B}} + i\,\hat{\sigma}_y^{\mathrm{A,\,B}}\right)
        \\ 
        \displaystyle \hat{L}_{\varphi}^{\mathrm{A,\,B}} = \frac{1}{\sqrt{2\,T_{\varphi}^{\mathrm{Q}}}} \, \hat{\sigma}_z^{\mathrm{A,\,B}}\\
        \displaystyle \hat{L}_{1}^{\mathrm{C}} = \frac{1}{2\sqrt{T_1^{\mathrm{C}}}}\, \left(\hat{\sigma}_x^{\mathrm{C}} + i\,\hat{\sigma}_y^{\mathrm{C}}\right)
        \\ 
        \displaystyle \hat{L}_{\varphi}^{\mathrm{C}} = \frac{1}{\sqrt{2\,T_{\varphi}^{\mathrm{C}}}}\,\hat{\sigma}_z^{\mathrm{C}}.
    \end{array}
    \label{eq:model_jump_operators}
\end{equation}
To calculate decoherence impacts, we compute CZ gate error with Eq.~(\ref{eq:error_me}), subtracting the coherent error part:
\begin{equation}
    \Delta\varepsilon_{\mathrm{ME}}(T) = \varepsilon_{\mathrm{ME}}(T) - \varepsilon_{\mathrm{ME}}(\infty).
\end{equation}

\begin{figure}[h!]
    \center{\includegraphics[width=\linewidth]{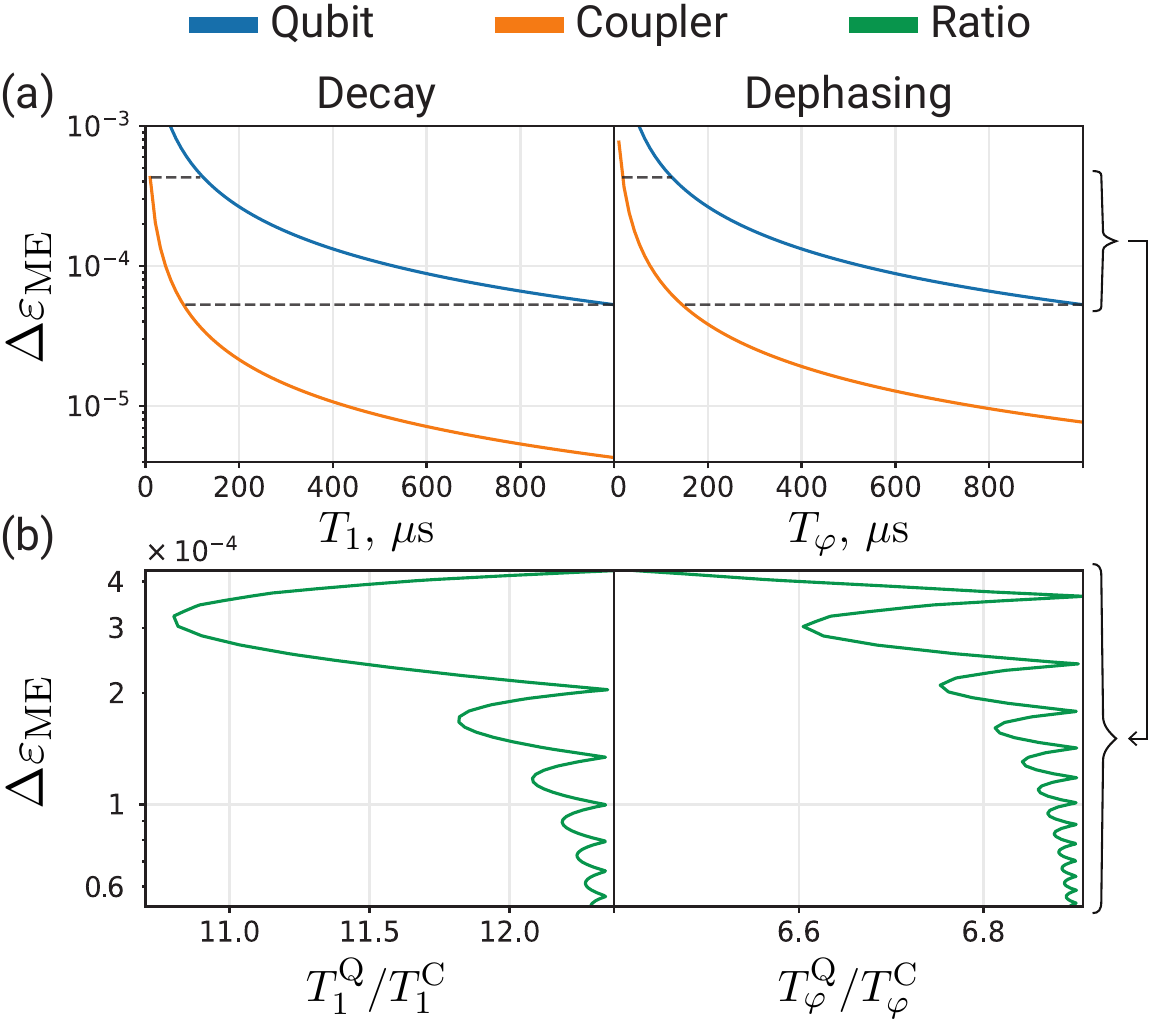}}
    \caption{(a) Qubit and coupler decoherence impacts separated case of decay (left) and pure dephasing (right). (b) The ratio between qubit and coupler coherence times corresponding to the same error rate impacts.}
    \label{fig:decoherence_QvsC_impact}
\end{figure}

Since the impacts of distinct decoherence sources prove to be additive with high accuracy (with deviations below $0.1\%$), we evaluate each contribution independently. The results are summarized in Fig.~\ref{fig:decoherence_QvsC_impact}.

The curves in Fig.~\ref{fig:decoherence_QvsC_impact}(a) show that the impact of qubit decoherence dominates that of the coupler. Fig.~\ref{fig:decoherence_QvsC_impact}(b) illustrates the ratio of qubit-to-coupler coherence times that yield the same decoherence-induced error. This ratio is substantially larger for energy relaxation -- up to 12.4 -- than for pure dephasing -- up to 6.9. That harmonizes with the choice of high-frequency transmon couplers: their dephasing times $T_{\varphi}$ tend to be relatively long, while their relaxation times $T_{1}$ decrease at high frequencies.

It is worth noting that the qubit-to-coupler coherence time ratio curve presented in Fig. \ref{fig:decoherence_QvsC_impact}(b) depends on the drive pulse parameters (both oscillations and the limit). Thus, while the presented ratios already ensure that coupler decoherence does not limit fluxonium performance, these factors can be further increased through suitable pulse-shape optimization.

%% file: appendixes/qq_analysis.tex
In this section, we discuss both longitudinal and transverse parasitic interactions between the nearest (NN) and next-nearest (NNN) qubits. The longitudinal interaction is quantified by the ZZ rate:
\begin{equation}
    \zeta_{\mathrm{qq}}=f_{\ket{11}'} - f_{\ket{01}'} -f_{\ket{10}'}+f_{\ket{00}'};
\end{equation}
where $f_{\ket{\cdot}'}$ denotes the frequency of the indicated dressed state. The transverse parasitic interaction is estimated by the qubit states hybridization measure $D(\ket{10}, \ket{01})$ defined in Eq.~(\ref{eq:D_hybridization_definition}). This parameter is particularly useful because it directly characterizes the strength of the parasitic component in the qubit drive operators. For a simple model of two qubits (A and B), these operators are:
\begin{equation}
    \renewcommand{\arraystretch}{2}
    \begin{array}{lll}
    
        \displaystyle \hat{H}_{\mathrm{drive}}^{\mathrm{A}}(t) = \epsilon_{\mathrm{A}} \sin{\omega_{\mathrm{A}}t} \cdot \hat{V}_{\mathrm{A}} ,
        \\ \displaystyle \hat{H}_{\mathrm{drive}}^{\mathrm{B}}(t) = \epsilon_{\mathrm{B}} \sin{\omega_{\mathrm{B}}t} \cdot \hat{V}_{\mathrm{B}},
    
        \\\displaystyle \hat{V}_{\mathrm{A}} = \sqrt{1 - D}\cdot \hat{\sigma}_x^{\mathrm{A}} + \sqrt{D}\cdot \hat{\sigma}_z^{\mathrm{A}} \hat{\sigma}_x^{\mathrm{B}},
        
        \\\displaystyle \hat{V}_{\mathrm{B}} = \sqrt{1 - D}\cdot \hat{\sigma}_x^{\mathrm{B}} - \sqrt{D}\cdot \hat{\sigma}_x^{\mathrm{A}} \hat{\sigma}_z^{\mathrm{B}}.
        
    \end{array}
    \label{ZX_influence}
\end{equation}
where $\epsilon_{\mathrm{A, B}}$ are the pulse amplitudes, and $\omega_{\mathrm{A, B}}$ are the drive frequencies. To estimate the error induced by the parasitic entangling component, we consider a resonant drive and, for simplicity, neglect qubit detuning, thereby obtaining an upper bound on the induced error. Under these assumptions, a nominal single-qubit rotation $R_X(\pi/2)$ applied to qubit A is accompanied by unintended two-qubit entangling. Considering $\mathrm{D} \ll 1$, we roughly estimate the corresponding unitary as $U\sim\exp(-i \sqrt{D}/2 \, (\hat{\sigma}_x^{\mathrm{A}} + \hat{\sigma}_y^{\mathrm{A}}) \, \hat{\sigma}_x^{\mathrm{B}})$. Finally, using Eq.~\ref{eq:error}, we estimate the associated two-qubit error as $\varepsilon \approx \, 2/5\, D$.

\subsection{Nearest neighboring qubits}

\begin{figure}[b!]
    \center{\includegraphics[width=\linewidth]{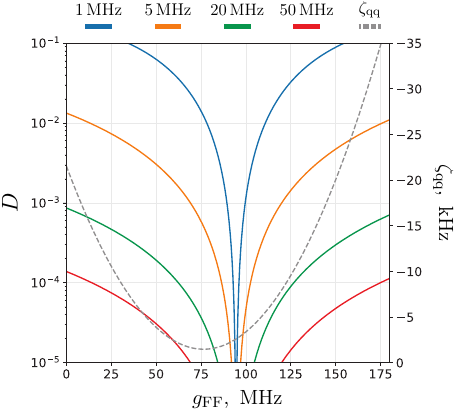}}
    \caption{Dependence of the NN-qubit hybridization $D$ and ZZ interaction $\mathrm{\zeta_{qq}}$ on the fluxonium coupling strength $g_{\mathrm{FF}}$ in the FTF system, evaluated for four different qubits detuning values $\Delta f_{\mathrm{qq}}$ (labeled above). The hybridization curves are color-coded by $\Delta f_{\mathrm{qq}}$ values, while $\mathrm{\zeta_{qq}}$ is shown as a single dashed line due to its insensitivity to qubits frequency shift. In the modeled FTF system, the effective mutual capacitive coupling between fluxoniums corresponds to $g_{\mathrm{FF}} \approx 87$ MHz.}
    \label{fig:FTF_dependency_on_df}
\end{figure}

The NN qubits are mediated by a transmon coupler, forming the FTF system. As demonstrated in \cite{MIT.FTF}, the corresponding ZZ interaction parameter $\mathrm{\zeta_{qq}}$ exhibits two key features: (1) relative insensitivity to qubit frequency variations and (2) a pronounced minimum as a function of the fluxonium–fluxonium coupling strength $g_{\mathrm{FF}}$. As for the qubit hybridization $D$, we examine its dependence on both $g_{\mathrm{FF}}$ and the qubit detuning $\Delta f_{\mathrm{qq}}$ through numerical diagonalization of the FTF Hamiltonian:
\begin{equation}
    \renewcommand{\arraystretch}{2}
    \begin{split}
        \hat{H}(g_{\mathrm{FF}}, \Delta f_{\mathrm{qq}}) = &\hat{H}_{\mathrm{T}} + \hat{H}_{\mathrm{F_1}}(+\Delta \mathrm{E_J}) + \hat{H}_{\mathrm{F_2}}(-\Delta \mathrm{E_J})+\\
        &g_{\mathrm{FT}}\,(\hat{n}_{\mathrm{F_1}} + \hat{n}_{\mathrm{F_2}})\,\hat{n}_{\mathrm{T}} + g_{\mathrm{FF}}\, \hat{n}_{\mathrm{F_1}} \hat{n}_{\mathrm{F_2}},
    \end{split}
    \label{QCQ_hybridization_model}
\end{equation}
where we take the parameters of the $\mathrm{C0_L}$-type coupler and A-type fluxonium (Table~\ref{tab:energy_table}) as references and adjust fluxonium $\Delta \mathrm{E_J}$ to achieve a specific qubit detuning $\Delta f_{\mathrm{qq}}$. We intentionally select the lowest-frequency coupler to examine the worst-case scenario for parasitic interaction suppression. The resulting dependencies shown in Fig.~\ref{fig:FTF_dependency_on_df} indicate 
that both $\mathrm{\zeta_{qq}}$ and $D$ exhibit minima as functions of $g_{\mathrm{FF}}$, albeit at different optimal values. Notably, the real effective fluxonium coupling, obtained via a complete capacitive circuit analysis, lies between these two optima. The $\mathrm{\zeta_{qq}}$ dependence is sufficiently smooth to yield robust suppression across a broad range of $g_{\mathrm{FF}}$, including the real effective value. By contrast, the hybridization parameter $D$ displays sharply peaked behavior and is highly sensitive to the qubit frequency detuning $\Delta f_{\mathrm{qq}}$. To maintain high-fidelity single-qubit gate operations, we target the suppression of the hybridization-induced error below $10^{-5}$. Based on the rough estimate made previously, this requires $D \lesssim 10^{-5}$ near the actual $g_{\mathrm{FF}}$, which in turn demands $\Delta f_{\mathrm{qq}} > 20$ MHz. To ensure robust suppression of hybridization errors, we implement fluxonium types A and B with a frequency detuning of $\Delta f_{\mathrm{qq}}\approx 50$ MHz.

\subsection{Next-nearest neighboring qubits}

\begin{figure}[h!]
    \center{\includegraphics[width=\linewidth]{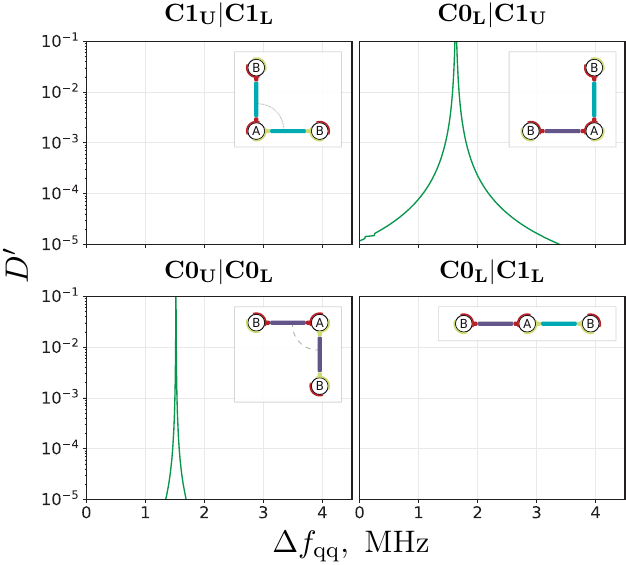}}
    \caption{Hybridization D of NNN qubits versus qubit-frequency detuning $\Delta f_{\mathrm{qq}}=|f_1 - f_2|$ for all four representative three-qubit circuits. The observed asymmetry (peaks shifted away from $\Delta f_{\mathrm{qq}}=0$) originates from the ordering of the couplers. In the complete square-grid architecture, each NNN qubit pair is connected via two coupling paths that sample both coupler orderings. This path degeneracy motivates us to use absolute $\Delta f_{\mathrm{qq}}$ values to capture the combined effect of both coupling pathways.}
    \label{fig:far_qq_dependency_on_df}
\end{figure}

Turning to the NNN qubits, their interactions are characterized by four representative Q-C-Q-C-Q configurations: $\mathrm{C1_U}|\mathrm{C1_L}$; $\mathrm{C0_U}|\mathrm{C0_L}$; $\mathrm{C0_L}|\mathrm{C1_U}$; $\mathrm{C0_L}|\mathrm{C1_L}$. (We do not consider $\mathrm{C0_U}|\mathrm{C1_U}$ and $\mathrm{C0_U}|\mathrm{C1_L}$, as in the context of NNN qubit coupling, they are similar to $\mathrm{C0_L}|\mathrm{C1_L}$ and $\mathrm{C0_L}|\mathrm{C1_U}$, respectively.) To examine the NNN qubit interactions, we numerically diagonalize the Hamiltonians for the characteristic circuits while varying $\Delta f_{\mathrm{qq}}$ through the $E_J$ modification. For each detuning, we compute $\mathrm{\zeta_{qq}}$ and $D'$, where the latter is a qubit-qubit hybridization modified to involve both states of the central qubit:
\begin{equation}
    \renewcommand{\arraystretch}{2}
    \begin{array}{lll}
    
        \displaystyle D' = \frac{1}{2} D(\ket{100}, \ket{001}) +\frac{1}{2} D(\ket{110}, \ket{011}).
        
    \end{array}
\end{equation}
This computation shows that, across all configurations, qubit-qubit ZZ interaction remains negligible ($\mathrm{\zeta_{qq}}<5$ Hz), while the hybridization reaches considerable values and exhibits a nontrivial dependence on $\Delta f_{\mathrm{qq}}$, as shown in Fig. \ref{fig:far_qq_dependency_on_df}.

Our simulation shows that non-trivial hybridization appears only in the $\mathrm{C0_U}|\mathrm{C0_L}$ and $\mathrm{C0_L}|\mathrm{C1_U}$ configurations. In the first case, it is caused by an additional oscillator mediating the couplers; however, due to the narrow $D$ peak, this effect proves practically negligible. Meanwhile, the $\mathrm{C0_L}|\mathrm{C1_U}$ configuration, being the building block of the strongly interconnected squares (Section~\ref{section:parasitic_interactions}), exhibits substantial hybridization due to numerous long-range capacitive couplings between its elements.

According to Fig.~\ref{fig:far_qq_dependency_on_df}, maintaining our target error threshold ($\varepsilon < 10^{-5}$) in this case requires a frequency detuning of $\Delta f_{\mathrm{qq}} > 3$ MHz for qubits located at opposite corners of strongly interconnected squares. In practice, such detuning might occur naturally from fabrication-induced variations and therefore does not require deliberate design compensation. However, to ensure the accuracy of our numerical modeling, we implement two fluxonium subtypes with a variation of $\Delta \mathrm{E_L}\approx1.5 \,\%$, arranging them in alternating rows throughout the grid.

%% file: appendixes/ZZ_CS.tex
\subsection{Impact to CZ gate performance}
\label{subappendix:ZZ_error}

We quantitatively examine how the $\zeta_{\mathrm{CS}}$ value affects CZ gate performance using a four-element model composed of two data qubits (A and B), a coupler (C), and a spectator qubit (S), with each component treated as a two-level system. When a near-resonant drive is applied to the coupler transition associated with the $\ket{11}_{\mathrm{AB}}$ computational state, the dynamics of the full system are governed by the Hamiltonian:
\begin{equation}
    \renewcommand{\arraystretch}{2.2}
    \begin{array}{lll}
        \displaystyle \hat{H}(t)/h = \mathcal{G} \,\left( \hat{\sigma}^-_{\mathrm{A}} \hat{\sigma}^+_{\mathrm{A}} + \hat{\sigma}^-_{\mathrm{B}} \hat{\sigma}^+_{\mathrm{B}}\right)\cdot \hat{\sigma}^+_{\mathrm{C}} \hat{\sigma}^-_{\mathrm{C}}
        
        \\\displaystyle \quad \quad \quad \,\,\,+ \, \zeta_{\mathrm{CS}} \cdot \hat{\sigma}^+_{\mathrm{S}} \hat{\sigma}^-_{\mathrm{S}} \cdot \hat{\sigma}^+_{\mathrm{C}} \hat{\sigma}^-_{\mathrm{C}}
        
        \\\displaystyle \quad \quad \quad \,\,\,+ \, \frac{1}{2} \mathcal{\epsilon}(t) \cdot \left( \hat{\sigma}^-_{\mathrm{C}} \, e^{-i \Delta t} + \hat{\sigma}^+_{\mathrm{C}} \, e^{i \Delta t} \right),
    \end{array}
    \label{eq:ZZ_simple_model}
\end{equation}
which is obtained in a manner similar to Eq.~\ref{eq:basic_model_ham}, with the sole modification being the term that defines the coupler–spectator ZZ interaction. As in Appendix \ref{appendix:model_dynamic}, we employ a truncated Gaussian envelope $\epsilon(t)$ defined in Eq.~\ref{eq:Gauss_envelope}. The model is therefore characterized by five parameters: two describing the system spectrum -- the spectral gap and the ZZ interaction strength ($\mathcal{G}$, $\zeta_{\mathrm{CS}}$); and three describing the drive -- duration, amplitude, and off-resonant detuning ($\tau$, $A$, $\Delta$).

We numerically solve the Schrödinger equation for the Hamiltonian (Eq.~\ref{eq:ZZ_simple_model}) and compute the reduced evolution operator $U$ for the three-qubit computational subspace $\mathrm{A \otimes B \otimes S}$, which includes the spectator. Then, we evaluate the error $\varepsilon$ of the CZ gate ($U_\text{ideal} = \mathrm{CZ_{AB}\otimes \mathbbm{1}_S}$) using Eq.~(\ref{eq:error}).

In our simulations, we fix the gap at $\mathcal{G}=100$ MHz, sweep $\zeta_{\mathrm{CS}}$ from 0 to 400 kHz, and vary the pulse duration $\tau$ from 30 to 100 ns. For each combination of these parameters, we optimize the amplitude $A$ and detuning $\Delta$, using SciPy to minimize the CZ error $\varepsilon(A, \Delta)$. This pulse calibration enables quantification of the error contribution from $\zeta_{\mathrm{CS}}$ for a given pulse duration and spectral gap.

\begin{figure}[h!]
    \center{\includegraphics[width=8cm]{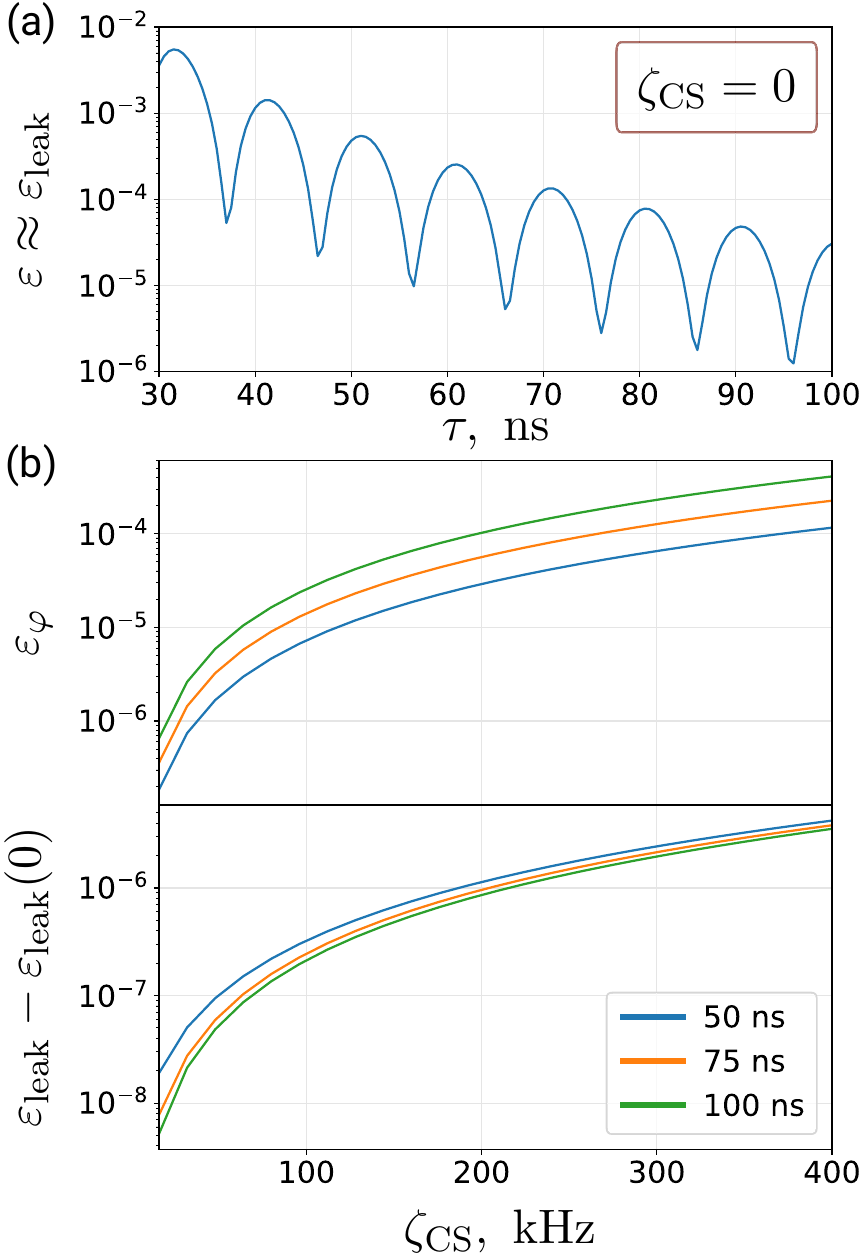}}
    \caption{CZ gate errors versus $\zeta_{\mathrm{CS}}$ in the system described by the Hamiltonian (\ref{eq:ZZ_simple_model}) with $\mathcal{G}=100$ MHz and truncated Gaussian drive pulses of varying durations $\tau$. For each combination of $\tau$ and $\zeta_{\mathrm{CS}}$, the pulse is optimized as a function of amplitude and detuning. The top panel (a) shows the dependence of gate errors on drive duration at $\zeta_{\mathrm{CS}}=0$, where phase errors are negligible and only leakage is shown. The bottom panel (b) depicts both phase errors and leakage variations versus $\zeta_{\mathrm{CS}}$ for three pulse durations.}
    \label{fig:spectator_error}
\end{figure}

We also decompose the total errors $\varepsilon$ into phase errors ($\varepsilon_{\mathrm{ph}}$) and leakage ($\varepsilon_{\mathrm{leak}}$) defined in Section~\ref{section:performance}. The resulting error relations, summarized in Fig. \ref{fig:spectator_error}, 
reveal the contrasting behavior between these two error channels. In the absence of coupler-spectator interaction ($\zeta_{\mathrm{CS}}=0$), the phase error remains negligible, and the total error is dominated by leakage. However, as $\zeta_{\mathrm{CS}}$ increases, $\varepsilon_{\mathrm{ph}}$ grows rapidly, while $\varepsilon_{\mathrm{leak}}$ shows minimal variation – remaining two orders of magnitude smaller. Notably, across nearly all pulse durations in the studied range, the leakage variation stays significantly lower than the $\varepsilon_{\mathrm{leak}}$ observed for $\zeta_{\mathrm{CS}}=0$. Thus, the coupler-spectator ZZ interaction generates phase errors in CZ gates, whereas the leakage is determined almost entirely by the drive pulse duration. Furthermore, our modeling demonstrates that the rate of phase error generation is directly proportional to the pulse duration.

\subsection{Approximate equation for CQCQ model}
\label{subappendix:ZZ_CS_equation}

As discussed in Section~\ref{section:parasitic_interactions}, the parasitic coupler-spectator ZZ interaction poses a significant challenge to scaling our architecture. While the numerical diagonalization of the relevant Hamiltonian provides accurate estimates of $\zeta_{\mathrm{CS}}$, developing systematic suppression methods requires a deeper analytical insight into its physical behavior. We therefore derive an approximate expression for the coupler-spectator ZZ rate (Eq.~\ref{eq:ZZ_equation}).

\begin{figure}[h!]
    \center{\includegraphics[width=0.8\linewidth]{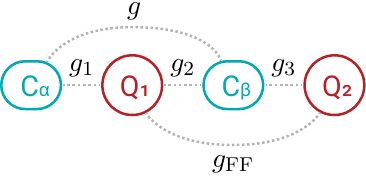}}
    \caption{Schematic of the C-Q-C-Q model Hamiltonian with purely capacitive couplings.}
    \label{fig:CQCQ_simple_model}
\end{figure}

First, we consider a simplified C-Q-C-Q model (Fig.~\ref{fig:CQCQ_simple_model}), where the coupler-spectator ZZ interaction 
strength is defined by Eq.~(\ref{eq:ZZ_definition}), depending on the type of $\mathrm{C_{\alpha}}$ coupler. The system's Hamiltonian is given by:
\begin{equation}
    \renewcommand{\arraystretch}{2}
    \begin{array}{llll}
        \displaystyle \hat{H} = \hat{H}_{\mathrm{T}}(\mathrm{E_C^{C_{\alpha}}, E_J^{C_{\alpha}}}) + \hat{H}_{\mathrm{T}}(\mathrm{E_C^{C_{\beta}}, E_J^{C_{\beta}}})
        
        \\\displaystyle \quad\,+\hat{H}_{\mathrm{F}}(\mathrm{E_C^{Q_1}, E_L^{Q_1}, E_J^{Q_1}}) + \hat{H}_{\mathrm{F}}(\mathrm{E_C^{Q_2}, E_L^{Q_2}, E_J^{Q_2}})

        \\\displaystyle \quad\,+g_1\,\hat{n}_{\mathrm{C_{\alpha}}}\hat{n}_{\mathrm{Q_1}} +g_2\,\hat{n}_{\mathrm{Q_1}} \hat{n}_{\mathrm{C_{\beta}}} +g_3\,\hat{n}_{\mathrm{C_{\beta}}} \hat{n}_{\mathrm{Q_2}}

        \\\displaystyle \quad\,+g\,\hat{n}_{\mathrm{C_{\alpha}}}\hat{n}_{\mathrm{C_{\beta}}} +g_{\mathrm{FF}}\,\hat{n}_{\mathrm{Q_1}} \hat{n}_{\mathrm{Q_2}},
    \end{array}
    \label{eq:CQCQ_model_Ham}
\end{equation}
where $\hat{H}_{\mathrm{F}}$ and $\hat{H}_{\mathrm{T}}$ are defined in Eq.~\ref{eq:elementary_Hams}, and $g_{i}$ denote the capacitive coupling strengths.

\begin{figure}[ht!]
    \center{\includegraphics[width=\linewidth]{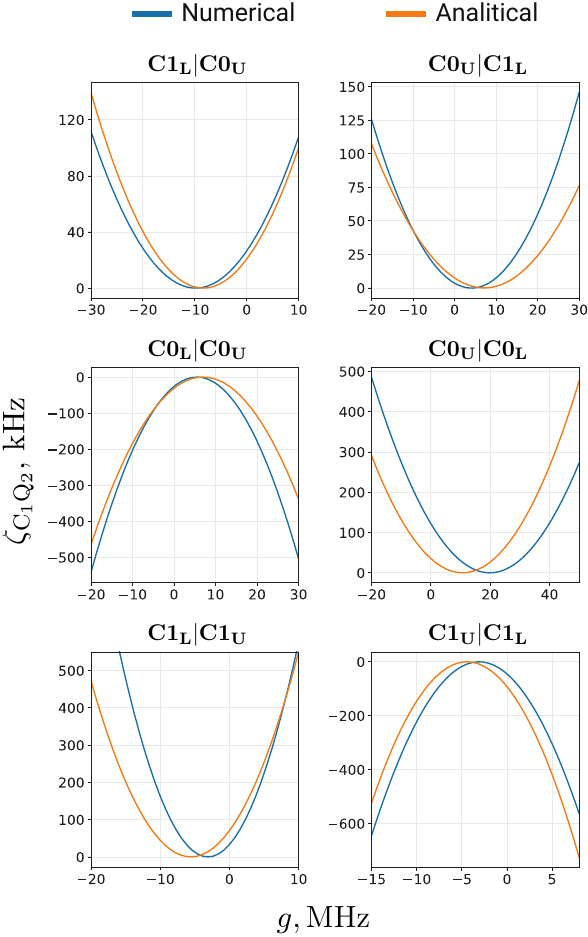}}
    \caption{Comparison of the $\zeta_{\mathrm{C_{\alpha}Q_2}}(g)$ obtained from the numerical computations and analytical approximations (\ref{eq:zz_equation_proof}) across different coupler configurations $\mathrm{C_{\alpha}|C_{\beta}}$ of the C-Q-C-Q system (see Fig.~\ref{fig:CQCQ_simple_model}).}
    \label{fig:zz_equation_proof}
\end{figure}

The parameters of the individual elements follow Table~\ref{tab:energy_table}. While $\mathrm{Q_1}$ and $\mathrm{Q_2}$ are fixed as type-A and type-B fluxoniums, respectively, we consider six distinct coupler configurations $\mathrm{C_{\alpha}|C_{\beta}}$ determined by coupler types and ordering: $\mathrm{C1_L|C0_U}$, $\mathrm{C0_L|C0_U}$, $\mathrm{C1_L|C1_U}$, and their reversed sequences (Fig.~\ref{fig:zz_equation_proof}).

The couplings are set as follows: $g_2 = g_3 = 300\,\mathrm{MHz}$ and $g_{\mathrm{FF}} = 80\,\mathrm{MHz}$ for $\mathrm{C0}$-type $\mathrm{C_{\beta}}$; $g_2 = g_3 = 200\,\mathrm{MHz}$ and $g_{\mathrm{FF}} =40\,\mathrm{MHz}$ for $\mathrm{C1}$-type $\mathrm{C_{\beta}}$; $g_1=300\,\mathrm{MHz}$ for $\mathrm{C0}$-type $\mathrm{C_{\alpha}}$ and $200\,\mathrm{MHz}$ for $\mathrm{C1}$-type $\mathrm{C_{\alpha}}$.

The remaining parameter $g$ serves as the sweep variable in the $\zeta_{\mathrm{C_{\alpha} Q_2}}(g)$ relation, which we subsequently fit with an analytical approximation. Notably, sweeping $g$ from positive to negative values encompasses both symmetric and antisymmetric couplers connections to $\mathrm{Q_1}$. As demonstrated in Appendix~\ref{appendix:differential_fluxonium}, the antisymmetric connection results in opposite signs for $g_1$ and $g_2$. This sign difference can be restored through the coordinate transformation $\hat{n}_{\mathrm{C_{\alpha}}} \rightarrow -\hat{n}_{\mathrm{C_{\alpha}}}$, which reverses the sign of $g$.

Using the defined Hamiltonian (Eq.~\ref{eq:CQCQ_model_Ham}), we numerically compute the $\zeta_{\mathrm{C_{\alpha} Q_2}}(g)$ relation. We first diagonalize each element separately, truncating the fluxonium spectra to 10 levels and the transmon spectra to 5 levels. Then we numerically diagonalize the full Hamiltonian (\ref{eq:CQCQ_model_Ham}) for all six coupler combinations without additional truncations, while sweeping $g$ and computing $\zeta_{\mathrm{C_{\alpha} Q_2}}$ via Eq. \ref{eq:ZZ_definition}.
% The resulting relations are shown in Fig. \ref{fig:zz_equation_proof}.

To obtain analytical approximations to these numerical results, we treat the capacitive coupling terms in Eq.~\ref{eq:CQCQ_model_Ham} as a perturbation $\hat{V}$ and apply perturbation theory to estimate the dressed-state energy shifts governing $\zeta_{\mathrm{C_{\alpha}Q_2}}$. To identify the dominant perturbation terms, we associate each with a virtual transition diagram (see Appendix~\ref{appendix:virtual_transitions}) and employ the following analysis.

Since the system involves only linear capacitive couplings, the condition $V_{ii} = 0$ holds, and every transition in a virtual transition diagram must correspond to either a virtual photon transfer between connected elements or a simultaneous photon creation/annihilation event.

In a first approximation, we restrict our analysis to photon transfers. Therefore, each transition diagram represents virtual photon propagation through the circuit. These diagrams are further constrained by the following rules. First, each photon transfer must obey parity conservation, which follows from the symmetry properties of both the fluxonium (biased at its sweet spot) and the transmon. For instance, a photon transfer $\ket{10} \rightarrow \ket{03}$ is allowed, whereas $\ket{10} \rightarrow \ket{02}$ is forbidden. Second, a transition diagram must form a continuous path connecting $\mathrm{C_{\alpha}}$ and $\mathrm{Q_2}$ in order to contribute to $\zeta_{\mathrm{C_{\alpha}Q_2}}$. This requirement reflects the nature of $\zeta_{\mathrm{C_{\alpha}Q_2}}$, which vanishes if the connection between the $\mathrm{C_{\alpha}}$ coupler and the $\mathrm{Q_2}$ qubit is broken. Finally, transition diagrams involving photon emission from a fluxonium $\ket{1}$ state contribute negligibly due to the small values of $n^{\mathrm{Q}}_{01}$ and $f^{\mathrm{Q}}_{01}$ (Table~\ref{tab:energy_table}).

The remaining diagrams include photon emission from the coupler $\mathrm{C_{\alpha}}$. Particularly, a photon propagates from $\mathrm{C_{\alpha}}$ to $\mathrm{Q_2}$ and back, forming a cycle. This restricts our analysis to one-cycle diagrams, as shorter diagrams -- corresponding to lower-order perturbations -- yield larger contributions.

Numerically comparing the impacts of the remaining virtual transition diagrams, we identify the dominant perturbation terms and derive analytical approximations for $\zeta_{\mathrm{C_{\alpha}Q_2}}$ in both $\mathrm{C0}$ and $\mathrm{C1}$-type cases. The resulting expressions, given in Eq. \ref{eq:zz_equation_proof}, are validated against numerical simulations in Fig.~\ref{fig:zz_equation_proof}. The factor of 2 multiplying the second term in each expression accounts for the two possible photon-propagation directions.

\begin{equation}
    \renewcommand{\arraystretch}{3.2}
    \begin{array}{lllllllll}

        \displaystyle \mathrm{C0}:\zeta_{\mathrm{C_{\alpha} Q_2}} = f_{\ket{1001}'} - f_{\ket{0001}'} - f_{\ket{1000}'} + f_{\ket{0000}'}
        \\\quad\approx \hspace{-0.2cm}\vcenter{\includegraphics[height=1.8cm]{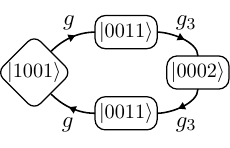}} \hspace{-5.35cm} +2\cdot \hspace{-0.3cm} \vcenter{\includegraphics[height=1.8cm]{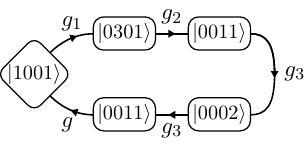}}
        
        \\\quad \,\,+ \hspace{-0.2cm} \vcenter{\includegraphics[height=1.8cm]{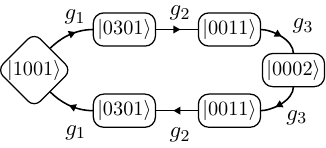}}
        
        \\\displaystyle \quad = \frac{(g_3 \, n^{\mathrm{Q_2}}_{12} n^{\mathrm{C_{\beta}}}_{01})^2}{f^{\mathrm{C_{\alpha}}}_{01} - f^{\mathrm{Q_2}}_{12}} \cdot \left( \frac{n^{C_{\alpha}}_{01} n^{\mathrm{C_{\beta}}}_{01}}{f^{\mathrm{C_{\alpha}}}_{01} - f^{\mathrm{C_{\beta}}}_{01}} \right)^2
        
        \\\displaystyle \,\quad \cdot \left(g + g_1 g_2 \frac{(n^{\mathrm{Q_1}}_{03})^2}{f^{\mathrm{C_{\alpha}}}_{01} - f^{\mathrm{Q_1}}_{03}}\right)^2

        \\

        \displaystyle \mathrm{C1}:\zeta_{\mathrm{C_{\alpha} Q_2}} = f_{\ket{1101}'} - f_{\ket{0101}'} - f_{\ket{1100}'} + f_{\ket{0100}'}
        \\\,\,\approx - \hspace{-0.2cm}\vcenter{\includegraphics[height=1.8cm]{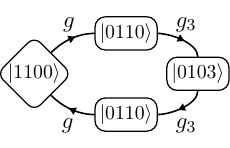}} \hspace{-5.35cm} -2 \cdot \hspace{-0.3cm} \vcenter{\includegraphics[height=1.8cm]{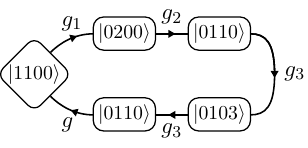}}
        
        \\\quad \,\,- \hspace{-0.2cm} \vcenter{\includegraphics[height=1.8cm]{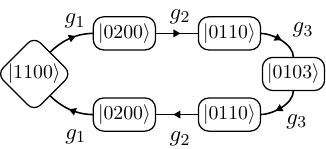}}
        
        \\\displaystyle \quad = - \frac{(g_3 \, n^{\mathrm{Q_2}}_{03} n^{\mathrm{C_{\beta}}}_{01})^2}{f^{\mathrm{C_{\alpha}}}_{01} - f^{\mathrm{Q_2}}_{03}} \cdot \left( \frac{n^{C_{\alpha}}_{01} n^{\mathrm{C_{\beta}}}_{01}}{f^{\mathrm{C_{\alpha}}}_{01} - f^{\mathrm{C_{\beta}}}_{01}} \right)^2
        
        \\\displaystyle \,\quad \cdot \left(g + g_1 g_2 \frac{(n^{\mathrm{Q_1}}_{12})^2}{f^{\mathrm{C_{\alpha}}}_{01} - f^{\mathrm{Q_1}}_{12}}\right)^2
        
    \end{array}
    \label{eq:zz_equation_proof}
\end{equation}

Regarding the suppression mechanisms for $\zeta_{\mathrm{C_{\alpha}Q_2}}$, as discussed in Section~\ref{section:parasitic_interactions}, particular attention must be paid to the $\mathrm{C0|C0}$ configuration. As shown in Fig.~\ref{fig:zz_equation_proof}, in this configuration $\zeta_{\mathrm{C_{\alpha}Q_2}}(g)$ exhibits an extremum at $g > 0$, which corresponds to a negative physical coupling strength due to the antisymmetric connection of $\mathrm{C0}$ couplers. Consequently, direct capacitive coupling between C0-type couplers cannot be used to suppress $\zeta_{\mathrm{C_{\alpha}Q_2}}$.

To address this issue, we introduce an additional oscillator between the C0-type couplers (Fig.~\ref{fig:CQOCQ_model}), thereby establishing a new suppression mechanism that exploits the structure of the transition diagrams in Eq.~\ref{eq:zz_equation_proof}. The diagrams contributing to each term in that equation differ only in the segment of the photon path connecting the $\mathrm{C_{\alpha}}$ and $\mathrm{C_{\beta}}$ couplers. In the $\mathrm{C0}$-type case, a photon can propagate between the couplers through two distinct pathways: either via the $\ket{3}$ state of the fluxonium $\mathrm{Q_1}$ or directly through the $g$ coupling. This path variation gives rise to the two terms in the last parenthesis of Eq.~\ref{eq:zz_equation_proof}. The oscillator provides a third path via its $\ket{1}$ state, thereby adding a compensatory term to the parentheses:
\begin{equation}
    \pm g_{\mathrm{O}}^2 \frac{(n^{\mathrm{O}})^2}{f^{\mathrm{C_{\alpha}}}_{01} - f^{\mathrm{O}}},
\end{equation}
where the sign is determined by the connection configuration. The structure of this term resembles that of the fluxonium, as both involve virtual photon propagation through an intermediate state.

Finally, we obtain a unified expression [Eq.~\ref{eq:ZZ_equation}] by combining the C0 and C1 analytic forms of Eq.~\ref{eq:zz_equation_proof} and incorporating the oscillator contribution. In this equation, we restrict $g$ to positive values and introduce $\pm$ variety to account for antisymmetric connections. Although the oscillator term is not validated against a numerical relation, its effectiveness is confirmed by the successful suppression of $\zeta_{\mathrm{C_{\alpha}Q_2}}$ in the $\mathrm{C0|C0}$ configuration (Table~\ref{tab:zz_antisym_results}).

%% file: appendixes/virtual_transitions.tex
In this appendix, we describe the virtual transition diagrams used to interpret terms arising in time-independent perturbation theory. This graphical approach is particularly valuable in the context of scaling, as it provides an intuitive picture of long-range ZZ interactions within a many-qubit lattice and simplifies the identification of the leading perturbative contributions to their rates. While a similar approach for the Schrieffer-Wolff transformation is introduced in \cite{Fors.ZZ_theory.2024}, here we focus on common perturbation theory, which yields simpler equations that are fully sufficient for our purposes.

For simplicity, we consider a perturbation $\hat{V}$ with $V_{ii}=0$. This condition is always satisfied for capacitive couplings and also holds for inductive couplings when circuit elements possess flux symmetry (e.g., fluxoniums at their symmetric sweetspots). Under this condition, the first non-zero energy correction for a state $\ket{i}$ is given by:
\begin{equation}
    \renewcommand{\arraystretch}{3}
    \begin{array}{ll}
        \displaystyle \Delta E^{(2)}_{i} = \sum_{j \neq i} \frac{|V_{ij}|^2}{E_{ij}},

        \hspace{1 cm} \displaystyle \frac{|V_{ij}|^2}{E_{ij}} = \hspace{-0.3 cm}\vcenter{\includegraphics[height=3 cm]{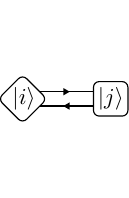}}
        \hspace{-6.2 cm} , \hspace{6.2 cm}
    \end{array}
    \label{eq:2th_order}
\end{equation}
where $E_{ij}=E_{i} - E_{j}$ is the difference between the energies of $\ket{i}$ and $\ket{j}$ bare states. The virtual transition diagram associated with a given perturbation term follows the next rules:
(1) the state $\ket{i}$, whose energy is being corrected, serves as the start and the end point, and is indicated by a diamond box;
(2) each matrix element $V_{nm}$ in the numerator corresponds to a virtual transition in the diagram;
(3) the sign of the term is given by $(-1)^{L-1}$, where $L$ – is the number of loops in the diagram;
(4) if the loop order matters, it is indicated by additional numerical labels (see Eq.~\ref{eq:4th_order} and Eq.~\ref{eq:6th_order});
(5) the denominator(s) consist of the energy differences ($E_{ij}$) between the initial state $\ket{i}$ and each virtual (mediating) states indicated by square boxes. The exact form of the denominator(s) can be taken from the original perturbation equation or can be inferred from the symmetry and the fixed energy scale of the whole term.

Using an iterative method \cite{Quantum_mechanics_LL3}, we compute perturbative corrections up to the 6th order and present the corresponding virtual transition diagrams to illustrate the applicability of our approach to the high-order terms used in Appendix \ref{subappendix:ZZ_CS_equation}:

\begin{equation}
    \renewcommand{\arraystretch}{3}
    \begin{array}{ll}
        \displaystyle \Delta E^{(3)}_{i} = \sum_{j, k \neq i} \frac{V_{ij} V_{jk} V_{ki}}{E_{ij} E_{ik}}, \,

        \\ \displaystyle \frac{V_{ij} V_{jk} V_{ki}}{E_{ij} E_{ik}} = \hspace{-0.3 cm}\vcenter{\includegraphics[height=3 cm]{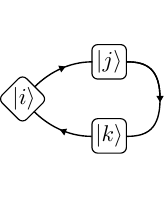}} \hspace{-5.63 cm} ; \hspace{5.63 cm} \quad

    \end{array}
    \label{eq:3th_order}
\end{equation}
% \vspace{-0.5 cm}
\begin{equation}
    \renewcommand{\arraystretch}{3}
    \begin{array}{ll}
        \displaystyle \Delta E^{(4)}_{i} = \sum_{j, k, n \neq i} \frac{V_{ij} V_{jk} V_{kn} V_{ni}}{E_{ij} E_{ik} E_{in}} \,-\,
        \frac{|V_{ij}|^2 |V_{ik}|^2}{E_{ij}^2 E_{ik}},

        \\\displaystyle \frac{V_{ij} V_{jk} V_{kn} V_{ni}}{E_{ij} E_{ik} E_{in}} = \hspace{-0.3 cm}\vcenter{\includegraphics[height=3 cm]{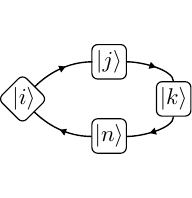}}
        \hspace{-5.25 cm} , \hspace{5.25 cm} \quad

        \\\displaystyle - \frac{|V_{ij}|^2 \cdot |V_{ik}|^2}{E_{ij}^2 E_{ik}} = \hspace{-0.3 cm}\vcenter{\includegraphics[height=3 cm]{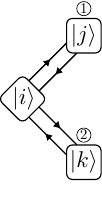}}
        \hspace{-6.5 cm} ; \hspace{6.5 cm} \quad
        
    \end{array}
    \label{eq:4th_order}
\end{equation}
\begin{equation}
    \renewcommand{\arraystretch}{3}
    \begin{array}{ll}
        \displaystyle \Delta E^{(5)}_{i} = \sum_{j, k, n, l \neq i} \frac{V_{ij} V_{jk} V_{kn} V_{nl} V_{li}}{E_{ij} E_{ik} E_{in} E_{il}} \,
        \\
        \displaystyle-\,
        \frac{|V_{ij}|^2 \cdot V_{ik} V_{kn} V_{ni}}{E_{ij} E_{ik} E_{in}} \left(\frac{1}{E_{ij}} + \frac{1}{E_{ik}} + \frac{1}{E_{in}} \right),

        \\\displaystyle \frac{V_{ij} V_{jk} V_{kn} V_{nl} V_{li}}{E_{ij} E_{ik} E_{in} E_{il}} = \hspace{-0.3 cm}\vcenter{\includegraphics[height=3 cm]{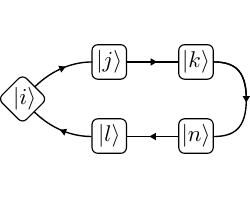}}
        \hspace{-4.4 cm} , \hspace{4.4 cm} \quad
        
        \\\displaystyle -\frac{|V_{ij}|^2 \cdot V_{ik} V_{kn} V_{ni}}{E_{ij} E_{ik} E_{in}} \left(\frac{1}{E_{ij}} + \frac{1}{E_{ik}} + \frac{1}{E_{in}} \right) \\ = \hspace{-0.3 cm}\vcenter{\includegraphics[height=3 cm]{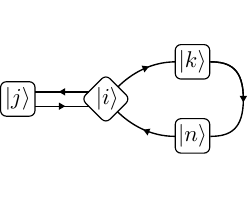}}
        \hspace{-4.46 cm} ; \hspace{4.46 cm} \quad
    
    \end{array}
    \label{eq:5th_order}
\end{equation}
\begin{equation}
    \renewcommand{\arraystretch}{3}
    \begin{array}{ll}
        \displaystyle \Delta E^{(6)}_{i} = \sum_{j, k, n, l, m \neq i} \frac{V_{ij} V_{jk} V_{kn} V_{nl} V_{lm} V_{mi}}{E_{ij} E_{ik} E_{in} E_{il} E_{im}} \,
        \\
        \displaystyle-\,
        \frac{|V_{ij}|^2 \cdot V_{ik} V_{kn} V_{nl} V_{li}}{E_{ij} E_{ik} E_{in} E_{il}} \left(\frac{1}{E_{ij}} + \frac{1}{E_{ik}} + \frac{1}{E_{in}} + \frac{1}{E_{il}} \right)
        \\
        \displaystyle-\,
        \frac{V_{ij} V_{jk} V_{ki} \cdot V_{in} V_{nl} V_{li} }{E_{ij} E_{ik} E_{in} E_{il}} \left(\frac{1}{E_{ij}} + \frac{1}{E_{ik}}\right)
        \\
        \displaystyle+\,
        \frac{|V_{ij}|^2 \cdot |V_{ik}|^2 \cdot |V_{in}|^2}{E_{ij} E_{ik} E_{in}} \left(\frac{1}{E_{ij}^2} + \frac{1}{E_{ik} E_{in}}\right),

        \\\displaystyle \frac{V_{ij} V_{jk} V_{kn} V_{nl} V_{lm} V_{mi}}{E_{ij} E_{ik} E_{in} E_{il} E_{im}} = \hspace{-0.3 cm}\vcenter{\includegraphics[height=3 cm]{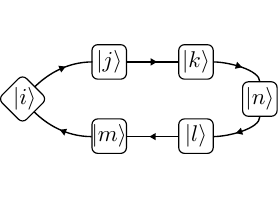}}
        \hspace{-3.98 cm} , \hspace{3.98 cm} \quad

        \\\displaystyle - \frac{|V_{ij}|^2 \cdot V_{ik} V_{kn} V_{nl} V_{li}}{E_{ij} E_{ik} E_{in} E_{il}} \left(\frac{1}{E_{ij}} + \frac{1}{E_{ik}} + \frac{1}{E_{in}} + \frac{1}{E_{il}} \right) \\ = \hspace{-0.3 cm}\vcenter{\includegraphics[height=3 cm]{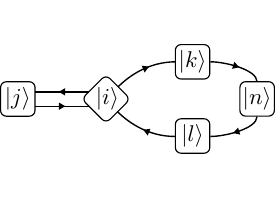}}
        \hspace{-4 cm} , \hspace{4 cm} \quad

        \\\displaystyle - \frac{V_{ij} V_{jk} V_{ki} \cdot V_{in} V_{nl} V_{li} }{E_{ij} E_{ik} E_{in} E_{il}} \left(\frac{1}{E_{ij}} + \frac{1}{E_{ik}}\right) \\ = \hspace{-0.3 cm}\vcenter{\includegraphics[height=3 cm]{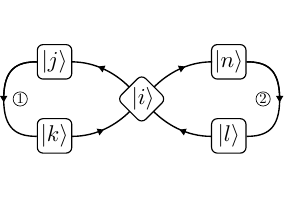}}
        \hspace{-3.9 cm} , \hspace{3.9 cm} \quad

        \\\displaystyle \frac{|V_{ij}|^2 \cdot |V_{ik}|^2 \cdot |V_{in}|^2}{E_{ij} E_{ik} E_{in}} \left(\frac{1}{E_{ij}^2} + \frac{1}{E_{ik} E_{in}}\right) = \hspace{-0.3 cm}\vcenter{\includegraphics[height=3 cm]{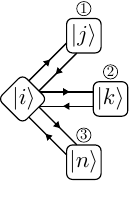}}
        \hspace{-6.27 cm} . \hspace{6.27 cm} \quad
        
    \end{array}
    \label{eq:6th_order}
\end{equation}

Finally, we note that this graphical framework can be generalized to the case of $V_{ii}\neq0$ by introducing "vortex" transitions:
\begin{equation}
    \renewcommand{\arraystretch}{3}
    \begin{array}{ll}
        \displaystyle \Delta E^{(1)}_{i} = V_{ii},
        
        \hspace{1cm} V_{ii} = \hspace{-0.3 cm}\vcenter{\includegraphics[height=0.947368 cm]{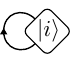}}
        \hspace{-7.1 cm} . \hspace{7.1 cm}
    \end{array}
    \label{eq:1th_order}
\end{equation}
However, this structure drastically increases the number of virtual transition diagrams, as the $n$-th order correction involves all possible combinations of $n$ appropriate virtual transitions.

%% file: appendixes/leakage.tex
\subsection{Reduced leakage model}
\label{appendix:leakage_model}

In our architecture, the spectrum near the coupler target transitions is relatively dense. As a consequence, the near-resonant microwave drives used to implement CZ gates can induce leakage both out of the computational subspace and from the excited target states. The dominant leakage channels arise from the side coupler transitions. They are inherent to the chosen microwave gate mechanism and exist even in the minimal Q-C-Q model (Fig.~\ref{fig:spectator_error}). Such leakage can be mitigated solely through pulse shaping and is not the focus of this appendix. 

Another class of leakage involves the transitions in elements surrounding the driven coupler. Such leakage is particularly pronounced when scaling microwave gates with fixed couplings. However, it can be suppressed by optimally tuning the transmon frequencies. To perform such calibrations numerically, one must evaluate leakage rates as a function of the coupler frequency and identify the relevant operating point. Accurate estimation of these rates requires solving the Schrödinger equation for large-circuit Hamiltonians, such as those of Q-C-Q-C-Q systems. However, in our simulations, the Hilbert-space dimension of these Hamiltonians typically exceeds 180, making precise modeling computationally expensive and time-consuming. To address this challenge, we construct a reduced leakage model that captures the essential physics at a dramatically lower computational cost.

Since leakages are induced by microwave drive pulses, each leakage channel can be described as a microwave transition between a source and a leakage state, $\ket{s}'\leftrightarrow\ket{l}'$, characterized by a frequency $f_\text{leak}$. For a target coupler transition $\ket{i}'\leftrightarrow\ket{t}'$ with frequency $f_{\text{target}}$, the near-resonant drive pulse applied to the coupler’s flux line is described by the perturbation operator of the form:
\begin{equation}
    \hat{V}(t)/h = \displaystyle\frac{\mathcal{\epsilon}(t)}{\bra{i}'\hat{\varphi}\ket{t}'}\cdot \sin{(f_{\text{target}}+\Delta)t} \cdot \hat{\varphi},
\end{equation}
where $\epsilon(t)$ is the time-dependent envelope, $\Delta$ is the signal detuning, and $\varphi$ is the flux operator of the driven coupler. The influence of this perturbation on the leakage channel depends on two key parameters: the detuning $\delta = f_\text{leak} - f_{\text{target}}$ and the effective matrix element $k = \abs{\bra{s}'\varphi\ket{l}'/\bra{i}'\varphi\ket{t}'}$.

Both $\delta$ and $k$ for any leakage channel can be computed via numerical diagonalization. However, obtaining the leakage rate under a non-trivial signal shape $\epsilon(t)$ requires time-domain simulations. To this end, we build a reduced model Hamiltonian to examine the rate of a certain $\ket{s}\leftrightarrow\ket{l}$ leakage channel characterized by $\delta$ and $k$. To account for side coupler transitions, which significantly influence CZ gate dynamics (Fig.~\ref{fig:spectator_error}), we base our model on an 8-dimensional Q-C-Q system comprising two qubits (A and B) and a two-level coupler (C). While the source state $\ket{s}$ of the leakage channel can be selected from the $\mathrm{A}\otimes\mathrm{B}\otimes\mathrm{C}$ basis, to define $\ket{l}$, we extend the 8-dimensional Hilbert space by a 1-dimensional $\{\ket{l}\}$, padding the operators with zeros to match the new dimension. Consequently, we derive the reduced leakage model Hamiltonian in the extended 9-dimensional space $\mathrm{A}\otimes\mathrm{B}\otimes\mathrm{C} \oplus \{\ket{l}\}$, which takes the form:
\begin{equation}
    \renewcommand{\arraystretch}{2.4}
    \begin{array}{lll}
        \displaystyle \hat{H}(t)/h = \mathcal{G} \,\left( \hat{\sigma}^-_{\mathrm{A}} \hat{\sigma}^+_{\mathrm{A}} + \hat{\sigma}^-_{\mathrm{B}} \hat{\sigma}^+_{\mathrm{B}}\right)\cdot \hat{\sigma}^+_{\mathrm{C}} \hat{\sigma}^-_{\mathrm{C}} + \delta \cdot \ket{l}\bra{l}
        
        \\\displaystyle \quad \quad \quad \,\,\,+ \, \frac{1}{2} \mathcal{\epsilon}(t) \cdot \left( \hat{\sigma}^-_{\mathrm{C}} \, e^{-i \Delta t} + \hat{\sigma}^+_{\mathrm{C}} \, e^{i \Delta t} \right)
        \\\displaystyle \quad \quad \quad \,\,\,+ \, \frac{1}{2} k\, \mathcal{\epsilon}(t) \cdot \left(\ket{s}\bra{l}\, e^{-i \Delta t} + \ket{l}\bra{s} \, e^{i \Delta t} \right),
    \end{array}
    \label{eq:leakage_model}
\end{equation}
where $\mathcal{G}$ denotes the coupler gap. In deriving, we follow an approach analogous to that for Eq.~\ref{eq:ZZ_simple_model}: switch to the frame of the target transition ($\ket{110}_{\mathrm{qqc}}\leftrightarrow\ket{111}_{\mathrm{qqc}}$) and apply RWA.

To compute the specific $\ket{s}\leftrightarrow\ket{l}$ leakage rate $\varepsilon_\text{leak}^{sl}$, we solve the Schrödinger equation with this Hamiltonian, additionally considering the case of $k=0$. Then, we compute the leakage errors following Eq.~(\ref{eq:error_leak}) and obtain the desired leakage rate as $\varepsilon_\text{leak}^{sl}=\varepsilon_\text{leak}(k) - \varepsilon_\text{leak}(k=0)$, excluding the contribution of the side transition leakages.

\begin{figure}[h!]
    \center{\includegraphics[width=\linewidth]{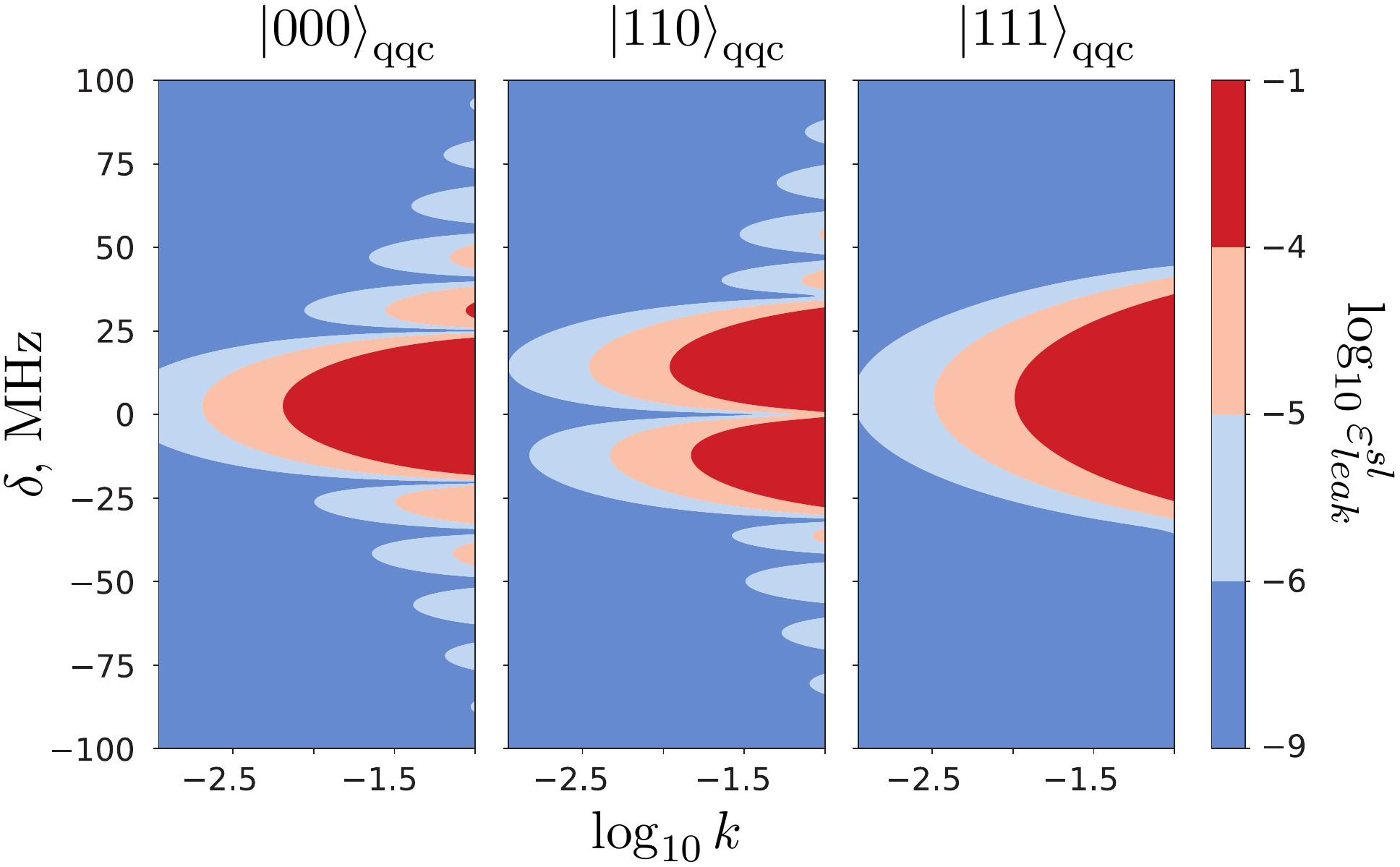}}
    \caption{Numerical mapping of leakage rates for the reduced leakage model. The leakage rates $\varepsilon_\text{leak}^{sl}$ for the three representative source states are plotted as functions of $k$ and $\delta$. For clarity, the rates are divided into four distinct regions according to their order of magnitude.}
    \label{fig:leakages_map}
\end{figure}

Applying this method to our architecture, we set $\mathcal{G}=100$ MHz and employ a Gaussian (\ref{eq:leakage_model}) 66 ns pulse optimized for the $k=0$ case. We then sweep $k$ across $[10^{-3}, 10^{-1}]$ and $\delta$ over $[-100, +100]$ MHz, computing leakage rates for three representative source states: two within the target transition manifold ($\ket{110}_{\mathrm{qqc}}$ and $\ket{111}_{\mathrm{qqc}}$) and one outside it ($\ket{000}_{\mathrm{qqc}}$). The resulting leakage maps (Fig.~\ref{fig:leakages_map}) 
enable us to estimate leakage rates for all transitions beyond the side ones, solely based on their respective $k$ and $\delta$ values. Note that in this work, we consider a leakage negligible if its rate satisfies $\varepsilon_\text{leak}^{sl}<10^{-5}$.

\subsection{Selected leakage channels}
\label{appendix:leakage_channels}

Here, we summarize the primary leakage channels beyond the side states that pose risks to the CZ gates.

The most significant leakage processes (with $k$ up to 0.3) involve transitions into high-energy fluxonium states: specifically, $\ket{01}_{\mathrm{C0Q}}\overset{f_{\mathrm{C0}}}{\rightarrow} \ket{02}_{\mathrm{C0Q}}$ and $\ket{00}_{\mathrm{C1Q}}\overset{f_{\mathrm{C1}}}{\rightarrow}\ket{03}_{\mathrm{C1Q}}$. These leakage channels arise from the gap-building interactions between the couplers and the qubits. Moreover, owing to the hybridization of high-energy fluxonium transitions, such leakage can occur even between non-adjusted coupler and qubit. Consequently, this necessitates maintaining a sufficient frequency detuning between couplers and their associated gap-building fluxonium transitions, which, under our parameters, exceed several spectral gaps.

The described leakage appears in a distinctive manner for the CZZ gate. When one coupler is excited, the drive on the second coupler can simultaneously excite two high-energy fluxonium transitions. The most problematic case ($k \sim 0.1$) is $\ket{1001}_{\mathrm{C0_L QC1_U Q}}\overset{f_{\mathrm{C1_U}}}{\rightarrow}\ket{0302}_{\mathrm{C0_L QC1_U Q}}$, for which the leakage resonance condition:
\begin{equation}
    f^{\mathrm{C1_U}}_{01} - f^{\mathrm{Q}}_{03} = -(f^{\mathrm{C0_L}}_{01} - f^{\mathrm{Q}}_{12}),
\end{equation}
can be satisfied due to the definitions of the U- and L-subtype couplers, enforcing $f^{\mathrm{C0_L}}_{01} < f^{\mathrm{Q}}_{12}$ and $f^{\mathrm{C1_U}}_{01} > f^{\mathrm{Q}}_{03}$. This mechanism imposes an additional requirement on coupler calibration to avoid such leakage.

Lastly, we consider leakage related to the excitation of the additional oscillators. Although direct leakage from the couplers into the oscillators is negligible -- given their frequencies are well separated from the oscillator frequencies -- a process of the form $\ket{100}_{\mathrm{C0_{\mathrm{U}}QO}}\overset{f_{\mathrm{C0_U}}}{\rightarrow} \ket{031}_{\mathrm{C0_{\mathrm{U}}QO}}$ is possible with an effective matrix element $k\sim0.07$. Nevertheless, this leakage can be effectively suppressed by the appropriate choice of the oscillator frequencies and the coupler anharmonicity.

%% file: appendixes/parameters.tex
We begin by defining the Hamiltonians of the fluxoniums, transmons, and oscillators as:
\begin{equation}
    \renewcommand{\arraystretch}{2.5}
    \begin{array}{lll}
        \displaystyle \hat{H}_{\mathrm{T}} = 4\,\mathrm{E_C}\,\hat{n}^2 - \mathrm{E_J}\,\cos{\hat{\varphi}}
        \\
        \displaystyle \hat{H}_{\mathrm{F}} = 4\,\mathrm{E_C}\,\hat{n}^2 + \frac{1}{2}\,\mathrm{E_L}\,\hat{\varphi}^2 - \mathrm{E_J}\,\cos{\hat{\varphi}}
        \\
        \displaystyle \hat{H}_{\mathcal{O}} = 4\,\mathrm{E_C}\,\hat{n}^2 + \frac{1}{2}\,\mathrm{E_L}\,\hat{\varphi}^2,
        
    \end{array}
    \label{eq:elementary_Hams}
\end{equation}
and establishing their optimal energy parameters and inter-element connections.

For the fluxoniums, we impose several conditions: 
\begin{enumerate}
    \item Strong detuning among the $\ket{1}\leftrightarrow\ket{2}$, $\ket{0}\leftrightarrow\ket{3}$, and $\ket{1}\leftrightarrow\ket{4}$ transitions, which enables effective localization of the target states (see Section \ref{section:alternating_couplers}); 
    \item A high $\ket{1}\leftrightarrow\ket{2}$ transition frequency, ensuring robust suppression of the qubit–qubit interactions (see Appendix \ref{appendix:qq_analysis}); 
    \item Strong detuning of the $\ket{3}\leftrightarrow\ket{4}$ transition from the $\ket{1}\leftrightarrow\ket{2}$ transition, as well as of the $\ket{5}$ state from the sum of the $\ket{2}$ and $\ket{3}$ state frequencies to prevent gap shrinking during CZZ gates (see Sec.~\ref{section:CZZ_gates}).
\end{enumerate}

For the transmons, we choose the optimal $\ket{0}\leftrightarrow\ket{1}$ transition frequencies compatible with the fluxoniums' spectra while avoiding leakage into the $\ket{2}$ state by maintaining an anharmonicity $\alpha$ larger than the spectral gap $\mathcal{G}$. To simplify the overall design, we set the charging energy to a fixed value $\mathrm{E_C = 150 \,\, MHz}$ for all couplers and adjust $\mathrm{E_J}$ to achieve the desired frequencies.

Finally, the couplings between the fluxoniums and transmons
\begin{table}[h!]
    \centering
    \begin{tabularx}{\linewidth}{ 
    l
    >{\centering\arraybackslash}X
    >{\centering\arraybackslash}X
    >{\centering\arraybackslash}X
    >{\centering\arraybackslash}X }
        \toprule
        \toprule
        \multicolumn{5}{c}{\textbf{Fluxoniums}} \\[0.2cm]
        & A1 & A2 & B1 & B2\\
        \hline
        $\mathrm{E_C/h\,\,(GHz)}$ & 1.22 & 1.22 & 1.251 & 1.251 \\
        $\mathrm{E_L/h\,\,(GHz)}$ & 1.3 & 1.28 & 1.3 & 1.28 \\
        $\mathrm{E_J/h\,\,(GHz)}$ & 6.982 & 6.982 & 6.788 & 6.788 \\
        $|n_{01}|$ & 0.077 & 0.076 & 0.086 & 0.084 \\
        $|n_{12}|$ & 0.595 & 0.595 & 0.591 & 0.591 \\
        $|n_{03}|$ & 0.517 & 0.519 & 0.499 & 0.501 \\
        \hline
        \multicolumn{5}{c}{\textbf{Transmons}} \\[0.2cm]
        & $\mathrm{C0_L}$ & $\mathrm{C0_U}$ & $\mathrm{C1_L}$ & $\mathrm{C1_U}$ \\
        \hline
        $\mathrm{E_C/h\,\,(GHz)}$ & 0.15 & 0.15 & 0.15 & 0.15 \\
        $\mathrm{E_J/h\,\,(GHz)}$ & 21.0 & 32.4 & 57.2 & 67.4 \\
        $|n_{01}|$ & 1.424 & 1.592 & 1.841 & 1.919 \\
        \hline
        \multicolumn{5}{c}{\textbf{Oscillators}} \\[0.2cm]
        & \multicolumn{2}{c}{$\mathcal{O}_0$} & \multicolumn{2}{c}{$\mathcal{O}_1$} \\
        \hline
        $\mathrm{E_C/h\,\,(GHz)}$ & \multicolumn{2}{c}{1.103} & \multicolumn{2}{c}{1.32} \\
        $\mathrm{E_L/h\,\,(GHz)}$ & \multicolumn{2}{c}{2} & \multicolumn{2}{c}{4} \\
        $|n_{01}|$ & \multicolumn{2}{c}{0.488} & \multicolumn{2}{c}{0.555} \\
        \hline
        \multicolumn{5}{c}{$\mathbf{Connections \,\, g_{ij}/h \,\, (MHz)}$} \\[0.2cm]
        & $\mathrm{C0_L}$ & $\mathrm{C0_U}$ & $\mathrm{C1_L}$ & $\mathrm{C1_U}$ \\
        \hline
        A & 336 & 267 & 207 & 260 \\
        B & 307 & 287 & 233 & 238 \\
        $\mathcal{O}_0$ & 270 & 270 & - & - \\
        $\mathcal{O}_1$ & - & - & 63 & 90 \\
        $\mathrm{C1_L}$ & - & - & - & 4 \\
        \bottomrule
        \bottomrule
    \end{tabularx}
    \caption{Optimal energy parameters for the architecture elements and the main couplings between them. Here we also introduce fluxonium subtypes differing in $\Delta \mathrm{E_L/h=20\,\,MHz}$, which are required for accurate numerical modeling (see Appendix~\ref{appendix:qq_analysis}).}
    \label{tab:energy_table}
\end{table}
are adjusted to produce spectral gaps of approximately $\mathrm{\sim 100\,\,MHz}$. The additional oscillators and couplings mediating the couplers are parameterized to compensate for parasitic interactions (see Section \ref{section:parasitic_interactions}), which are evaluated by accounting for long-range parasitic capacitive couplings computed via the inverse capacitive matrix of the 13-qubit system shown in Fig.~\ref{fig:architecture_model}.

The optimal energy parameters are summarized in Table \ref{tab:energy_table}, excluding long-range capacitive couplings (e.g. between the nearest fluxoniums) computed separately.

Next, we determine the capacitances of elements and couplings, optimizing them to achieve the target $\mathrm{E_C}$ and coupling strengths $g$ in the 13-qubit system (Fig.~\ref{fig:architecture_model}). For design simplicity, we minimize the total number of parameters by standardizing these values. The final values are listed in Table~\ref{tab:capacitance_table}.

\begin{table}[h!]
    \centering
    \begin{tabularx}{\linewidth}{ 
    >{\centering\arraybackslash}X
    >{\centering\arraybackslash}X
    >{\centering\arraybackslash}X
    >{\centering\arraybackslash}X
    >{\centering\arraybackslash}X }
        \toprule
        \toprule
        \multicolumn{5}{c}{\textbf{Fluxoniums}} \\[0.2cm]
        & $\text{A}_X$ & $\text{A}_Y$ & $\text{B}_X$ & $\text{B}_Y$ \\
        \midrule
        2+2 & 30.91 & 31.28 & 30.23 & 30.3 \\
        2+0 & 31.07 & 30.26 & 30.24 & 29.49 \\
        1+1 & 30.74 & 30.65 & 29.92 & 29.85 \\
        1+0 & 30.44 & 30.54 & 29.69 & 29.71 \\
        \midrule
        \multicolumn{5}{c}{\textbf{Transmons}} \\[0.2cm]
        \multicolumn{5}{c}{136} \\
        \midrule
        \multicolumn{5}{c}{\textbf{Oscillators}} \\[0.2cm]
        $\mathcal{O}_{0X}$ & $\mathcal{O}_{0Y}$ & & $\mathcal{O}_{1X}$ & $\mathcal{O}_{1Y}$ \\
        34.69 & 34.69 & & 28.42 & 28.42 \\
        \midrule
        \multicolumn{5}{c}{\textbf{Connections}} \\[0.2cm]
        & $\mathrm{C0_L}$ & $\mathrm{C0_U}$ & $\mathrm{C1_L}$ & $\mathrm{C1_U}$ \\
        A                 & 9.09 & 7.17 & 5.55 & 6.99 \\
        B                 & 8.07 & 7.53  & 6.1 & 6.23 \\
        $\mathcal{O}_0$   & 8.04 & 8.04 & - & - \\
        $\mathcal{O}_1$   & - & -  & 1.55 & 2.21 \\
        $\mathrm{C1_L}$   & - & - & - & 0.37 \\
        \midrule
        \multicolumn{5}{c}{\textbf{Inner capacitance}} \\[0.2cm]
        JJ & 0.5 & & JJ array & 1 \\
        \bottomrule
        \bottomrule
    \end{tabularx}
    \caption{Standardized capacitances (in fF) of the circuit elements and their couplings, optimized for the 13-qubit lattice in Fig. \ref{fig:architecture_model}. Distinct islands of the differential elements are indicated by $X$ and $Y$. Fluxoniums are additionally grouped with regard to the number of couplers connected to their islands.}
    \label{tab:capacitance_table}
\end{table}

\begin{figure*}[!htb]
    \center{\includegraphics[width=\linewidth]{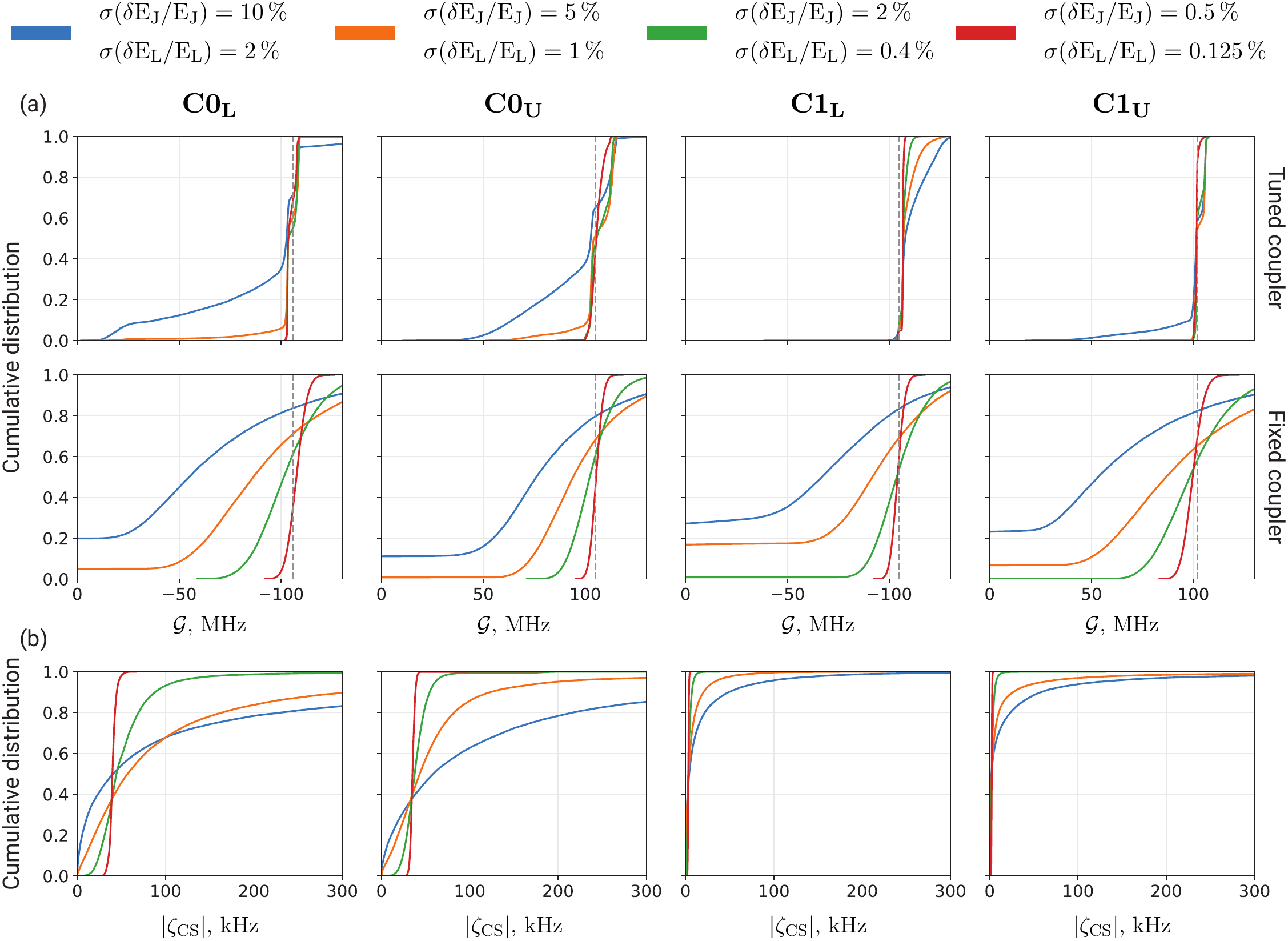}}
    \caption{Cumulative distributions of $\mathcal{G}$ (a) and $\zeta_{\mathrm{CS}}$ (b) for the $\mathrm{C0_U|C0_L}$ and $\mathrm{C1_U|C1_L}$ configurations under random variation of $\delta E_{\mathrm{J}}/E_{\mathrm{J}}$ and $\delta E_{\mathrm{L}}/E_{\mathrm{L}}$ with distinct standard deviations. For the former characteristic (a), we show results with both calibrated and fixed coupler $E_{\mathrm{J}}$, while for the latter (b) only with the calibrated ones.}
    \label{fig:robustness_research}
\end{figure*}

%% file: appendixes/fabrication_robustness.tex
In this appendix, we probe the robustness of our architecture to fabrication imperfections, which induce deviations in the element parameters. \second{Since fluxonium qubits are naturally well isolated in our layout (Appendix~\ref{appendix:qq_analysis}), we consider single-qubit operations to be unconditionally robust. Two-qubit operations, in contrast, rely strongly on higher fluxonium transitions, the couplers' frequency allocation, and the additional oscillators, which makes them sensitive to fabrication imperfections.}

\second{Unfortunately, a straightforward simulation of $\mathrm{CZ}$ gates in a large circuit while varying parameters is computationally expensive. Instead, we focus on the spectral gap $\mathcal{G}$, which approximately sets the gate duration as $\sim 1/\mathcal{G}$ for fixed leakage errors, and on the coupler-spectator ZZ ($\zeta_{\mathrm{CS}}$), which induces the phase error.}

To examine the robustness of these characteristics, we \second{consider $\mathrm{C0U|C0L}$ and $\mathrm{C1U|C1L}$ configurations with $\zeta_{\mathrm{CS}}$-suppression elements, and}  numerically diagonalize their C-Q-C-Q subcircuits while randomly varying the parameters of the involved elements, including the oscillator \second{$\mathcal{O}_0$ ($\mathcal{O}_1$ is not considered due to complexity of C-Q-C-Q-C diagonalization).} Specifically, we \second{consider independent and} normally distributed fluctuations in $E_{\mathrm{J}}$ and $E_{\mathrm{L}}$, \second{and neglect deviations in capacitances (note, that robustness of the direct capacitive coupling, suppressing $\zeta_{\mathrm{CS}}$ in $\mathrm{C1U|C1L}$, is supported by plots in Fig.\ref{eq:zz_equation_proof})}. The corresponding standard deviations follow $\sigma( \delta E_{\mathrm{L}}/E_{\mathrm{L}}) = \sigma( \delta E_{\mathrm{J}}/E_{\mathrm{J}})/5$, reflecting that a fluxonium \second{and oscillator} inductance is formed by an array of independently fluctuating junctions. \second{In addition}, we consider a compensatory calibration mechanism based on the coupler flux tuning. For each randomly generated parameter set, we numerically adjust the coupler $E_{\mathrm{J}}$ \second{in range of $\pm15\%$} to restore the target gap value. The resulting cumulative distributions of $\mathcal{G}$ and $\zeta_{\mathrm{CS}}$ for different $\sigma$ are shown in Fig. \ref{fig:robustness_research}.

\second{The crucial event that disables a coupler is a transmon crossing the fluxonium transition associated with the coupler type ($\ket{0} \leftrightarrow \ket{3}$ or $\ket{1} \leftrightarrow \ket{2}$). This changes the subtype of the coupler (U/L), breaking the architecture pattern and thereby spoiling the long-range interaction suppression mechanisms. In the simulation, these events flip the sign of $\mathcal{G}$, which is clearly seen in the $\mathcal{G}$-distribution for fixed-flux transmons. To suppress the probability of this, either maintaining $\sigma(\delta E_{\mathrm{J}}/E_{\mathrm{J}}) < 2\%$ or introducing adaptive transmon tuning is needed.}

\second{Since considering $\zeta_{\mathrm{CS}}$ is meaningful only for valid couplers of the proper subtypes, we compute its cumulative distributions as the coupler $E_{\mathrm{J}}$ is tuned (Fig.~\ref{fig:robustness_research}(b)). Notably, because of the variation in the fluxoniums and the oscillator, tuning the coupler alone does not guarantee effective suppression of $\zeta_{\mathrm{CS}}$. Consequently, the two characteristic values of 300~kHz and 100~kHz, which correspond to phase errors $\varepsilon_{\varphi}$ of $\sim10^{-4}$ and $\sim10^{-5}$ respectively (Fig.~\ref{fig:spectator_error}(a)), require fabrication accuracies of $\sigma(\delta E_{\mathrm{J}}/E_{\mathrm{J}}) < 2\%$ and $\sigma(\delta E_{\mathrm{J}}/E_{\mathrm{J}}) < 0.5\%$ respectively. Such precise calibration of Josephson junctions can be achieved using the post-fabrication techniques described in \cite{zhang2022high, wang2024precision}.}

%% file: ref.bib
@article{moskalenko.tunable_FF_gate.2021,
    author = {Moskalenko, I. N. and Besedin, I. S. and Simakov, I. A. and Ustinov, A. V.},
    title = {Tunable coupling scheme for implementing two-qubit gates on fluxonium qubits},
    journal = {Applied Physics Letters},
    volume = {119},
    number = {19},
    pages = {194001},
    year = {2021},
    month = {11},
    issn = {0003-6951},
    doi = {10.1063/5.0064800},
    url = {https://doi.org/10.1063/5.0064800}
}

@misc{tasler.CZZ_for_surface.2025,
      title={Optimizing Superconducting Three-Qubit Gates for Surface-Code Error Correction}, 
      author={Stephan Tasler and Josias Old and Lukas Heunisch and Verena Feulner and Timo Eckstein and Markus Müller and Michael J. Hartmann},
      year={2025},
      eprint={2506.09028},
      archivePrefix={arXiv},
      primaryClass={quant-ph},
      url={https://arxiv.org/abs/2506.09028}, 
}

@article{mundada.ZZ_suppression_via_bus.2023,
  title = {Suppression of Qubit Crosstalk in a Tunable Coupling Superconducting Circuit},
  author = {Mundada, Pranav and Zhang, Gengyan and Hazard, Thomas and Houck, Andrew},
  journal = {Phys. Rev. Appl.},
  volume = {12},
  issue = {5},
  pages = {054023},
  numpages = {10},
  year = {2019},
  month = {Nov},
  publisher = {American Physical Society},
  doi = {10.1103/PhysRevApplied.12.054023},
  url = {https://link.aps.org/doi/10.1103/PhysRevApplied.12.054023}
}

@article{chitta2022numerical_methods,
  doi = {10.1088/1367-2630/ac94f2},
  title={Computer-aided quantization and numerical analysis of superconducting circuits},
  author={Chitta, Sai Pavan and Zhao, Tianpu and Huang, Ziwen and Mondragon-Shem, Ian and Koch, Jens},
  journal={New Journal of Physics},
  volume={24},
  number={10},
  pages={103020},
  year={2022},
  publisher={IOP Publishing}
}

@BOOK{Quantum_mechanics_LL3,
  title     = "Quantum Mechanics: {Non-Relativistic} Theory",
  author    = "Landau, L D and Lifshitz, E M",
  publisher = "Pergamon",
  year      =  2013
}

@article{Dumas2024, title={Measurement-Induced Transmon Ionization}, volume={14}, ISSN={2160-3308}, url={http://dx.doi.org/10.1103/PhysRevX.14.041023}, DOI={10.1103/physrevx.14.041023}, number={4}, journal={Physical Review X}, publisher={American Physical Society (APS)}, author={Dumas, Marie Frédérique and Groleau-Paré, Benjamin and McDonald, Alexander and Muñoz-Arias, Manuel H. and Lledó, Cristóbal and D’Anjou, Benjamin and Blais, Alexandre}, year={2024}, month=oct }

@article{lange.transmon_coupler_research.2025,
  title = {Cross-talk in superconducting qubit lattices with tunable couplers---comparing transmon and fluxonium architectures},
  author = {Lange, Florian and Heunisch, Lukas and Fehske, Holger and DiVincenzo, David P. and Hartmann, Michael J.},
  journal = {Quantum Science and Technology},
  volume = {11},
  number = {1},
  pages = {015020},
  year = {2026},
  publisher = {IOP Publishing},
  issn = {2058-9565},
  doi = {10.1088/2058-9565/ae2358},
  url = {https://doi.org/10.1088/2058-9565/ae2358},
  eprint = {2504.10298},
  archivePrefix = {arXiv},
  primaryClass = {quant-ph}
}

@article{Rol2019,
  title = {Fast,  High-Fidelity Conditional-Phase Gate Exploiting Leakage Interference in Weakly Anharmonic Superconducting Qubits},
  volume = {123},
  ISSN = {1079-7114},
  url = {http://dx.doi.org/10.1103/PhysRevLett.123.120502},
  DOI = {10.1103/physrevlett.123.120502},
  number = {12},
  journal = {Physical Review Letters},
  publisher = {American Physical Society (APS)},
  author = {Rol,  M.A. and Battistel,  F. and Malinowski,  F.K. and Bultink,  C.C. and Tarasinski,  B.M. and Vollmer,  R. and Haider,  N. and Muthusubramanian,  N. and Bruno,  A. and Terhal,  B.M. and DiCarlo,  L.},
  year = {2019},
  month = sep 
}

@article{Foxen2020,
  title = {Demonstrating a Continuous Set of Two-qubit Gates for Near-term Quantum Algorithms},
  volume = {125},
  ISSN = {1079-7114},
  url = {http://dx.doi.org/10.1103/PhysRevLett.125.120504},
  DOI = {10.1103/physrevlett.125.120504},
  number = {12},
  journal = {Physical Review Letters},
  publisher = {American Physical Society (APS)},
  author = {Foxen,  B. and Neill,  C. and Dunsworth,  A. and Roushan,  P. and Chiaro,  B. and Megrant,  A. and Kelly,  J. and Chen,  Zijun and Satzinger,  K. and Barends,  R. and Arute,  F. and Arya,  K. and Babbush,  R. and Bacon,  D. and Bardin,  J.C. and Boixo,  S. and Buell,  D. and Burkett,  B. and Chen,  Yu and Collins,  et al.},
  year = {2020},
  month = sep 
}

@article{zhao.scalable_FTF_architecture_2.2025,
  title = {Scalable native multiqubit gates via engineered noncomputational-state interactions in superconducting fluxonium qubits},
  author = {Zhao, Peng and Xu, Peng and Xue, Zheng-Yuan},
  journal = {Phys. Rev. A},
  volume = {113},
  issue = {2},
  pages = {022604},
  year = {2026},
  publisher = {American Physical Society},
  doi = {10.1103/xwny-vkft},
  url = {https://link.aps.org/doi/10.1103/xwny-vkft},
  eprint = {2507.18984},
  archivePrefix = {arXiv},
  primaryClass = {quant-ph}
}

@misc{Heunisch.hybrid_FT_scaling_2025,
      title={Scalable Fluxonium-Transmon Architecture for Error Corrected Quantum Processors}, 
      author={Lukas Heunisch and Longxiang Huang and Stephan Tasler and Johannes Schirk and Florian Wallner and Verena Feulner and Bijita Sarma and Klaus Liegener and Christian M. F. Schneider and Stefan Filipp and Michael J. Hartmann},
      year={2025},
      eprint={2508.09267},
      archivePrefix={arXiv},
      primaryClass={quant-ph},
      url={https://arxiv.org/abs/2508.09267}, 
}

@article{Lin.cool_CNOT_on_fluxoniums,
  title = {24 Days-Stable CNOT Gate on Fluxonium Qubits with Over 99.9\% Fidelity},
  author = {Lin, Wei-Ju and Cho, Hyunheung and Chen, Yinqi and Vavilov, Maxim G. and Wang, Chen and Manucharyan, Vladimir E.},
  journal = {PRX Quantum},
  volume = {6},
  issue = {1},
  pages = {010349},
  numpages = {20},
  year = {2025},
  month = {Mar},
  publisher = {American Physical Society},
  url = {https://link.aps.org/doi/10.1103/PRXQuantum.6.010349}
}

@article{singh.FTF_CZ_experimental.2025,
  title = {Fast microwave-driven two-qubit gates between fluxonium qubits with a transmon coupler},
  author = {Singh, Siddharth and Huang, Eugene Y. and Hu, Jinlun and Yilmaz, Figen and Zwanenburg, Martijn F. S. and Kumaravadivel, Piranavan and Wang, Siyu and Stefanski, Taryn V. and Andersen, Christian Kraglund},
  journal = {Phys. Rev. Appl.},
  volume = {25},
  issue = {2},
  pages = {024020},
  year = {2026},
  month = {Feb},
  publisher = {American Physical Society},
  doi = {10.1103/yxf3-jtx5},
  url = {https://link.aps.org/doi/10.1103/yxf3-jtx5},
  eprint = {2504.13718},
  archivePrefix = {arXiv},
  primaryClass = {quant-ph}
}

@article{rower.CoolFluxoniumSQGates.2024,
  title = {Suppressing Counter-Rotating Errors for Fast Single-Qubit Gates with Fluxonium},
  author = {Rower, David A. and Ding, Leon and Zhang, Helin and Hays, Max and An, Junyoung and Harrington, Patrick M. and Rosen, Ilan T. and Gertler, Jeffrey M. and Hazard, Thomas M. and Niedzielski, Bethany M. and Schwartz, Mollie E. and Gustavsson, Simon and Serniak, Kyle and Grover, Jeffrey A. and Oliver, William D.},
  journal = {PRX Quantum},
  volume = {5},
  issue = {4},
  pages = {040342},
  numpages = {17},
  year = {2024},
  month = {Dec},
  publisher = {American Physical Society},
  doi = {10.1103/PRXQuantum.5.040342},
  url = {https://link.aps.org/doi/10.1103/PRXQuantum.5.040342}
}

@article{pedersenFidelityQuantumOperations2007,
    title = {Fidelity of quantum operations},
    journal = {Physics Letters A},
    volume = {367},
    number = {1},
    pages = {47-51},
    year = {2007},
    issn = {0375-9601},
    doi = {https://doi.org/10.1016/j.physleta.2007.02.069},
    url = {https://www.sciencedirect.com/science/article/pii/S0375960107003271},
    author = {Line Hjortshøj Pedersen and Niels Martin Møller and Klaus Mølmer}
}

@article{Rosenfeld.FOF,
  title = {High-Fidelity Two-Qubit Gates between Fluxonium Qubits with a Resonator Coupler},
  author = {Rosenfeld, Emma L. and Hann, Connor T. and Schuster, David I. and Matheny, Matthew H. and Clerk, Aashish A.},
  journal = {PRX Quantum},
  volume = {5},
  issue = {4},
  pages = {040317},
  numpages = {35},
  year = {2024},
  month = {Nov},
  publisher = {American Physical Society},
  doi = {10.1103/PRXQuantum.5.040317},
  url = {https://link.aps.org/doi/10.1103/PRXQuantum.5.040317}
}

@article{Nguyen.Scaling,
  title = {Blueprint for a High-Performance Fluxonium Quantum Processor},
  author = {Nguyen, Long B. and Koolstra, Gerwin and Kim, Yosep and Morvan, Alexis and Chistolini, Trevor and Singh, Shraddha and Nesterov, Konstantin N. and J\"unger, Christian and Chen, Larry and Pedramrazi, Zahra and Mitchell, Bradley K. and Kreikebaum, John Mark and Puri, Shruti and Santiago, David I. and Siddiqi, Irfan},
  journal = {PRX Quantum},
  volume = {3},
  issue = {3},
  pages = {037001},
  numpages = {38},
  year = {2022},
  month = {Aug},
  publisher = {American Physical Society},
  doi = {10.1103/PRXQuantum.3.037001},
  url = {https://link.aps.org/doi/10.1103/PRXQuantum.3.037001}
}

@article{Simakov.FFF,
  title = {Coupler Microwave-Activated Controlled-Phase Gate on Fluxonium Qubits},
  author = {Simakov, Ilya A. and Mazhorin, Grigoriy S. and Moskalenko, Ilya N. and Abramov, Nikolay N. and Grigorev, Alexander A. and Moskalev, Dmitry O. and Pishchimova, Anastasiya A. and Smirnov, Nikita S. and Zikiy, Evgeniy V. and Rodionov, Ilya A. and Besedin, Ilya S.},
  journal = {PRX Quantum},
  volume = {4},
  issue = {4},
  pages = {040321},
  numpages = {11},
  year = {2023},
  month = {Nov},
  publisher = {American Physical Society},
  doi = {10.1103/PRXQuantum.4.040321},
  url = {https://link.aps.org/doi/10.1103/PRXQuantum.4.040321}
}

@article{zhao.scalable_FTF_architecture.2025,
  title = {Scalable fluxonium-qubit architecture with tunable interactions between noncomputational levels},
  author = {Zhao, Peng and Zhao, Guming and Li, Shaowei and Zha, Chen and Gong, Ming},
  journal = {Phys. Rev. Appl.},
  volume = {25},
  issue = {4},
  pages = {044072},
  year = {2026},
  month = {Apr},
  publisher = {American Physical Society},
  doi = {10.1103/mfjl-5rk6},
  url = {https://link.aps.org/doi/10.1103/mfjl-5rk6},
  eprint = {2504.09888},
  archivePrefix = {arXiv},
  primaryClass = {quant-ph}
}

@article{moskalenko.gate.2022,
   title={High fidelity two-qubit gates on fluxoniums using a tunable coupler},
   volume={8},
   ISSN={2056-6387},
   url={http://dx.doi.org/10.1038/s41534-022-00644-x},
   DOI={10.1038/s41534-022-00644-x},
   number={1},
   journal={npj Quantum Information},
   publisher={Springer Science and Business Media LLC},
   author={Moskalenko, Ilya N. and Simakov, Ilya A. and Abramov, Nikolay N. and Grigorev, Alexander A. and Moskalev, Dmitry O. and Pishchimova, Anastasiya A. and Smirnov, Nikita S. and Zikiy, Evgeniy V. and Rodionov, Ilya A. and Besedin, Ilya S.},
   year={2022},
   month=nov }

@article{MIT.FTF,
  title = {High-Fidelity, Frequency-Flexible Two-Qubit Fluxonium Gates with a Transmon Coupler},
  author = {Ding, Leon and Hays, Max and Sung, Youngkyu and Kannan, Bharath and An, Junyoung and Di Paolo, Agustin and Karamlou, Amir H. and Hazard, Thomas M. and Azar, Kate and Kim, David K. and Niedzielski, Bethany M. and Melville, Alexander and Schwartz, Mollie E. and Yoder, Jonilyn L. and Orlando, Terry P. and Gustavsson, Simon and Grover, Jeffrey A. and Serniak, Kyle and Oliver, William D.},
  journal = {Phys. Rev. X},
  volume = {13},
  issue = {3},
  pages = {031035},
  numpages = {24},
  year = {2023},
  month = {Sep},
  publisher = {American Physical Society},
  url = {https://link.aps.org/doi/10.1103/PhysRevX.13.031035}
}

@misc{Fors.ZZ_theory.2024,
      title={Comprehensive explanation of ZZ coupling in superconducting qubits}, 
      author={Simon Pettersson Fors and Jorge Fernández-Pendás and Anton Frisk Kockum},
      year={2024},
      eprint={2408.15402},
      archivePrefix={arXiv},
      primaryClass={quant-ph},
      url={https://arxiv.org/abs/2408.15402}, 
}

@article{Ciani.Scaling,
  title = {Microwave-activated gates between a fluxonium and a transmon qubit},
  author = {Ciani, A. and Varbanov, B. M. and Jolly, N. and Andersen, C. K. and Terhal, B. M.},
  journal = {Phys. Rev. Res.},
  volume = {4},
  issue = {4},
  pages = {043127},
  numpages = {20},
  year = {2022},
  month = {Nov},
  publisher = {American Physical Society},
  doi = {10.1103/PhysRevResearch.4.043127},
  url = {https://link.aps.org/doi/10.1103/PhysRevResearch.4.043127}
}

@article{nielsen2002simple,
    title = {A simple formula for the average gate fidelity of a quantum dynamical operation},
    journal = {Physics Letters A},
    volume = {303},
    number = {4},
    pages = {249-252},
    year = {2002},
    doi = {https://doi.org/10.1016/S0375-9601(02)01272-0},
    url = {https://www.sciencedirect.com/science/article/pii/S0375960102012720},
    author = {Michael A Nielsen},
}

@article{mazhorin2025impact,
    author = {Mazhorin, Grigoriy S. and Kugut, Andrei A. and Polyanskiy, Artyom M. and Moskalenko, Ilya N. and Simakov, Ilya A.},
    title = {Impact of qubit anharmonicity on near-resonant Rabi oscillations},
    journal = {Applied Physics Letters},
    volume = {126},
    number = {20},
    pages = {204002},
    year = {2025},
    month = {05},
    issn = {0003-6951},
    doi = {10.1063/5.0263783},
    url = {https://doi.org/10.1063/5.0263783},
}

@article{marques2022logical,
  title={Logical-qubit operations in an error-detecting surface code},
  author={Marques, J Ferreira and Varbanov, BM and Moreira, MS and Ali, Hany and Muthusubramanian, Nandini and Zachariadis, Christos and Battistel, Francesco and Beekman, Marc and Haider, Nadia and Vlothuizen, Wouter and others},
  journal={Nature Physics},
  volume={18},
  number={1},
  pages={80--86},
  year={2022},
  publisher={Nature Publishing Group UK London},
  doi={10.1038/s41567-021-01423-9}
}

@article{erhard2021entangling,
  title={Entangling logical qubits with lattice surgery},
  author={Erhard, Alexander and Poulsen Nautrup, Hendrik and Meth, Michael and Postler, Lukas and Stricker, Roman and Stadler, Martin and Negnevitsky, Vlad and Ringbauer, Martin and Schindler, Philipp and Briegel, Hans J and others},
  journal={Nature},
  volume={589},
  number={7841},
  pages={220--224},
  year={2021},
  publisher={Nature Publishing Group UK London},
  doi={10.1038/s41586-020-03079-6}
}

@article{besedin2025realizinglatticesurgerydistancethree,
      title={Realizing lattice surgery on two distance-three repetition codes with superconducting qubits}, 
      author={Besedin, Ilya and Kerschbaum, Michael and Knoll, Jonathan and Hesner, Ian and Bödeker, Lukas and Colmenarez, Luis and Hofele, Luca and Lacroix, Nathan and Hellings, Christoph and Swiadek, François and Flasby, Alexander and Bahrami Panah, Mohsen and Colao Zanuz, Dante and Müller, Markus and Wallraff, Andreas},
      journal={Nature Physics},
      year={2026},
      publisher={Nature Publishing Group UK London},
      doi={10.1038/s41567-025-03090-6},
      url={https://doi.org/10.1038/s41567-025-03090-6},
      eprint={2501.04612},
      archivePrefix={arXiv},
      primaryClass={quant-ph}
}

@article{acharya2024quantumerrorcorrectionsurface,
    title={Quantum error correction below the surface code threshold}, 
    author={Rajeev Acharya and Laleh Aghababaie-Beni and Igor Aleiner and Trond I. Andersen and Markus Ansmann and Frank Arute and Kunal Arya and Abraham Asfaw and Nikita Astrakhantsev and Juan Atalaya and Ryan Babbush and others},
    collaboration = {Google AI Quantum},
    journal={Nature},
    volume={638},
    pages={920–926},
    year={2025},
    publisher={Nature Publishing Group},
    doi={10.1038/s41586-024-08449-y}
}

@article{PhysRevLett.129.030501,
  title = {Realization of an Error-Correcting Surface Code with Superconducting Qubits},
  author = {Zhao, Youwei and Ye, Yangsen and Huang, He-Liang and Zhang, Yiming and Wu, Dachao and Guan, Huijie and Zhu, Qingling and Wei, Zuolin and He, Tan and Cao, Sirui and others},
  journal = {Phys. Rev. Lett.},
  volume = {129},
  issue = {3},
  pages = {030501},
  numpages = {7},
  year = {2022},
  month = {Jul},
  publisher = {American Physical Society},
  doi = {10.1103/PhysRevLett.129.030501},
  url = {https://link.aps.org/doi/10.1103/PhysRevLett.129.030501}
}

@article{Krinner_2022,
	doi = {10.1038/s41586-022-04566-8},
	year = 2022,
	month = {may},
	publisher = {Springer Science and Business Media {LLC}},
	volume = {605},
	number = {7911},
	pages = {669--674},
	author = {Sebastian Krinner and Nathan Lacroix and Ants Remm and Agustin Di Paolo and Elie Genois and Catherine Leroux and Christoph Hellings and Stefania Lazar and Francois Swiadek and Johannes Herrmann and others},
	title = {Realizing repeated quantum error correction in a distance-three surface code},
	journal = {Nature}
}

@article{google2023suppressing,
  title={Suppressing quantum errors by scaling a surface code logical qubit},
  author={Rajeev Acharya and Igor Aleiner and Richard Allen and Trond I. Andersen and Markus Ansmann and Frank Arute and Kunal Arya and Abraham Asfaw and Juan Atalaya and Ryan Babbush and others},
  journal={Nature},
  volume={614},
  number={7949},
  pages={676--681},
  year={2023},
  publisher={Nature Publishing Group UK London},
  url={https://doi.org/10.1038/s41586-022-05434-1}
}

@article{Somoroff_Millisecond_2023,
  title = {Millisecond Coherence in a Superconducting Qubit},
  author = {Somoroff, Aaron and Ficheux, Quentin and Mencia, Raymond A. and Xiong, Haonan and Kuzmin, Roman and Manucharyan, Vladimir E.},
  journal = {Phys. Rev. Lett.},
  volume = {130},
  issue = {26},
  pages = {267001},
  numpages = {6},
  year = {2023},
  month = {Jun},
  publisher = {American Physical Society},
  doi = {10.1103/PhysRevLett.130.267001},
  url = {https://link.aps.org/doi/10.1103/PhysRevLett.130.267001}
}

@article{Koch_Charge_2007,
  title = {Charge-insensitive qubit design derived from the Cooper pair box},
  author = {Koch, Jens and Yu, Terri M. and Gambetta, Jay and Houck, A. A. and Schuster, D. I. and Majer, J. and Blais, Alexandre and Devoret, M. H. and Girvin, S. M. and Schoelkopf, R. J.},
  journal = {Phys. Rev. A},
  volume = {76},
  issue = {4},
  pages = {042319},
  numpages = {19},
  year = {2007},
  month = {Oct},
  publisher = {American Physical Society},
  doi = {10.1103/PhysRevA.76.042319},
  url = {https://link.aps.org/doi/10.1103/PhysRevA.76.042319}
}

@article{Manucharyan_2009,
	doi = {10.1126/science.1175552},
  
	url = {https://doi.org/10.1126/science.1175552},
  
	year = 2009,
	month = {oct},
  
	publisher = {American Association for the Advancement of Science ({AAAS})},
  
	volume = {326},
  
	number = {5949},
  
	pages = {113--116},
  
	author = {Vladimir E. Manucharyan and Jens Koch and Leonid I. Glazman and Michel H. Devoret},
  
	title = {Fluxonium: Single Cooper-Pair Circuit Free of Charge Offsets},
  
	journal = {Science}
}

@article{Bao_2021,
  title = {Fluxonium: An Alternative Qubit Platform for High-Fidelity Operations},
  author = {Bao, Feng and Deng, Hao and Ding, Dawei and Gao, Ran and Gao, Xun and Huang, Cupjin and Jiang, Xun and Ku, Hsiang-Sheng and Li, Zhisheng and Ma, Xizheng and Ni, Xiaotong and Qin, Jin and Song, Zhijun and Sun, Hantao and Tang, Chengchun and Wang, Tenghui and Wu, Feng and Xia, Tian and Yu, Wenlong and Zhang, et al.},
  journal = {Phys. Rev. Lett.},
  volume = {129},
  issue = {1},
  pages = {010502},
  numpages = {6},
  year = {2022},
  month = {Jun},
  publisher = {American Physical Society},
  doi = {10.1103/PhysRevLett.129.010502},
  url = {https://link.aps.org/doi/10.1103/PhysRevLett.129.010502}
}

@article{Ficheux2021,
  title = {Fast Logic with Slow Qubits: Microwave-Activated Controlled-Z Gate on Low-Frequency Fluxoniums},
  author = {Ficheux, Quentin and Nguyen, Long B. and Somoroff, Aaron and Xiong, Haonan and Nesterov, Konstantin N. and Vavilov, Maxim G. and Manucharyan, Vladimir E.},
  journal = {Phys. Rev. X},
  volume = {11},
  issue = {2},
  pages = {021026},
  numpages = {16},
  year = {2021},
  month = {May},
  publisher = {American Physical Society},
  doi = {10.1103/PhysRevX.11.021026},
  url = {https://link.aps.org/doi/10.1103/PhysRevX.11.021026}
}

@article{Arute_2019,
	title={Quantum supremacy using a programmable superconducting processor},
	volume={574},
	ISSN={1476-4687},
	DOI={10.1038/s41586-019-1666-5},
	number={7779},
	journal={Nature},
	publisher={Springer Science and Business Media LLC},
	author={Arute, Frank and Arya, Kunal and Babbush, Ryan and Bacon, Dave and Bardin, Joseph C. and Barends, Rami and Biswas, Rupak and Boixo, Sergio and Brandao, Fernando G. S. L. and Buell, David A. and others},
	year={2019}, pages={505–510} }

@article{Strong_Wu_2021,
  title = {Strong Quantum Computational Advantage Using a Superconducting Quantum Processor},
  author = {Wu, Yulin and Bao, Wan-Su and Cao, Sirui and Chen, Fusheng and Chen, Ming-Cheng and Chen, Xiawei and Chung, Tung-Hsun and Deng, Hui and Du, Yajie and Fan, Daojin and other},
  journal = {Phys. Rev. Lett.},
  volume = {127},
  issue = {18},
  pages = {180501},
  numpages = {7},
  year = {2021},
  month = {Oct},
  publisher = {American Physical Society},
  doi = {10.1103/PhysRevLett.127.180501},
  url = {https://link.aps.org/doi/10.1103/PhysRevLett.127.180501}
}

@article{morvan2024phase,
  title={Phase transitions in random circuit sampling},
  author={Morvan, Alexis and Villalonga, B and Mi, X and Mandr{\`a}, S and Bengtsson, A and Klimov, PV and Chen, Z and Hong, S and Erickson, C and Drozdov, IK and others},
  journal={Nature},
  volume={634},
  number={8033},
  pages={328--333},
  year={2024},
  publisher={Nature Publishing Group UK London},
  doi={10.1038/s41586-024-07998-6}
}

@article{gao2025establishing,
  title={Establishing a new benchmark in quantum computational advantage with 105-qubit zuchongzhi 3.0 processor},
  author={Gao, Dongxin and Fan, Daojin and Zha, Chen and Bei, Jiahao and Cai, Guoqing and Cai, Jianbin and Cao, Sirui and Chen, Fusheng and Chen, Jiang and Chen, Kefu and others},
  journal={Phys. Rev. Lett.},
  volume={134},
  number={9},
  pages={090601},
  year={2025},
  publisher={APS},
  doi={10.1103/PhysRevLett.134.090601}
}

@article{google2025observation,
  author={Dmitry A. Abanin and Rajeev Acharya and Laleh Aghababaie-Beni and Georg Aigeldinger and Ashok Ajoy and Ross Alcaraz and Igor Aleiner and Trond I. Andersen and Markus Ansmann and Frank Arute, et al.},
  title={Observation of constructive interference at the edge of quantum ergodicity},
  journal={Nature},
  volume={646},
  number={8086},
  pages={825--830},
  year={2025},
  publisher={Nature Publishing Group UK London},
  doi={10.1038/s41586-025-09526-6}
}

@inproceedings{wang2024precision,
  title={Precision frequency tuning of tunable transmon qubits using alternating-bias assisted annealing},
  author={Wang, Xiqiao and Howard, Joel and Sete, Eyob A and Stiehl, Greg and Kopas, Cameron and Poletto, Stefano and Wu, Xian and Field, Mark and Sharac, Nicholas and Eckberg, Christopher and others},
  booktitle={2024 IEEE International Conference on Quantum Computing and Engineering (QCE)},
  volume={1},
  pages={1315--1323},
  year={2024},
  organization={IEEE}
}

@article{zhang2022high,
  title={High-performance superconducting quantum processors via laser annealing of transmon qubits},
  author={Zhang, Eric J and Srinivasan, Srikanth and Sundaresan, Neereja and Bogorin, Daniela F and Martin, Yves and Hertzberg, Jared B and Timmerwilke, John and Pritchett, Emily J and Yau, Jeng-Bang and Wang, Cindy and others},
  journal={Science Advances},
  volume={8},
  number={19},
  pages={eabi6690},
  year={2022},
  publisher={American Association for the Advancement of Science}
}

@misc{mazhorin2026nativecczgatefluxonium,
      title={Native CCZ Gate with Fluxonium Qubits and a Microwave-Driven Coupler}, 
      author={Grigoriy S. Mazhorin and Tatyana A. Chudakova and Alena S. Kazmina and Nikolai G. Berezkin and Arina V. Zotova and Artyom M. Polyanskiy and Nikolay N. Abramov and Mikhail A. Tarkhov and Alexander M. Mumlyakov and Igor V. Trofimov and Elizaveta A. Krivko and Nikita Yu. Rudenko and Maxim V. Chichkov and Vladimir I. Chichkov and Ilya A. Simakov},
      year={2026},
      eprint={2607.27094},
      archivePrefix={arXiv},
      primaryClass={quant-ph},
      url={https://arxiv.org/abs/2607.27094}, 
}

@misc{chan2026system,
  title={System-Level Design of Scalable Fluxonium Quantum Processors with Double-Transmon Couplers},
  author={Guo Xuan Chan and Wangwei Lan and Tenghui Wang and Xizheng Ma and Chunqing Deng and Lijing Jin},
  year={2026},
  eprint={2604.26373},
  archivePrefix={arXiv},
  primaryClass={quant-ph},
  url={https://arxiv.org/abs/2604.26373},
}

@misc{zwanenburg2026crosstalkmultiqubitfluxoniumarchitectures,
      title={Crosstalk in Multi-Qubit Fluxonium Architectures with Transmon Couplers}, 
      author={Martijn F. S. Zwanenburg and Christian Kraglund Andersen},
      year={2026},
      eprint={2603.09870},
      archivePrefix={arXiv},
      primaryClass={quant-ph},
      url={https://arxiv.org/abs/2603.09870}, 
}

@misc{huang2026explorationfluxoniumparameterscapacitive,
      title={Exploration of Fluxonium Parameters for Capacitive Cross-Resonance Gates}, 
      author={Eugene Y. Huang and Christian Kraglund Andersen},
      year={2026},
      eprint={2603.17936},
      archivePrefix={arXiv},
      primaryClass={quant-ph},
      url={https://arxiv.org/abs/2603.17936}, 
}
